\documentclass[11pt,preprint]{aastex}

\shorttitle{SF Activity and LSS of Galaxy ODs at $z \sim 1$}
\shortauthors{Lee, S.-K. et al.}

\begin{document}

\title{More Connected, More Active:\\Galaxy Clusters and Groups at $z \sim 1$ and The Connection between Their Quiescent Galaxy Fractions and Large-scale Environments}

\author{Seong-Kook Lee\altaffilmark{1,5}, Myungshin Im\altaffilmark{1,6}, 
Minhee Hyun\altaffilmark{1}, Bomi Park\altaffilmark{2}, 
Jae-Woo Kim\altaffilmark{3}, Dohyeong Kim\altaffilmark{4}, and 
Yongjung Kim\altaffilmark{1}}

\altaffiltext{1}{Center for the Exploration of the Origin of the Universe, 
Department of Physics and Astronomy, Seoul National University, 
Seoul 08826, Korea}
\altaffiltext{2}{Department of Astronomy and Space Science, 
Kyung Hee University, Yongin 17104, Korea}
\altaffiltext{3}{Korea Astronomy and Space Science Institute, Daejeon 34055, Korea}
\altaffiltext{4}{Kavli Institute for Astronomy and Astrophysics, Peking University, Beijing 100871, China}
\altaffiltext{5}{s.joshualee@gmail.com}
\altaffiltext{6}{mim@astro.snu.ac.kr}

\begin{abstract}
High-redshift galaxy clusters, unlike local counterparts, show 
diverse star formation activities. 
However, it is still unclear what keeps some of the high-redshift 
clusters active in star formation.
To address this issue, we performed a multi-object spectroscopic 
(MOS) observation of 226 high-redshift ($0.8 < z < 1.3$) galaxies 
in galaxy cluster candidates and the areas surrounding them.
Our spectroscopic observation reveals six to eight clusters/groups 
at $z \sim 0.9$ and $z\sim 1.3$.
The redshift measurements 
demonstrate the reliability of our photometric redshift 
measurements, which in turn gives credibility 
for using photometric redshift members for the analysis of 
large-scale structures (LSSs).  
Our investigation of the 
large-scale environment ($\sim$10 Mpc) surrounding each galaxy 
cluster reveals LSSs --- structures up to 
$\sim 10$ Mpc scale --- around many of, but not all, 
the confirmed overdensities and the cluster candidates.
We investigate the correlation between quiescent galaxy fraction of 
galaxy overdensties and their surrounding LSSs, with a larger sample 
of $\sim 20$ overdensities including photometrically selected 
overdensities at $0.6 < z < 0.9$.
Interestingly, galaxy overdensities embedded within these extended 
LSSs show a lower fraction of quiescent galaxies ($\sim 20\%$) 
than isolated ones at similar redshifts 
(with a quiescent galaxy fraction of $\sim 50 \%$).
Furthermore, we find a possible indication that clusters/groups with a high quiescent 
galaxy fraction are more centrally concentrated. 
Based on these results, we suggest that LSSs are the main 
reservoirs of gas and star-forming galaxies to keep galaxy 
clusters fresh and extended in size at $z \sim 1$.
\end{abstract}


\section{Introduction}

In the local universe, galaxies show distinct properties in 
different environments, such as morphology 
\citep[e.g.,][]{bam09,ko12}, color, 
star-formation rate (SFR), 
and size \citep[e.g.,][]{coo12,pap12,str13,yoo17}.
Especially, star-formation (SF) activities are clearly 
distinguished depending on their environment 
\citep[e.g.,][]{lew02,kau04,bla05,bam09}, 
in the sense that galaxies in field 
are on average bluer or more actively forming stars than those in clusters. 

This picture of environmental dependence of galaxy properties becomes 
more complicated at higher redshifts. 
The SF activities in galaxy clusters are found to be 
evolving fast from $z=0$ to 1 with a higher SF rate (SFR) at 
higher redshifts \citep{sai08,bai09,tra10,web13}, and 
the evolution seems faster in clusters than in the field \citep{alb14}.
Consequently, the environmental dependence becomes weaker or even 
reversed with increasing redshifts  
\citep[e.g.,][]{but84,elb07,coo08,muz12,sco13}, 
although on average, the environmental dependence persists 
out to $z \sim 1$ \citep{pat09a,qua12,van13,jia18}.

The weakened environmental dependence is due to dense environments 
having a variety of SF activities. 
Recently, \citet[][hereafter, L15]{lee15} analyzed the evolution 
in the fraction of quiescent galaxies in clusters and in field at 
$0.5 < z < 2$ using the multi-wavelength data of the Ultra Deep 
Survey \citep[UDS;][]{alm07}. 
Figure~\ref{qfrac} shows the quiescent galaxy fraction of individual 
galaxy clusters found in the UDS field 
as well as the mean values for field galaxies from L15.
The quiescent fraction is measured for galaxies 
with a stellar mass greater than $10^{9.1}$ M$_{\odot}$. 
While local galaxy clusters have similar quiescent galaxy fractions 
of $\sim 0.7$ \citep{bal06}, we can see a broad range in the quiescent 
galaxy fraction of galaxy clusters at $z > 0.6$ with values 
ranging from 0.1 to 0.8. 
This cluster-by-cluster variation in the quiescent galaxy 
fraction persists up to redshifts as high as $z \sim 1.6$, 
at which point the difference between clusters and the field starts 
to disappear \citep[see also,][]{hay11,bro13,alb16,cook16}.
\citet{hay19}, using the data from Hyper Suprime-Cam (HSC) Subaru Strategic 
Program (SSP) survey \citep{aih18}, 
also found that there is a variation in the quiescent 
galaxy fraction among $z \sim 0.9$ clusters which are embedded in the 
large-scale structure (LSS) of the CL1604 supercluster \citep{lem12}.
The evolution in the quiescent fraction, especially out to 
$z \sim 1$, is closely related to the SF quenching 
mechanism \citep[e.g., L15;][]{pen10}, and close investigation 
of these high redshift clusters can possibly offer insight 
on what controls SF activities in dense environments.
At $z > 1$, more clusters are being identified 
with strong SF activity \citep{bro13,alb14,alb16,shi18}.
The insight gathered from $z \sim 1$ clusters can help understand 
the underlying mechanism for the SF activity in clusters at $z > 1$.

The large-scale environment of clusters, 
which is well beyond the scale of individual galaxy clusters, 
can also be considered as a key factor that regulates 
SF in clusters. 
At $z \sim 0.23$, \citet{fad08} found an enhancement in the 
fraction of SF galaxies within filamentary structures outside 
the Abell 1763 cluster, speculating that these filaments 
are feeding the cluster with the infall of galaxies and galaxy groups.
This higher SF fraction may be due to cold gas retention 
or enhanced galaxy-galaxy interaction within these structures. 
Studying the cluster A3266, \citet{bai09} suggested 
that the similarity of infrared (IR) luminosity function between 
field and clusters may due to the continual replenishment of 
star-forming galaxies from the field to the cluster.
\citet{mah12} also found the increase in the SF activity on 
the outskirts of nearby ($0.02 \leq z \leq 0.15$) clusters.
Interestingly, this increase in SF is more commonly seen in 
dynamically unrelaxed clusters, which are in the course of 
assembly through falling from surrounding filamentary structures.
At intermediate redshifts, \citet{gea11} studied large-scale 
(15 Mpc scale) environments around $z \sim 0.55$ galaxy clusters.  
They showed that, outside of the cluster, 
there are intermediate density regions where the fraction of 
SF galaxies is enhanced, while the mean SFR shows no significant 
environmental dependence. 

At higher redshifts, \citet{lub09}, studying the large-scale 
environment of two clusters at similar redshift $z \sim 0.8$, 
found that one of them is more isolated and has 
a large quiescent galaxy population, while another is in the 
process of four-way group-group merger with a high fraction 
of star forming galaxies ($\sim 80\%$).
And, \citet{koy08} found that the fraction of the 
IR bright galaxies, supposedly star-forming galaxies, 
is enhanced at the medium density environment of a 
$z = 0.81$ cluster, 
suggesting enhanced SF activities in filaments and groups.  
At $z \sim 0.85$, \citet{pin13} showed that the star-forming 
galaxy fraction is higher at intermediate density regions 
--- the outskirts of clusters, groups, and filaments --- 
compared to higher (cluster-core region) or lower (field) density 
regions. 
\citet{dar14} also showed that the fraction of H$\alpha$ emitters 
is enhanced in the filamentary structures at $z \sim 0.84$. 
In summary, intermediate density regions, such as filaments around 
clusters, seem to provide an environment that is 
favourable to sustain SF activities, possibly via cold gas 
accretion/replenishment \citep[e.g.,][]{kle17}, 
than field or cluster environments.
Connected with these filamentary, intermediate density structures, 
galaxy clusters can be supplied with fresh SF galaxies from these 
structures, which increases the fraction of SF galaxies in clusters 
\citep[e.g.,][]{ell01,ebe04,tra05,sai08,koy18}.

In this regard, \citet{ara16} suggested a cosmic web detachment 
(CWD) model as a mechanism to quench SF in clusters. 
According to this model, the SF activities are maintained to some 
extent in clusters due to the supply of cold gas and SF galaxies 
through the cosmic web. 
However, after the cosmic web is detached, 
galaxies in clusters lose the driver for continuing SF. 
This model can be tested effectively at $z \sim 1$ clusters, 
where we can find clusters with many distinct levels 
of quiescent galaxy fraction (e.g., L15). 

To understand what makes the observed cluster-by-cluster variation 
in the SF property, 
we performed multi-object spectroscopy of 226 galaxies in the UDS 
field to study clusters/groups at $z \sim 1$ and LSSs 
associated with them. 
With the data, we study the quiescent fraction of galaxies in 
overdensities at $0.6 < z < 1.3$ and examine how the presence of 
surrounding LSSs influences the SF activities in clusters/groups.
The quiescent galaxy fraction is a useful 
indicator of SF activity when there is a lack of deep, high 
resolution IR data, by not being affected by missing a small 
number of highly obscured (therefore high SFR) galaxies in 
the sample, nor uncertainties in deriving the exact values 
of SFRs in the absence of IR data.

In Section~\ref{datas}, we explain the data and sample and the 
spectroscopic observation 
with the Magellan telescope as well as the data reduction 
and redshift measurement. 
Based on these spectroscopic data, we show the reliability of 
our photometric redshift measurement and stellar population 
estimation from spectral energy distribution (SED)-fitting in 
Section~\ref{accu}. 
We present properties of spectroscopically studied 
galaxy clusters/groups 
as well as LSSs near these clusters in Sections~\ref{spiclgr} 
and \ref{ESS}, respectively. 
Some of these are newly discovered structures. 
In Section~\ref{cause}, we discuss the causes of variation 
in SF properties of individual overdensities. 
Then, we summarize our results in Section~\ref{summa}.
We adopt the standard flat $\rm{\Lambda}$CDM 
cosmology with ($\Omega_{m}, \Omega_{\Lambda}$) = (0.3, 0.7) 
and $H_{0}$ = 70 {\rm km s$^{-1}$ Mpc$^{-1}$}, 
which is supported by observations in the past 
decades \citep[e.g.,][]{im97}.
All magnitudes are given in the AB magnitude system \citep{oke74}.

\section{Data}\label{datas}

\subsection{Sample}\label{sample}

Our sample is drawn from high-redshift ($0.5 < z < 2.0$) cluster candidates 
that are selected from a 0.77 deg$^2$ area of the entire UDS field. 
Our cluster-finding algorithm is described in L15 \citep[also in][]{kan15}. 
Briefly, galaxy cluster candidates are selected 
as overdense regions, where the galaxy surface number density 
is higher than the field value by more than 4 $\sigma$ at 
each redshift bin. Photometric redshifts ($z_{phot}$s) of galaxies are 
estimated to the accuracy of 
$\Delta z$/($1+z$) = 0.028, using EAZY \citep{bra08}.  
The redshift bins have the width of $\pm 0.028 \times (1+z)$, and 
the step size is 0.02.  
Stellar masses and SFRs of galaxies are 
measured through SED-fitting, 
assuming the \citet{cha03} IMF and delayed star-formation 
histories (SFHs) \citep[e.g.,][]{lee10}. 
We applied the stellar mass cut of log 
(M$_{*}$/M$_{\odot}$) $\geq 9.1$.
Cluster member galaxies are those within a 1 Mpc radius and 
within the redshift range of the mean photometric redshift uncertainty 
(i.e., $\pm 0.028 \times (z_{cl}+1)$), where $z_{cl}$ is the 
cluster redshift. 
Refer to L15 for a more detailed explanation about the 
procedure of galaxy cluster selection and the estimation of 
stellar population property through SED-fitting. 
We note that, unlike the methods that rely on the existence 
of a well-developed red sequence, our cluster finding method is 
not biased against active star-forming clusters, 
which is crucial in this study (also see \citet{tre07} 
and \citet{eis08} for other photometric redshift-based cluster 
finding methods).
There are 46 cluster candidates, and they have halo masses of 
log ($M_{200}/M_{\odot}$) in the range of [13.4,14.2] (L15), 
where the $M_{200}$ values are obtained by converting 
the sum of stellar masses of the member galaxy candidates by 
dividing it with a factor of 0.013, where the factor of 0.013 is the 
mean ratio of the sum of the stellar mass to the X-ray based cluster 
mass of 13 clusters that are in common with L15 and the sample of 
\citet[][hereafter, F10]{fin10}. 

\subsection{Spectroscopy Targets}\label{starget}

\begin{figure}
 \includegraphics{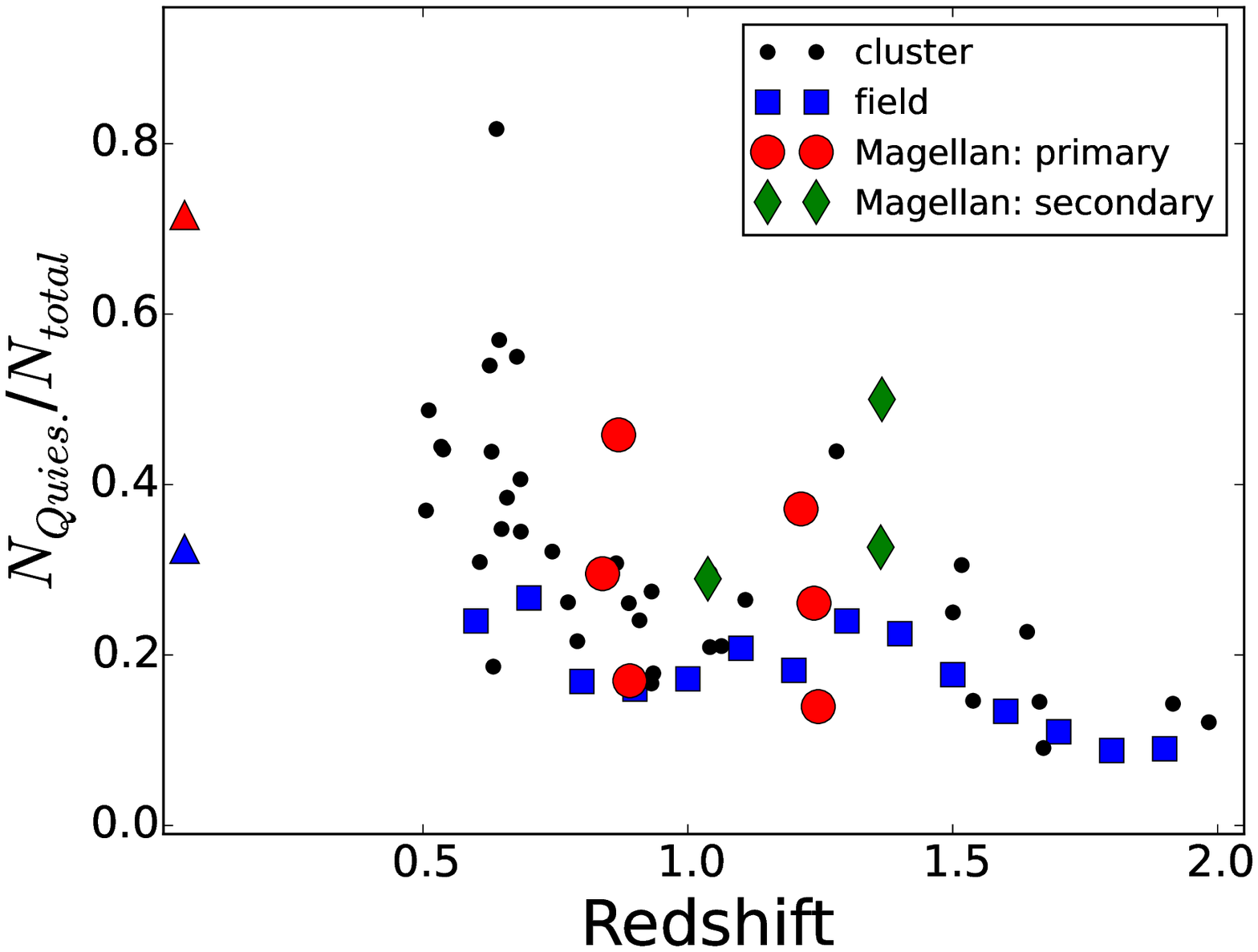}
 \caption{The fraction of quiescent galaxies in clusters (small black circles) and field 
(blue squares). The main target clusters of our Magellan observation are shown as large 
red circles, and green diamonds show secondary Magellan targets. 
The quiescent galaxy fraction is defined as the fraction of 
galaxies whose sSFR (SFR/M$_{*}$) values are smaller than 
1/($3 \times t(z)$) yr$^{-1}$ and whose stellar masses are greater 
than $10^{9.1}$ M$_{\odot}$. 
The red and blue triangles are the corresponding local values for 
the cluster and field, respectively from \citet{bal06}.}
 \label{qfrac}
\end{figure}

Among the 46 cluster candidates, we selected targets for 
the multi-object spectroscopy using the Magellan Baade 6.5 m Telescope 
at Las Campanas Observatory, Chile, based on 
(1) the SF property --- covering a wide range in quiescent 
galaxy fraction, 
(2) redshift --- residing at similar redshifts at $z \sim 1$, and 
(3) their vicinity with each other --- to be within the field-of-view 
(FoV) of the Inamori Magellan Areal Camera and Spectrograph 
\citep[IMACS;][]{dre11} instrument. 
Based on these criteria, we select six cluster candidates 
that show different quiescent galaxy fractions 
while residing at similar redshifts, $z \sim 0.9$ and 1.2 
as our main targets. 
These target cluster candidates are shown as red circles in Figure~\ref{qfrac}. 
As can be seen in this figure, some of the target cluster candidates   
have a similar fraction of quiescent galaxies with field galaxies 
at a similar epoch, 
while some have significantly larger quiescent fractions.   
Then, three additional cluster candidates, which are 
at $z \sim 1.0$ and 1.4, 
are also selected as spectroscopic targets because these 
clusters are within the FoV 
of Magellan/IMACS ($27 \arcmin$ diameter circle). 
These cluster candidates are shown as the green diamonds in Figure~\ref{qfrac}.  
In addition, field galaxies that are at similar photometric 
redshift ranges with the selected cluster targets and within 
the same Magellan FoV are also included.
In summary, we observed 6 clusters at $z \sim 0.9$ and 1.2 with on average 
$\sim 25$ slits assigned per cluster, and 3 additional clusters at 
$z \sim 1.0$ and 1.4 with much less slit assignments ($\sim 13$ per cluster). 
The remaining slits were assigned to field galaxies and other objects 
of interest.

\begin{figure*}
 \centering
 \includegraphics[scale=0.5]{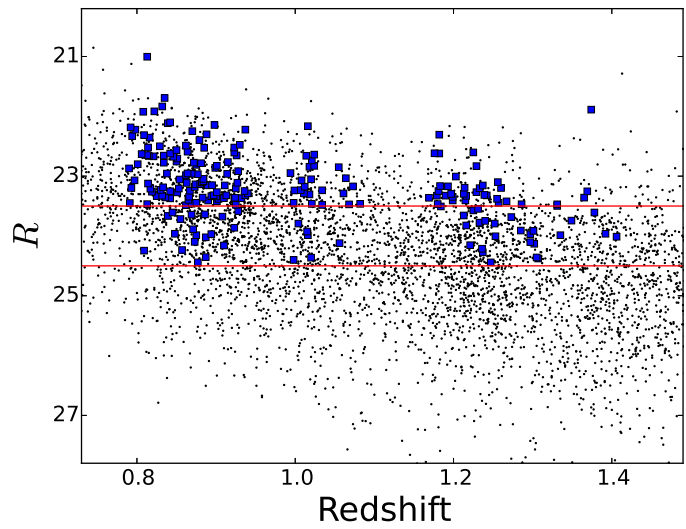}
\caption{$R$-band magnitude versus redshift of galaxies in the 
Magellan/IMACS field of view (the black dots). 
The target galaxies for the spectroscopic observation 
are shown as the blue squares. 
The magnitude limits for the target selection, $R < 23.5$ 
(the first priority) and $R < 24.5$ (the second priority) are 
indicated as the red lines.}
 \label{zrmg}
\end{figure*}

Target galaxies are selected based on their $R$-band magnitude.
First, only galaxies brighter than $R \leq 24.5$ are selected.
Then, among these, higher priority is assigned to galaxies 
that are brighter than $R \leq 23.5$ in the mask generation procedure.     
Figure~\ref{zrmg} shows the photometric redshifts and 
$R$-band magnitudes of galaxies within the FoV of Magellan/IMACS 
(shown as black points) 
as well as the target galaxies (shown as blue squares). 

In Table~\ref{cltarget}, we list our target cluster candidates, 
along with the number of member galaxy candidates observed and the number 
of the observed galaxies with spectroscopic redshift identification.

\subsection{Spectroscopic Observation and Data Reduction}

We prepared two separate masks for our observation 
of the same sky region with slightly different slit assignments. 
Basically, the same faint ($R \geq 23.5$) galaxies are included 
in both masks, while different bright galaxies are targeted in 
different masks.
In addition, $i$-dropout candidate galaxies are included 
with the lowest priority for serendipitous discovery of $z \sim 6$ 
bright star-forming galaxies or quasars. 
About 250 slits are assigned for the targets above in each mask, 
among which $43\%$ are for the cluster galaxy candidates, $39\%$ 
for the field galaxies, and $17\%$ for the $i$-band dropout candidates. 
One slit in each mask is assigned for a standard (K-type) star.

The multi-object spectroscopic (MOS) observations 
were carried out using the IMACS f/2 camera on the Magellan/Baade telescope 
in September 2014 and in September 2015. 
We prepared two masks for our observation as explained 
in the above paragraph.
The slit length and width were set at 
$6 \arcsec \times 1 \arcsec$, respectively. 
The total exposure time for each mask was 2.5 hours, 
which was divided into five 30-minute exposures. 
The average seeing was $0.8 \arcsec$ and $1.1 \arcsec$, 
for the 2014 and 2015 runs, respectively.
We used the Grism with 200 lines per millimeter and the 
WB6300-9500 filter to provide a wavelength coverage of 
630 to 950 nm at the spectral resolution of 
R$\sim 650$, which is suitable to catch major spectroscopic 
features of galaxies of our interest. 

Data reduction was done using the Carnegie Observatories 
System for Multi Object Spectroscopy (COSMOS), which is a 
software package offered from Carnegie Observatories to 
reduce spectroscopic data taken with IMACS and LDSS3 instruments. 
For wavelength calibration, He, Ne, Ar arc frames were used. 
Following the standard COSMOS routine, we did flat fielding and 
bias subtraction from science frames. 
After the sky subtraction, we extracted two-dimensional (2D) spectra 
and stacked them onto one combined 2D spectrum. 
The bulk of cosmic rays were removed in the process. 
For the faint ($R \geq 23.5$) objects, 
we stacked all 10 single-exposed 2D spectra 
from two masks for higher S/N. 
More details about the data reduction are described in 
the COSMOS cookbook (Villanueva 2014).
One dimensional (1D) spectra for each source were extracted 
from the combined 2D spectrum using the extraction 
width that corresponds to twice the average seeing value in 2014 
and 2015 each. 
With a standard star included in the mask, 
we did flux calibration. 

\subsection{Spectroscopic Redshift}

\begin{figure*}
 \centering
 \includegraphics[scale=0.8]{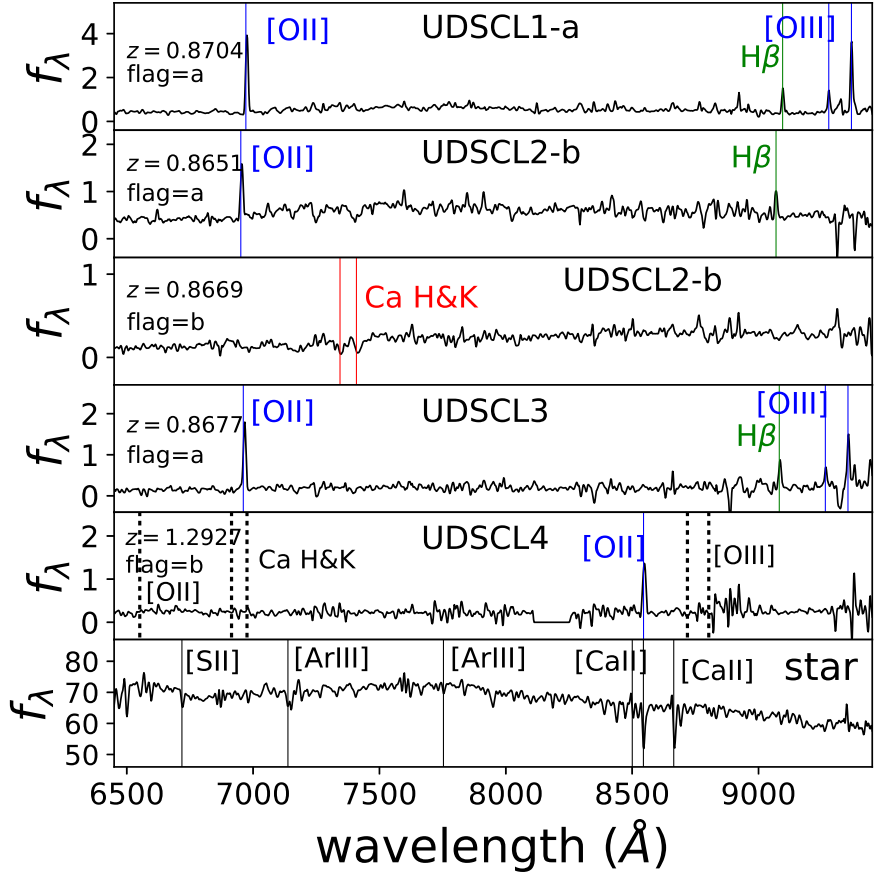}
\caption{Examples of Magellan/IMACS spectra of cluster member galaxies 
as well as a star. 
The redshift in each panel is the measured spectroscopic redshift for each galaxy. 
Several emission or absorption spectral features are shown as 
colored vertical lines. 
In the fifth panel, we show the spectrum of a $z = 1.293$ galaxy, 
based on the strong [\ion{O}{2}] emission line at $\lambda \sim 8545$ \AA. 
If this line was H$\beta$ (meaning the redshift of this galaxy 
was $z \sim 0.758$), other lines should have appeared at the wavelengths 
marked with black dotted lines (and labeled in black). 
The absence of these lines makes the redshift determination 
of this galaxy more reliable. 
In the bottom panel, we show a stellar spectrum.}
 \label{specexam}
\end{figure*}

From the extracted 1D spectra, 
we determined the redshifts of galaxies using the SpecPro software \citep{mas11}. 
The main spectral features for the redshift identification are 
[\ion{O}{2}] 3727, H${\beta}$, 
H${\gamma}$, and [\ion{O}{3}] emission lines, Ca H$\&$K and G band 
absorption lines, 
and the 4000 $\rm{\AA}$ break. 
We show example spectra of several confirmed cluster member 
galaxies in Figure~\ref{specexam}. 
In each panel, we show the determined spectroscopic redshift 
as well as the line features used in the redshift determination.  

We assigned quality flag `a' for galaxies whose redshift 
measurements are based on two or more emission or 
absorption lines. 
If the redshift determination is based on a single line feature 
(we consider Ca H$\&$K lines as a single feature), the 
quality flag `b' is given.  
For majority of galaxies at $z < 1.2$, we can determine the redshift 
based on multiple spectral features as well as the clear continuum shape 
(quality flag `a'). 
For many $z > 1.2$ galaxies, the redshift measurement was 
done based on a single feature, [\ion{O}{2}] $\lambda 3727$, 
while we can also see the Ca H$\&$K absorption lines in some cases. 
However, even when only a single emission line is seen in the 
spectrum, our redshift determination at $z > 1.2$ is reliable 
because (1) we do not see other significant emission features 
that should be seen if the given line is not [\ion{O}{2}], and 
(2) the photometric redshift measured from multi-band SEDs also 
matches well within the photometric redshift uncertainty. 
For example, for a galaxy shown in the fifth panel of 
Figure~\ref{specexam}, if the emission line at 
$\lambda \sim 8545$ \AA, which is identified as [\ion{O}{2}] (giving 
this galaxy a redshift of 1.293), were H${\beta}$ (i.e., at 
$z \sim 0.758$), [\ion{O}{2}] and [\ion{O}{3}] lines should have been seen 
at $\lambda \sim 6551$ \AA, and at 
$\lambda \sim 8720$ and 8804 \AA, as shown with the black 
dotted lines in this panel. 

\begin{figure*}
 \centering
 \includegraphics[scale=0.45]{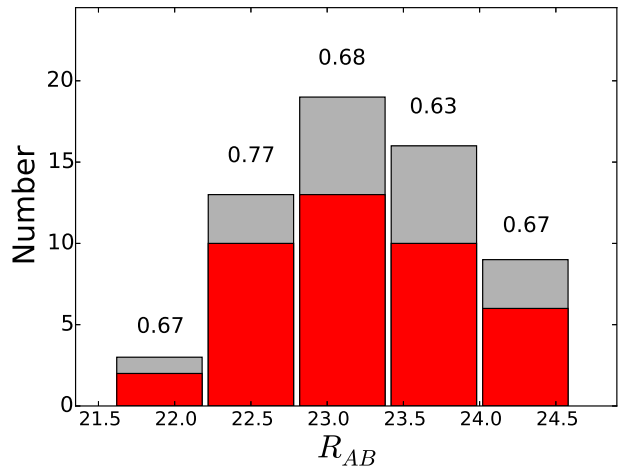}
\caption{The $R$-band magnitude dependent number distribution of cluster 
member galaxy candidates.
Gray bars show the number of member galaxy candidates 
of which redshifts are measured from our Magellan observation presented 
in this work. 
Red bars show the number of galaxies that are confirmed as the 
members of galaxy clusters. 
Numbers above gray bars are the fraction of confirmed galaxies as cluster 
members at each magnitude bin. 
As can be seen here, the $R$-band dependence is not significant.}
\label{confmg}
\end{figure*}

We determined the redshifts of 173 galaxies out of 274 targets, 
which gives the redshift determination rate (or success rate) 
of $\sim 63 \%$. 
If we exclude $i$-dropout candidates, the success rate increases 
to $\sim 76 \%$. 
Figure~\ref{confmg} shows the cluster member confirmation rate from 
spectroscopic redshift as a function of $R$-band magnitude.
Table~\ref{tab1} shows all the target galaxies of our Magellan 
observation, including those whose redshifts were not measured, 
as well as stars. 

\begin{figure*}
 \centering
  \includegraphics[scale=0.65]{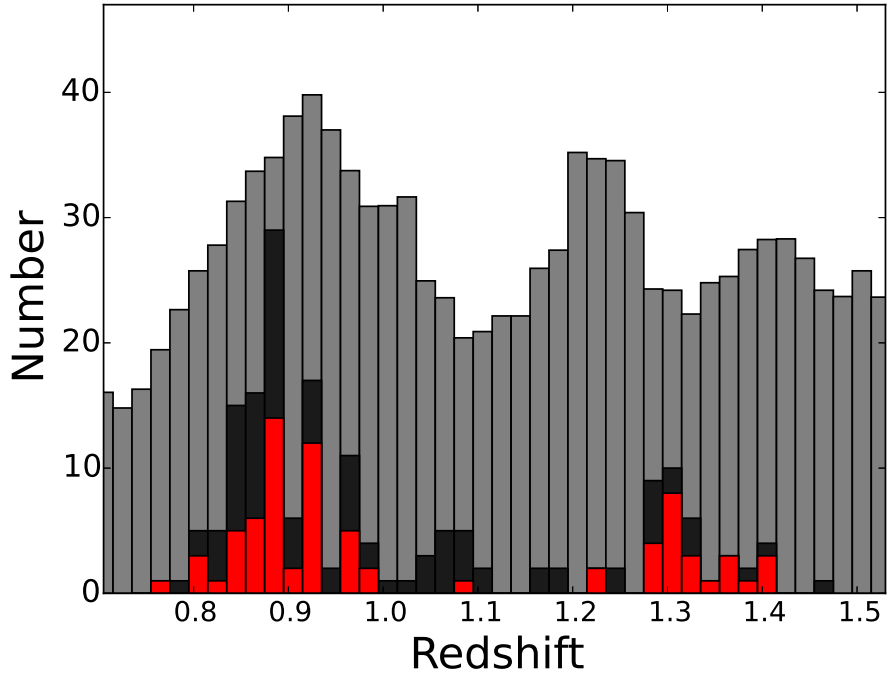}
\caption{Distribution of spectroscopic redshifts from Magellan observation. 
The red histogram shows the distribution of candidate cluster galaxies 
while the black histogram is for total (cluster$+$field) galaxies. 
Both distributions show broad peaks at $z \sim 0.9$ and $z \sim 1.3$ where our six main 
cluster targets reside (refer to Section~\ref{sample} for detail.). 
The light gray histogram shows the photometric redshift distribution of 
the mass-limited (log ($M_{*}/M_{\odot}$)$\geq 9.1$) 
UDS galaxies, whose numbers are scaled down for visual purpose.}
 \label{spzdst}
\end{figure*}

Among 173 galaxies with secure redshifts from our observation, 
about $44 \%$ are cluster galaxy candidates, while the remaining 
galaxies are field ones. 
Figure~\ref{spzdst} shows the distribution of the spectroscopic redshift 
of all 173 galaxies in black. 
The red histogram is for cluster galaxy candidates. 
As explained in Section~\ref{starget}, field galaxy targets 
are selected for being at 
similar redshifts with cluster galaxies. 
Therefore, the redshift distribution of these two populations 
show similar trends with broad peaks at $z \sim 0.9$ and $z \sim 1.3$. 

\section{Accuracy of Photometric Redshift and SED-Fitting}\label{accu}

\begin{figure*}
 \centering
 \includegraphics[scale=0.65]{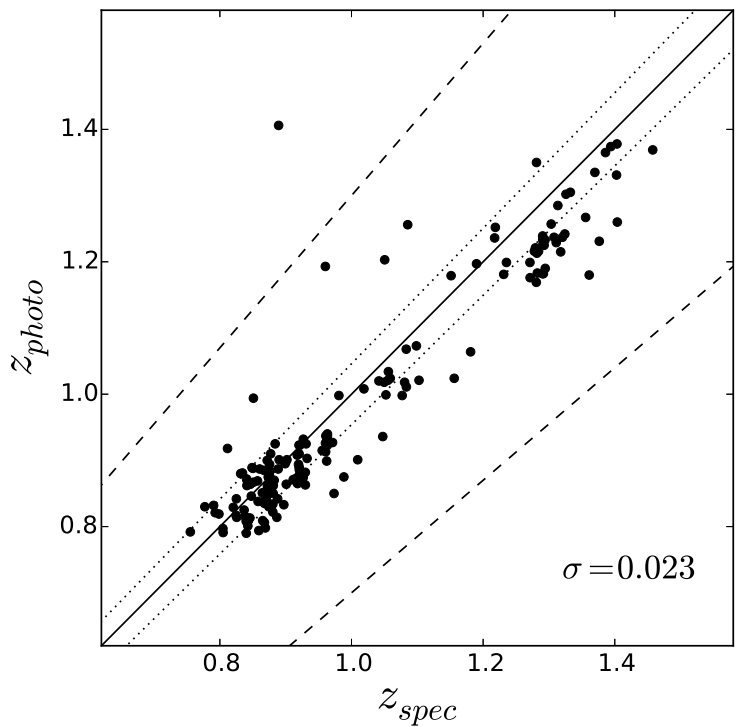}
\caption{Comparison of spectroscopic redshifts from our Magellan observation 
to photometric redshifts. The photometric redshift uncertainty, is 
$|\Delta z|$/($1+z_{spec}) = 0.023$, where $|\Delta z|$ = $|z_{phot} - z_{spec}|$.
Dotted and dashed lines correspond to the $|\Delta z|$/$(1+z_{spec}) = 0.023$ 
and 0.15, respectively.} 
 \label{szpz}
\end{figure*}

Figure~\ref{szpz} shows the comparison of the spectroscopic redshifts 
obtained from our Magellan observation and the photometric 
redshifts from L15, which were measured using the broadband 
photometric data from optical to mid-IR. 
The normalized median absolute deviation (NMAD) between 
photometric and spectroscopic redshifts, 
$|\Delta z|$/($1+z_{spec}$), based on our Magellan observation, 
is 0.023, 
which is similar to the value reported in L15 (0.028). 
Here, $\Delta z$ is the difference between spectroscopic and 
photometric redshifts. 
There is only one outlier with $|\Delta z| / (1 + z_{spec})$ greater 
than 0.15. 

\begin{figure*}
 \centering
 \includegraphics[scale=0.55]{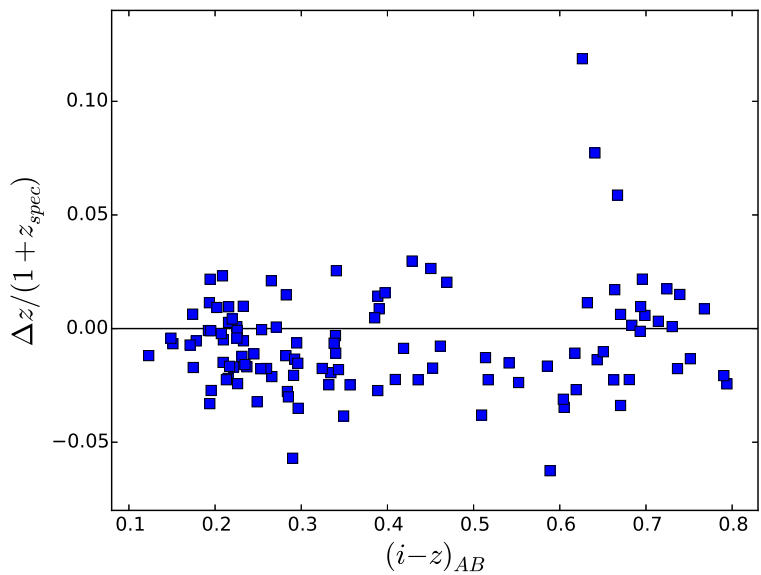}
\caption{$(i-z)$ colour dependence of normalized photometric redshift error, 
$\Delta z$/($1+z_{spec}$) for $z \sim 0.9$ galaxies. There is no clear 
colour dependence in redshift error. We exclude one outlier 
($|\Delta z|$/$(1+z_{spec}) > 0.15$) in this figure.} 
 \label{dziz}
\end{figure*}

We check if there is any color dependence in photometric redshift error 
as a function of $(i-z)$ colour. 
Here, we focus on $z \sim 0.9$ galaxies, and for these 
galaxies, $(i-z)$ colour brackets the $4000 \AA$ break, a common strong 
spectral feature for photometric redshift estimates.
As can be seen in Figure~\ref{dziz}, photometric redshift error does 
not show any clear colour dependence. The values of mean and standard deviation 
of normalized redshift error are $-0.008 \pm 0.017$ for blue galaxies 
(i.e., galaxies with $(i-z) < 0.5$), and $-0.003 \pm 0.033$ for red 
($(i-z) \geq 0.5$) galaxies. Note that we exclude one red outlier (with 
$\Delta z / (1 + z_{spec}) > 0.15$ and $(i-z) > 0.5$) in this calculation 
and in Figure~\ref{dziz}.

For the spectroscopically identified overdensities (refer to 
the next section for the detailed 
explanation about the spectroscopic identification and 
properties of the clusters/groups), 
$31.7 \%$ of the spectroscopically identified galaxy candidates 
turned out to be non-members --- giving the member 
confirmation rate of $68.3 \%$.
The high confirmation rate suggests that a photometric redshift 
can trace large-scale galaxy overdensities reasonably well.
This confirmation fraction does not show a clear dependence on the 
magnitude of galaxies as shown in Figure~\ref{confmg}. 

\begin{figure*}
 \centering
 \includegraphics[scale=0.32]{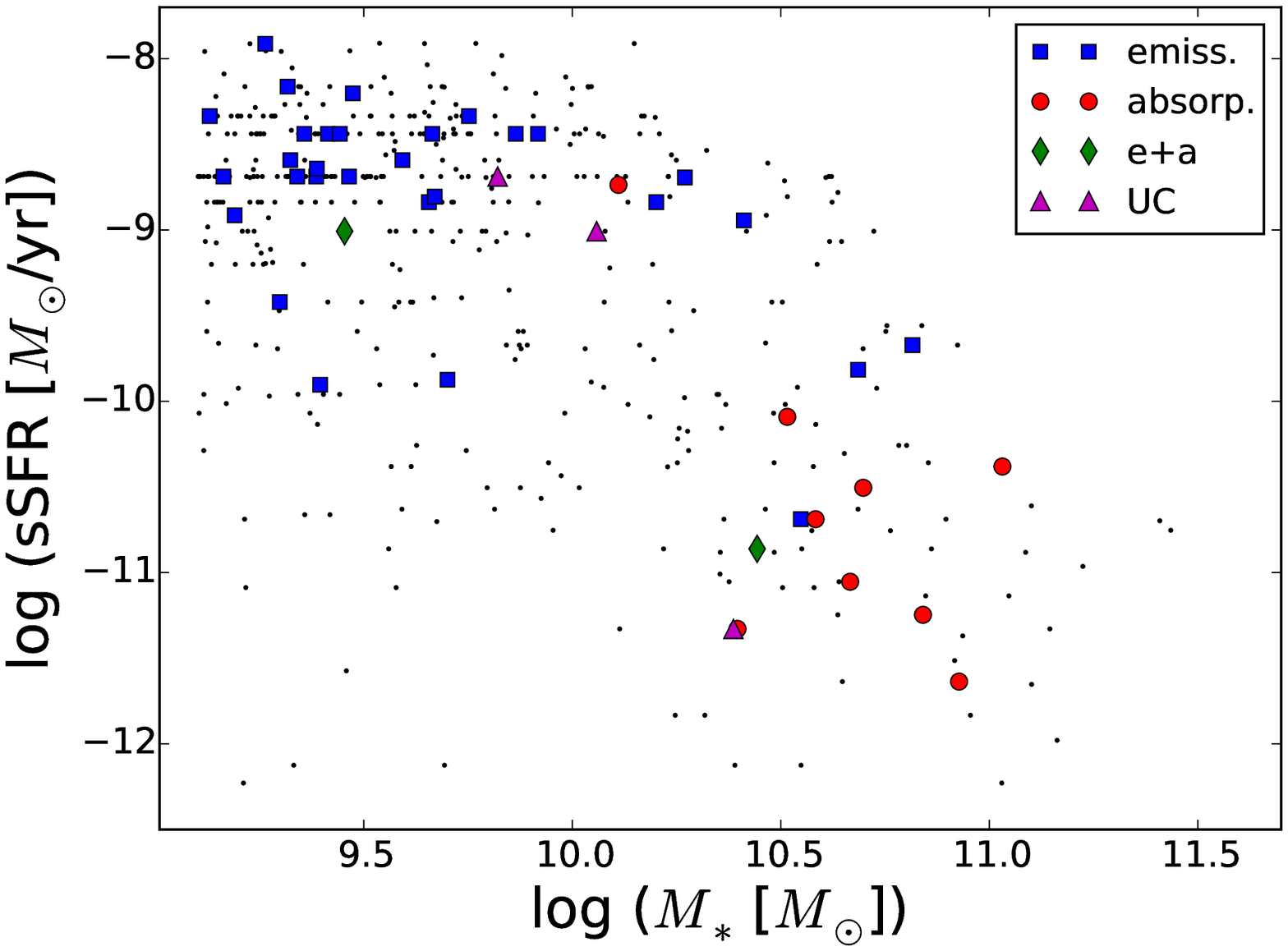}
\includegraphics[scale=0.32]{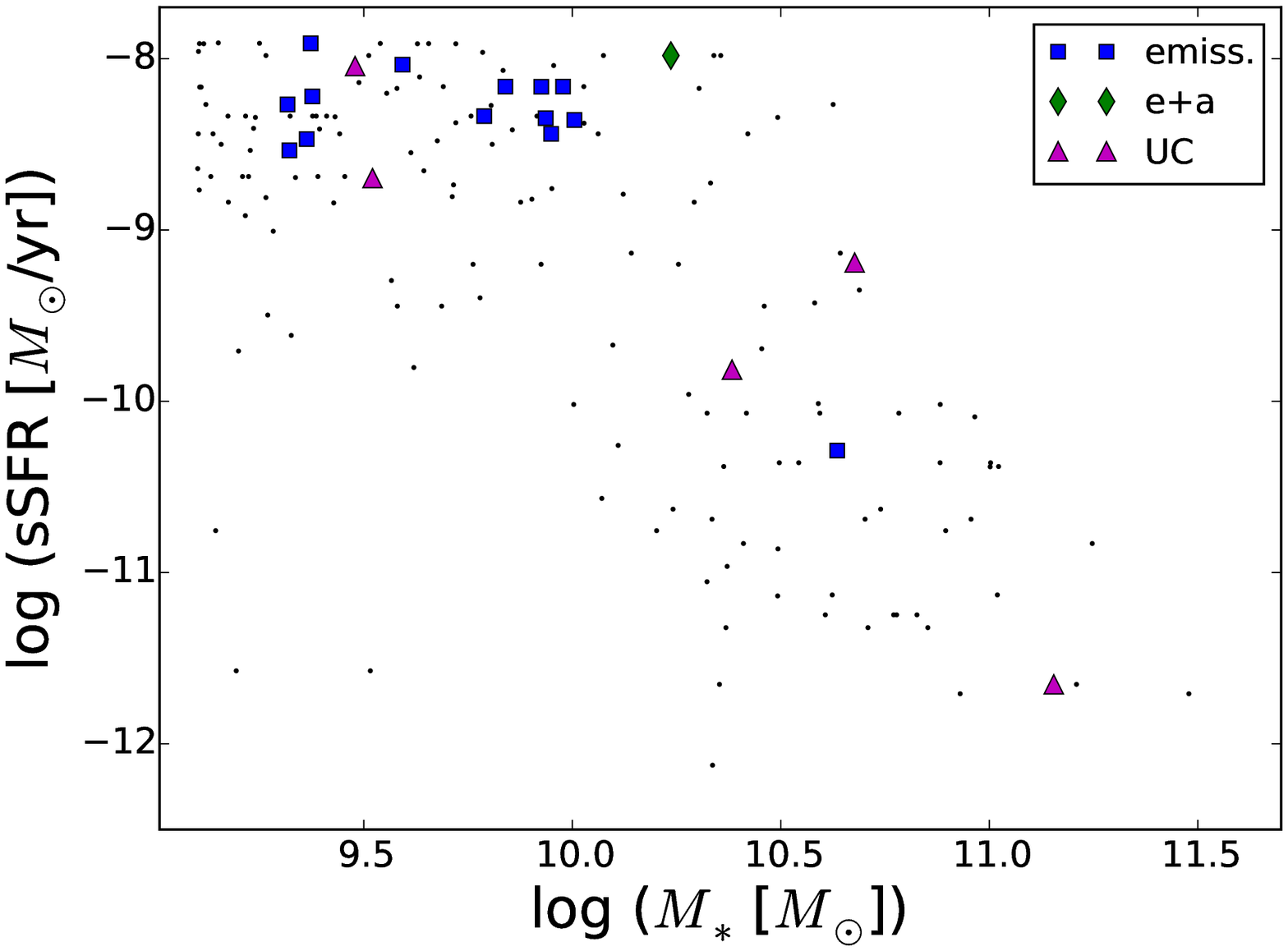}\\
\includegraphics[scale=0.32]{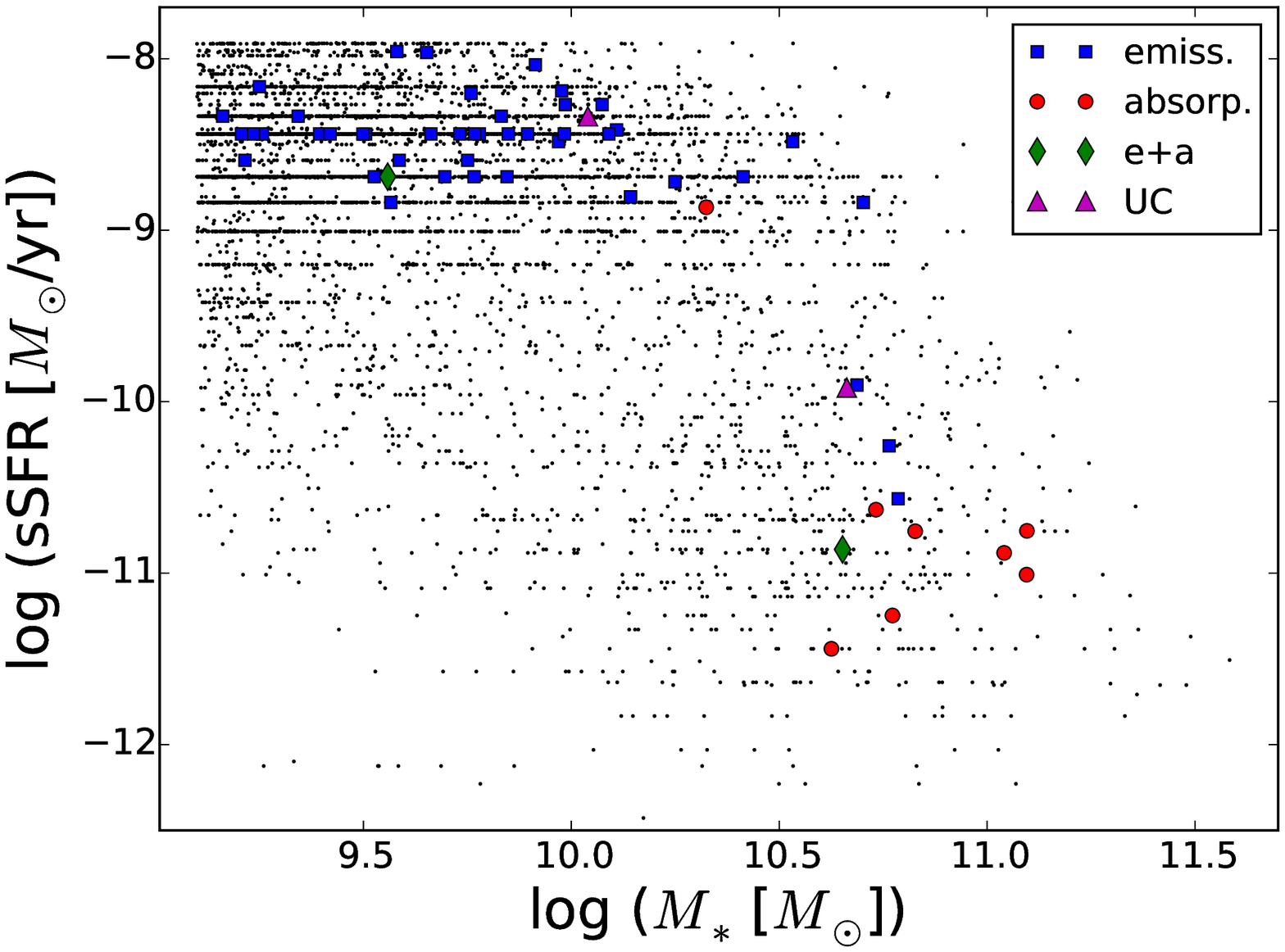}
\includegraphics[scale=0.32]{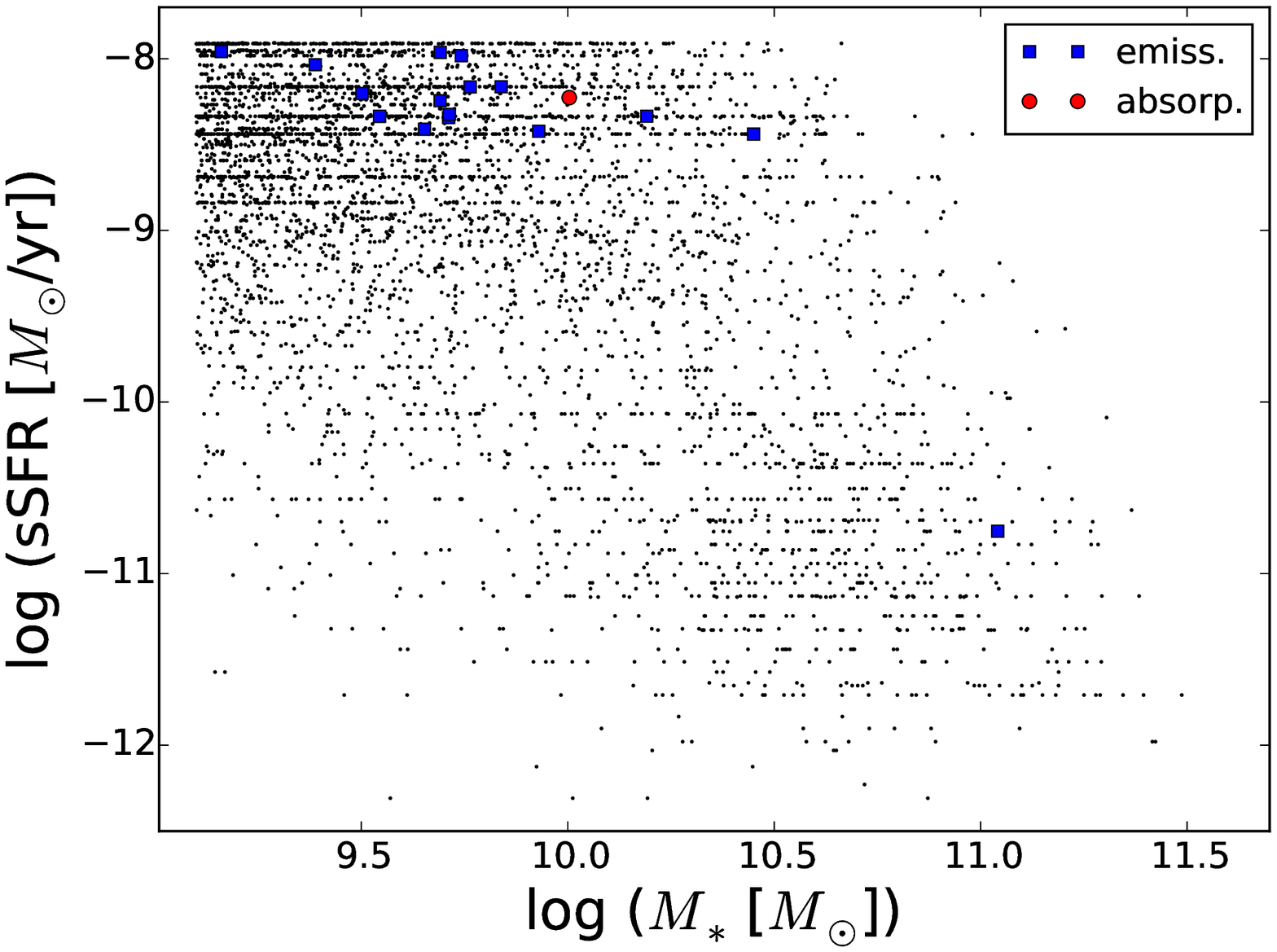}\\
\caption{Stellar-mass--sSFR diagram of galaxies at $z \sim 0.9$ ({\bf Left}) 
and $z \sim 1.2$ ({\bf Right}). Panels in the upper row are for cluster 
candidate galaxies, and the lower row is for field galaxies. Blue squares represent 
galaxies with emission features only in their spectrum, while red 
circles are the ones with absorption features only. 
Green diamonds are the galaxies with both emission and absorption features, 
and purple triangles are the galaxies whose spectral features are weak. 
Black dots represent the galaxies with photometric redshifts only.} 
 \label{zmsssfr}
\end{figure*}

In Figure~\ref{zmsssfr}, we show the stellar-mass--specific SFRs (sSFRs) 
of galaxies at redshifts 
$\sim 0.9$ (left column) and $\sim 1.2$ (right column) for clusters 
(upper row) as well as field (lower row). 
Here, we also show whether the emission or absorption features are 
seen in their spectra when spectroscopic redshifts are available. 
We can see that almost all galaxies with clear emission-line 
(shown as the blue squares in each panel of 
Figure~\ref{zmsssfr}) are also classified 
as star-forming galaxies in our SED-fitting, while all galaxies with 
only the absorption feature (shown as the red circles) are quiescent 
galaxies based on the classification from the SED-fitting. 

\begin{figure*}
\centering
 \includegraphics[scale=0.8]{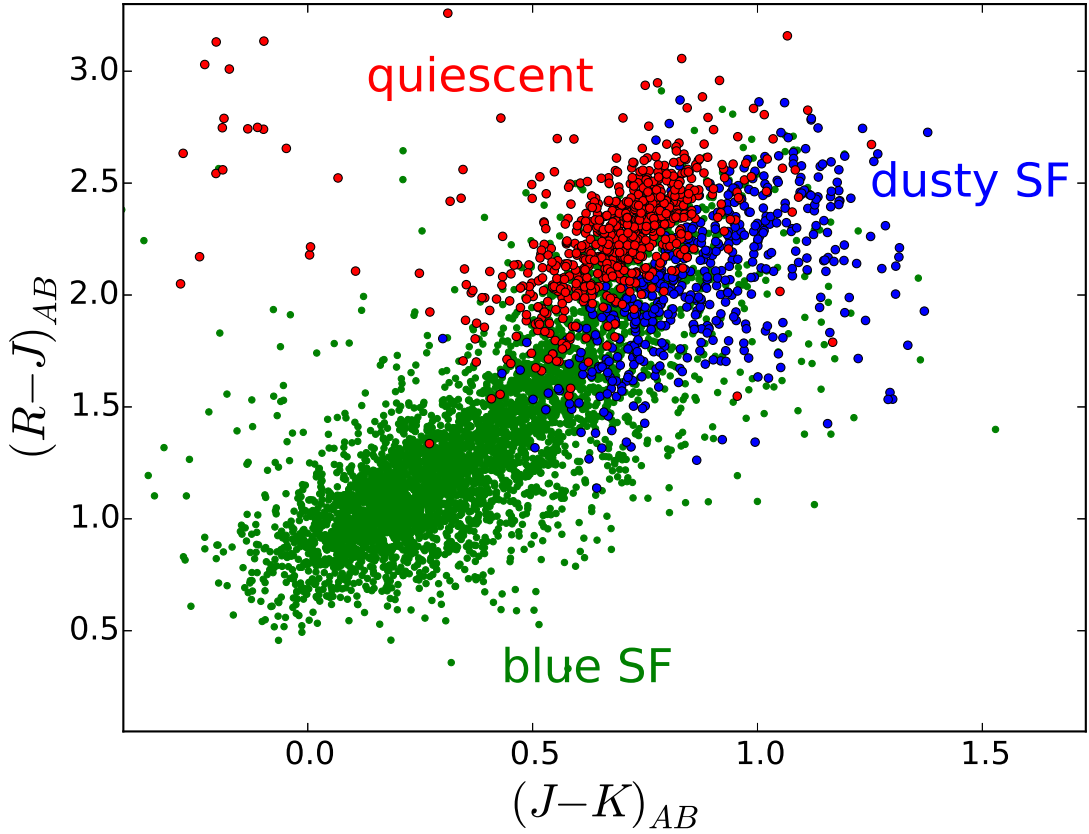}
\caption{$(R-J)$ versus $(J-K)$ colour-colour diagram of 
$0.8 \leq z \leq 0.95$ galaxies. Red points are quiescent galaxies, 
while green and blue points are SF galaxies with low ($E(B-V) < 0.45$) 
and high ($E(B-V) \geq 0.45$) extinction, which are classified in 
our SED-fitting. The location of each type of galaxies in this 
diagram shows the reliability of galaxy type classification from 
our SED-fitting.} 
 \label{rjk}
\end{figure*}

Figure~\ref{rjk} shows the location of galaxies with the 
photometric redshift, $0.8 \leq z_{ph} \leq 0.95$, in the $(R-J) - (J-K)$ 
diagram, which corresponding to the rest-frame $UVJ$ diagram at 
this redshift range. 
In this figure, red and blue points are quiescent and dusty 
($E(B-V) \geq 0.45$) SF galaxies classified in our SED-fitting. 
Figures~\ref{zmsssfr} and \ref{rjk} demonstrate the reliability of 
SED-fitting in classifying them as being star-forming or quiescent.

In L15, we have already shown that our SED-fitting (combined with the 
deep optical data) can distinguish between quiescent galaxies 
and dusty SF galaxies (Figure 3 in L15). 
Also, we checked the MIPS 24 $\micron$m detection of our sample galaxies 
using SpUDS MIPS catalog, finding that all 24 $\micron$m-detected galaxies 
are correctly classified as dusty SF galaxies based on our SED-fitting. 

\section{Spectroscopic Identification of Clusters/Groups at $z \sim 1$}\label{spiclgr}

\subsection{Spectroscopically Identified Clusters/Groups}

To identify galaxy clusters and groups from the spectroscopic data, 
we apply the following procedure. 

\begin{figure*}
\centering
 \includegraphics[scale=0.6]{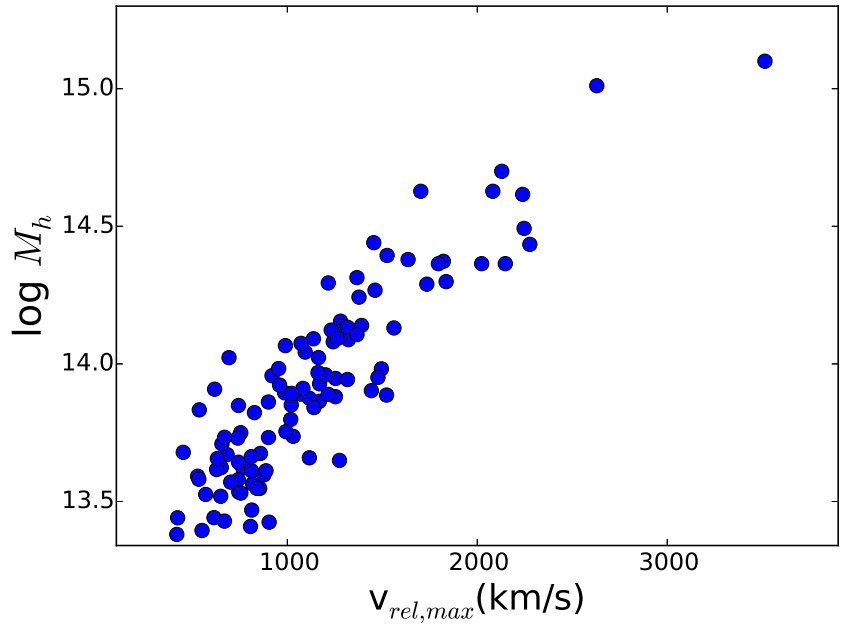}
\caption{Correlation between the maximum rest-frame relative 
velocities (v$_{rel,max}$) of mock galaxy clusters from GALFORM 
simulation and their halo masses ($M_{h}$ in $M_{\odot}$). 
This shows that velocity cut of $\sim 3000$ km s$^{-1}$ is needed 
to properly detect and measure the velocity dispersions of galaxy 
clusters with their mass up to $\sim 10^{15}$ $M_{\odot}$.} 
 \label{rvmhmodel}
\end{figure*}

First, from the spectroscopic redshift distribution of each 
cluster candidate, we discriminate galaxies 
as separate structures if the velocity gap (defined as the difference 
in the rest-frame velocities of galaxies adjacent in the velocity space) 
is larger than 1500 km/s.  
Next, for each separate structure, 
we apply 3-$\sigma$ clipping over the redshift distribution of 
cluster/group galaxy candidates to exclude obvious outliers. 
A tentative cluster/group redshift is determined to be the mean 
redshift of the remaining galaxies. 
Then, we further exclude any galaxies 
if their relative rest-frame velocity goes beyond 
$\pm$3000 km s$^{-1}$ from the cluster redshift. 
This velocity cut is set after investigating the correlation 
between the maximum rest-frame velocities and halo masses 
of mock galaxy clusters from the GALFORM simulation \citep{mer13} 
(Figure~\ref{rvmhmodel}). 
While we use a generous velocity cut of 3000 km s$^{-1}$, 
member galaxies of all confirmed clusters/groups are 
within $\pm$2000 km s$^{-1}$, except the one at $z \sim 1.3$ 
(we discuss about this structure in next section). 
And, for majority of the confirmed clusters/groups, rest-frame 
velocities of member galaxies are within $\pm$1000 km s$^{-1}$.

\begin{figure*}
\centering
 \includegraphics[scale=0.5]{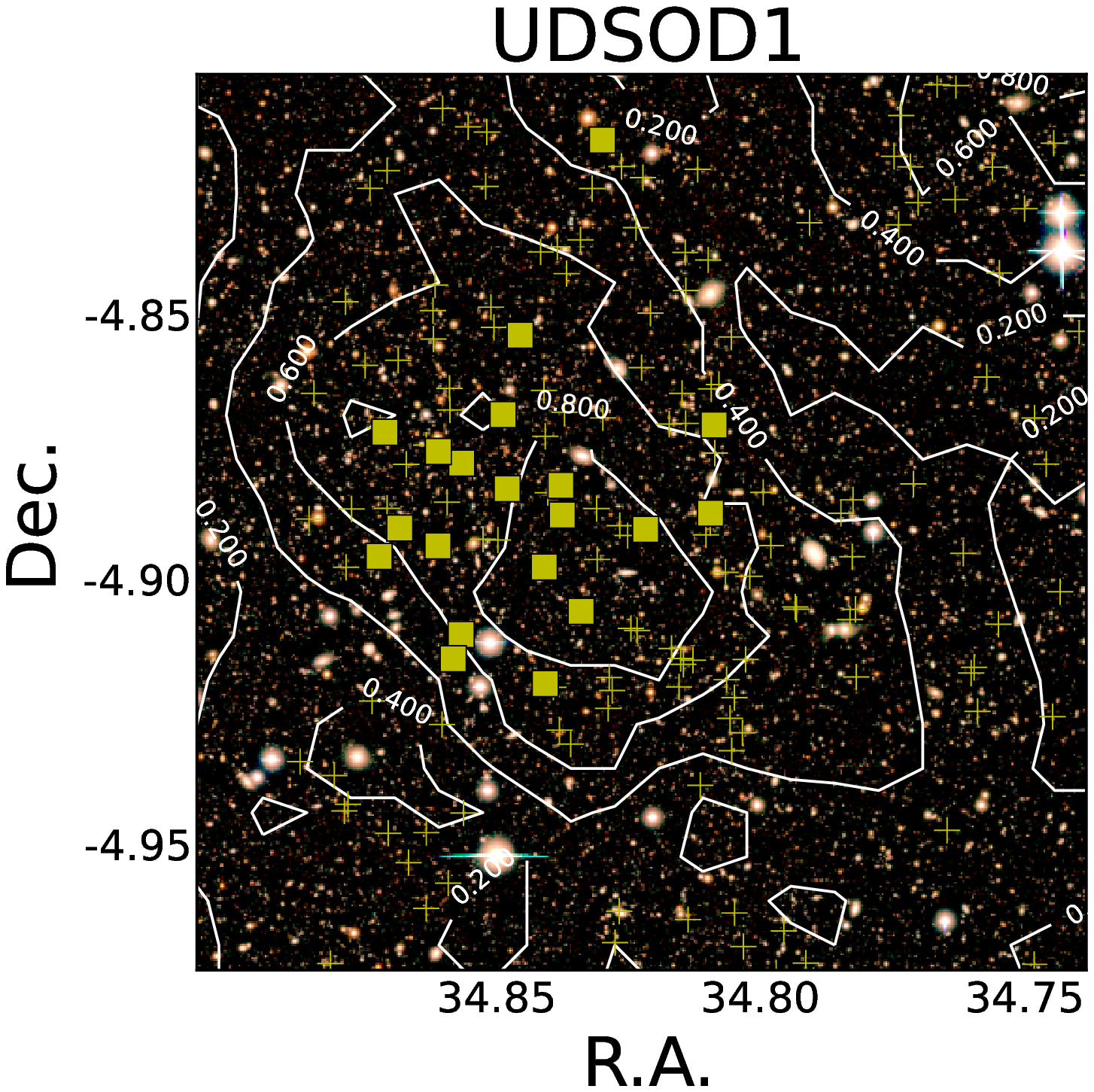}
\includegraphics[scale=0.5]{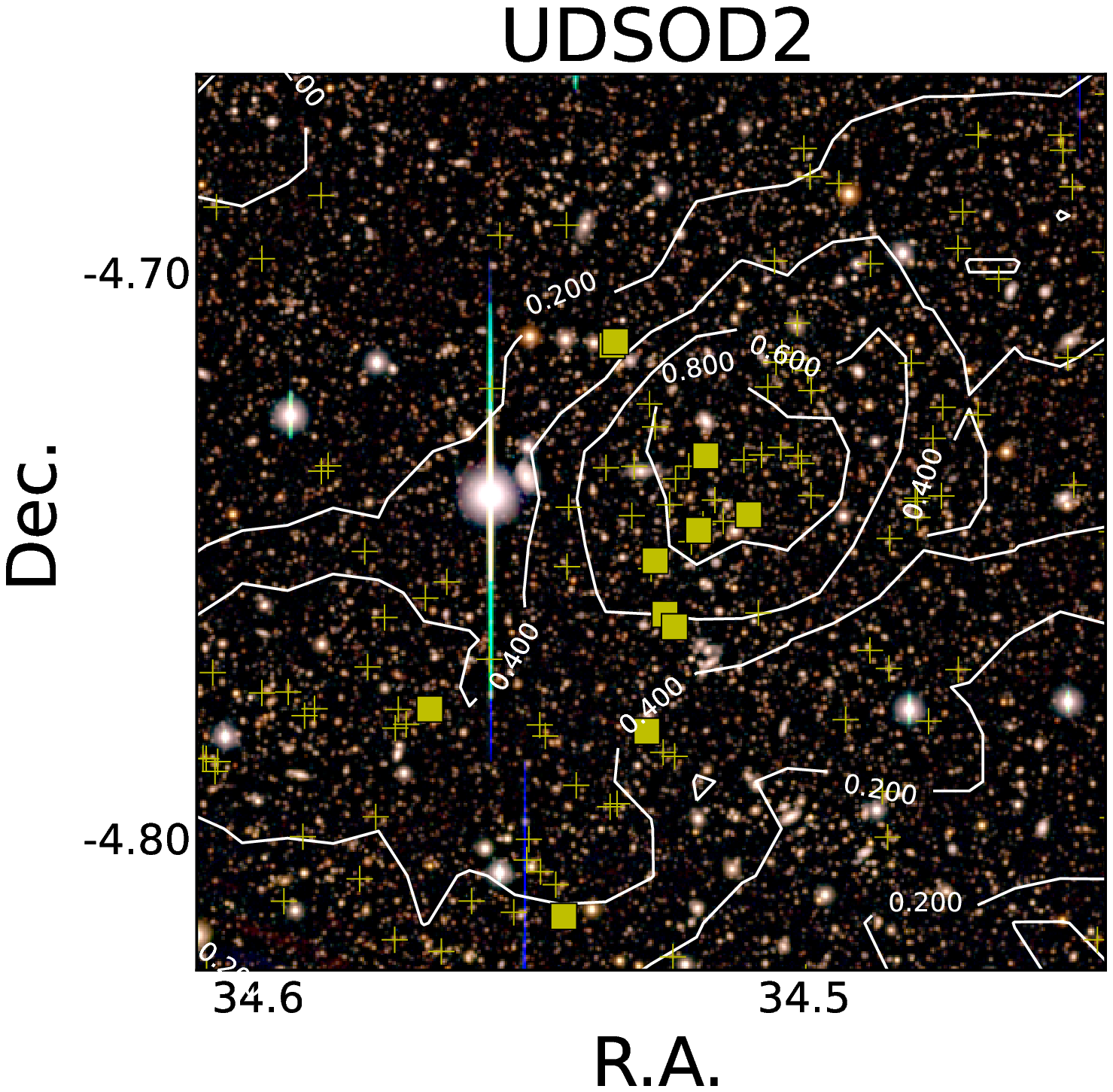}\\
\includegraphics[scale=0.5]{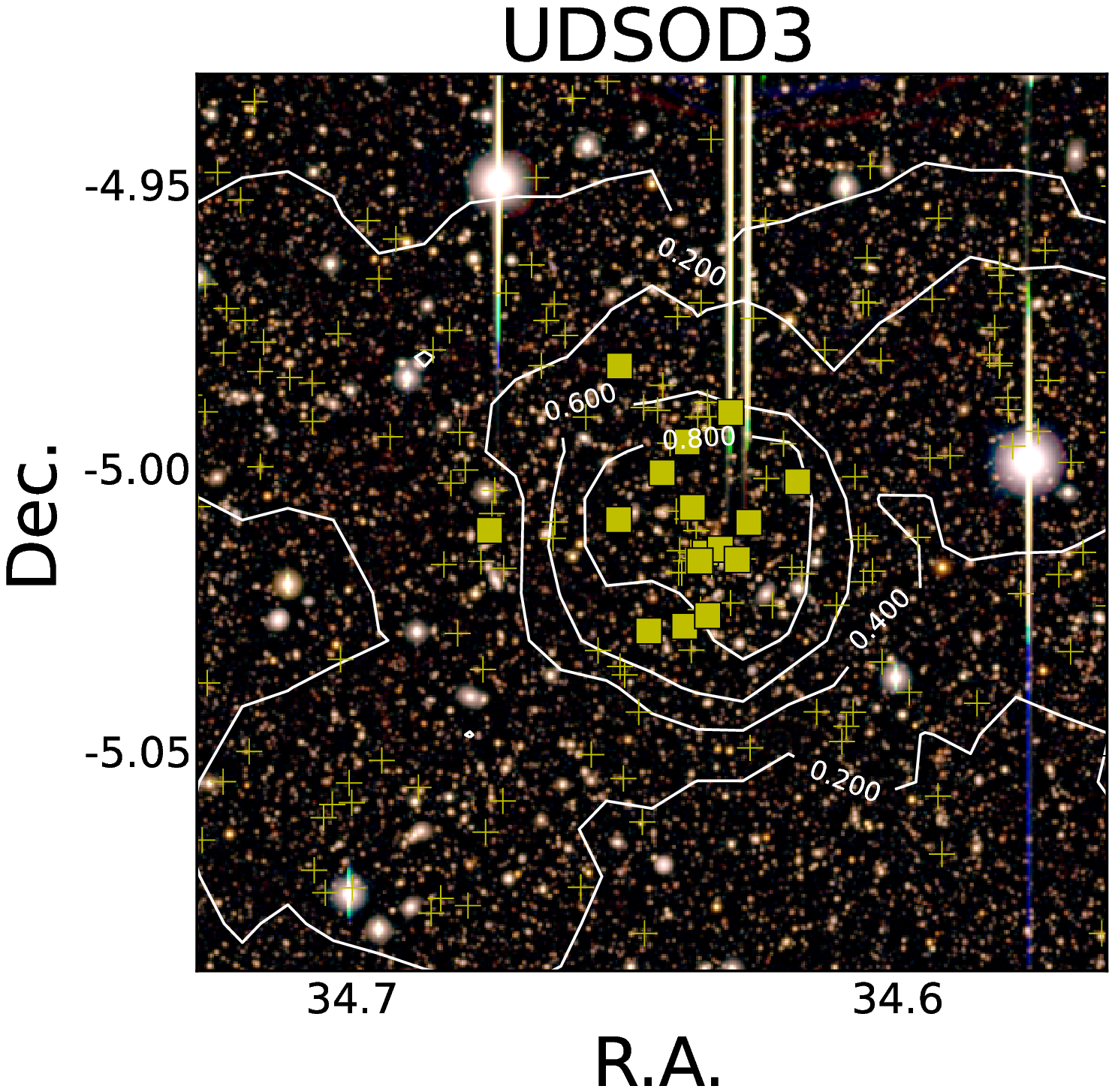}
\includegraphics[scale=0.5]{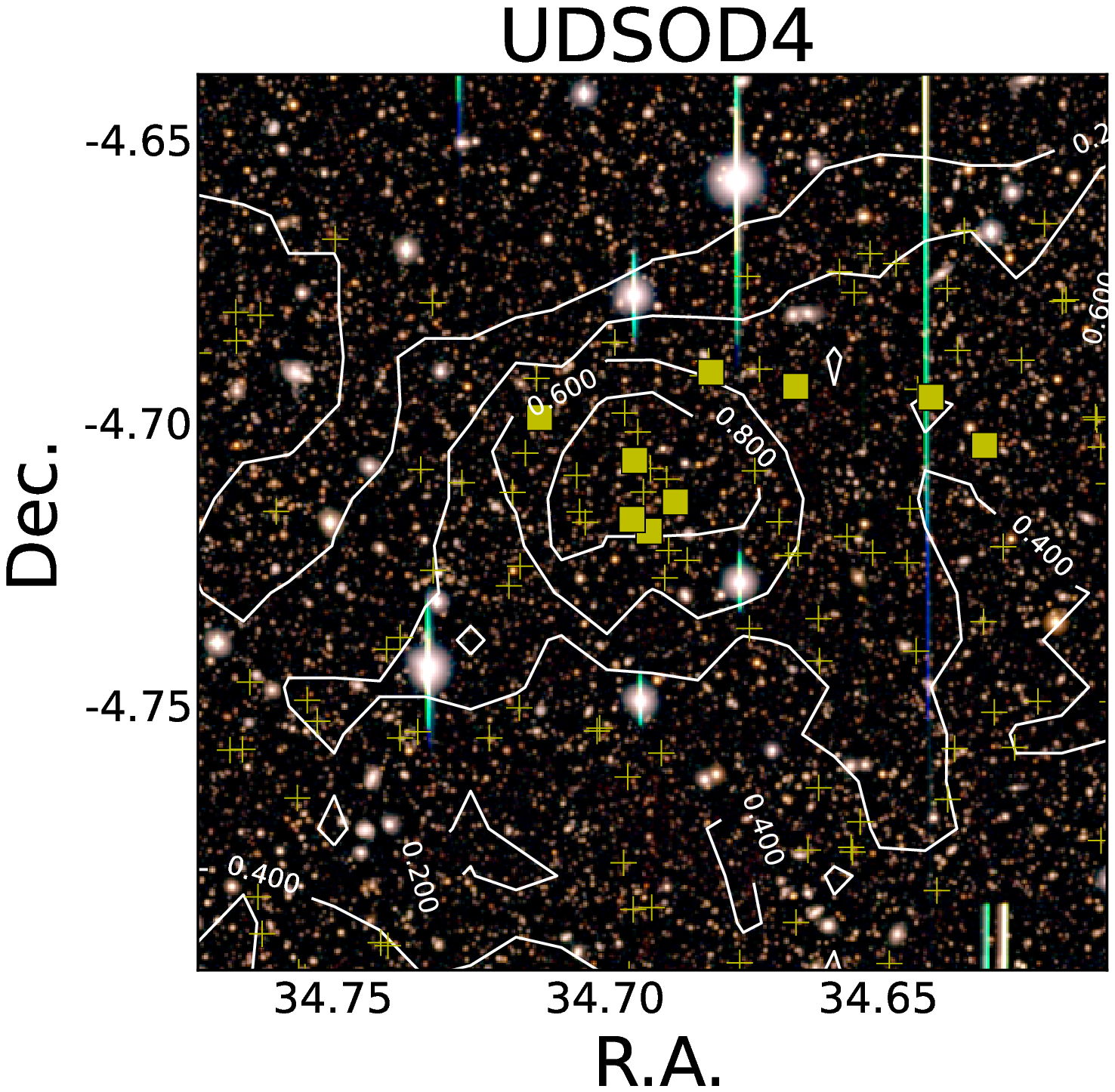}
\caption{Composite colour ($R$,$i'$,$z'$) images of the confirmed overdensities.
White contours show the galaxy surface number density levels, 
normalized to the maximum value in each region.
Yellow squares show the locations of the spectroscopically 
confirmed members belonging to each overdensity, while photometrically identified 
member candidates as well as galaxies in the same photometric redshift range 
are shown with small yellow crosses.
Each image tile size corresponds to $10\arcmin \times 10\arcmin$.}
 \label{climage}
\end{figure*}

Using this process, 
we confirm four candidates to 
contain significant overdensities that can be clusters/groups: 
three at $z \sim 0.9$ and one at $z \sim 1.3$. 
The colour composite images with the density contour of these 
overdensities  are shown in Figure~\ref{climage}, and 
their spectroscopic redshift distributions are given in 
Figure~\ref{clspzdst}. 
The redshift distribution shows that one cluster candidate, 
UDSOD1, is a system with 
three or four clusters/groups that are aligned along the 
line of sight, and another candidate, UDSOD2, is a system 
with two clusters/groups aligned along the line of sight. 
This kind of occurrence of multiple, spatially overlapped 
peaks within a single photometrically selected overdensity 
is also found in our test using the mock galaxy catalog from 
the GALFORM simulation \citep{mer13} as shown 
in APPENDIX A, as well as in literature \citep[e.g.][]{van07,van09}

UDSOD1, UDSOD2, and UDSOD3 have been identified as galaxy clusters 
or groups in F10 based on the extended X-ray emission and the 
red sequence of galaxies. 
Their counterparts in F10 are named as SXDF37XGG, SXDF21XGG, and 
SXDF46XGG. 
However, in F10, the number of $z_{spec}$ of member galaxies was small, 
1 for UDSOD1, 0 for UDSOD2, and 6 for UDSOD3. 
Our spectroscopic observation unambiguously confirms their cluster 
nature and also reveals multiple peaks which can be separate 
structures along the line of sight. 
UDSOD4 stands as a newly discovered structure, not discussed in 
previous studies.

For the other cluster candidates, the spectroscopic redshift 
distributions of candidate member galaxies are shown in 
Figure~\ref{uncandspzdst}, and we find that 
the number of measured redshifts is too small 
to confirm them as clusters/groups.
For the two cluster candidates at $z \sim 1.21$ and 1.25, 
four and three galaxies, are 
found to possibly form redshift peaks, 
out of five and seven redshift measured candidates, respectively. 
More spectroscopic data can 
tell whether they are truly clusters/groups in future. 
For the three remaining cluster candidates (one at $z \sim 1.0$ 
and two at $z \sim 1.4$), the numbers of spectroscopically 
confirmed galaxies are only 
four, two and three respectively. 
The cluster candidate at $z \sim 1.04$ is possibly an 
overlap of two structures: one at $z \sim 0.97$ and 
another at $z \sim 1.08$. 

\subsection{Properties of Spectroscopically Identified Galaxy Clusters/Groups}
\label{odprop}

We describe the properties of the spectroscopically identified 
clusters/groups here. 
The derived properties are summarized in Table 2.

\subsubsection{UDSOD1}

Using $z_{phot}$, this massive structure was initially identified 
as a single cluster candidate at $z \sim 0.89$ located at 
$\alpha \sim$ 02$^{h}$19$^{m}$22$^{s}$ and $\delta$ 
= -4$\,^{\circ}$52$\arcmin$57$\arcsec$. 
Our spectroscopic observation of the member galaxy candidates 
has revealed that three to four overdensities are spatially overlapped. 
The upper left panel of Figure~\ref{clspzdst} shows the 
distribution of spectroscopic redshift of these overdensities 
at redshifts $z \sim 0.875$ (UDSOD1-a), 0.920 
(UDSOD1-b), and 0.963 (UDSOD1-c) within a 2 Mpc projected radius 
circle from the group/cluster center. 
The velocity differences between peaks of each adjacent structure 
are about 7000 km s$^{-1}$.
The number of spectroscopically confirmed members of these 
clusters/groups are seven (UDSOD1-a), eight (UDSOD1-b), and five 
(UDSOD1-c) within a 2 Mpc of projected radius from each group/cluster center. 
Among these, six (UDSOD1-a), eight (UDSOD1-b), and five (UDSOD1-c) 
are within a 1 Mpc projected radius from each group/cluster center. 
In Figure~\ref{3dudsod1}, we show the spatial distributions 
of these spectroscopically confirmed member galaxies. 
As can be seen in this figure, the spatial positions of 
the spectroscopic member galaxies of UDSOD1-a and UDSOD1-b 
are closely overlapped, while the members of UDSOD1-c are 
more concentrated on the eastern side. 

We estimate the halo masses (throughout this paper, 
halo mass refers $M_{200}$ in most cases) from the 
total stellar masses within 1 Mpc, 
$M_{*,total}$, of these clusters/groups. 
To compute the total stellar mass, we use both 
the spectroscopically identified members and the member 
candidates based on the photometric redshift, but applying 
a weight, $w_{i}$, to each member candidate with $z_{phot}$ only.  
Here, $w_{i}$  is computed as the number of 
spectroscopically confirmed members divided by the sum of 
the numbers of spectroscopic members and outliers.
When a cluster candidate turns out to be spatially overlapped 
multiple structures based on spectroscopic redshifts, 
galaxies belong to different structures are considered as 
outliers --- e.g., spectroscopically identified members 
of UDSOD1-b and UDSOD1-c are considered as outliers in calculating 
$w_{i}$ for UDSOD1-a. 
For the spectroscopically confirmed members, this weight value 
is 1. 
Then, the total stellar mass is computed as 
\begin{equation}
M_{*,total} = \Sigma w_{i} \times M_{*,i}, 
\end{equation}
where log $M_{*,i} \geq 9.1$. 

Galaxies whose photometric redshifts are in the range of 
$z_{cl} \pm 0.023 \times (1+z_{cl})$ are included in this calculation.

The $M_{*,total}$ values are found to be 
$4.8 \times 10^{11}$ M$_{\odot}$, $2.3 \times 10^{11}$ M$_{\odot}$, 
and $3.5 \times 10^{10}$ M$_{\odot}$ for UDSOD1-a, UDSOD1-b, 
and UDSOD1-c, respectively, and their values are listed 
in Table~\ref{tab2}.
From the calibration between $M_{200}$ and $M_{*,total}$  
derived from our model test (APPENDIX B), 
$M_{200}$'s are estimated as $4.8 \times 10^{13}$ M$_{\odot}$, 
$2.3 \times 10^{13}$ M$_{\odot}$, and $3.5 \times 10^{12}$ M$_{\odot}$, each. 
In case of UDSOD1-a, the redshift distribution is made of two peaks, with 
the velocity gap to be about 1200 km s$^{-1}$.  
If UDSOD1-a is considered as two groups, their 
$M_{*,total}$'s are $9.6 \times 10^{10}$ M$_{\odot}$ and 
$2.0 \times 10^{11}$ M$_{\odot}$, giving the 
stellar-mass--calibrated $M_{200}$ as $9.6 \times 10^{12}$ M$_{\odot}$ and 
$2.0 \times 10^{13}$ M$_{\odot}$, each. 

In Table~\ref{tab2}, we summarize the derived properties of the UDSOD1 and 
other overdensities presented below. The X-ray-based 
$M_{200}$ of UDSOD1 is available with the value of 
$9.3 \times 10^{13}$ M$_{\odot}$ (SXDF37XGG; F10). This value is 
in a reasonable agreement with $M_{200}$ from $M_{*,total}$ 
with UDSOD1-a as a single structure.
Note that there was only one $z_{spec}$ 
available in F10 for SXDF37XGG with $z_{spec}=0.845$ 
which was suggested as the redshift of the structure. 
With our $z_{spec}$, we suggest that the redshift of the structure 
should be revised to either $z_{spec}=0.875$ of UDSOD1-a or 
$z_{spec}=0.920$ of UDSOD1-b.

From the stellar-mass--calibrated $M_{200}$, we estimate $R_{200}$ 
using the following relation \citep[e.g.,][]{kim16}.

\begin{equation}
R_{200} = 10^{-2/3} \frac{G^{1/3}}{H(z)^{2/3}} M_{200}^{1/3}, 
\end{equation}

where $H(z)$ is the Hubble parameter at redshift $z$. 

Then we also calculated $F_{q}$, the quiescent fraction of each overdensity, 
within $R_{200}$. 
Like $M_{*,total}$, we used both spectroscopic members and 
photometric members and adopted the same weights, $w_{i}$, to compute $F_{q}$. 
The large fraction of member galaxies in the overdensities UDSOD1-a, 
UDSOD1-b, and UDSOD1-c are actively forming stars with 
a quiescent galaxy fraction ($F_{q}$) of 
$\sim 0.09$, 0.27, and 0.00, respectively. 

\subsubsection{UDSOD2}

The redshift distribution of UDSOD2 candidate galaxies 
within 2 Mpc from the overdensity center is shown in 
the upper right panel in Figure~\ref{clspzdst}. 
For this cluster candidate, we 
identify two significant redshift peaks 
at $z = 0.840$ (UDSOD2-a with 6 spectroscopic redshift members) 
and 0.865 (UDSOD2-b with 5 spectroscopic redshift members), and 
the velocity gap between the peaks of these two structures 
is $\sim 4000$ km s$^{-1}$. 
As shown in Figure~\ref{3dudsod2}, the spatial distributions 
of member galaxies of these two structures 
overlap with each other to some degree, 
while UDSOD2-b is more extended toward the south-east.

We find that the total stellar masses of $2.8 \times 10^{11}$ M$_{\odot}$ 
and $4.8 \times 10^{11}$  M$_{\odot}$ for UDSOD2-a and UDSOD2-b respectively, 
which translate into $M_{200}$ of $2.8 \times 10^{13}$ M$_{\odot}$ and 
$4.8 \times 10^{13}$  M$_{\odot}$ each based on our $M_{200}-M_{*,total}$ calibration. 
The derived values of $M_{200}$ of UDSOD2-b, the main structure, or the sum of 
the masses of the two substructures, agree well 
with the X-ray derived $M_{200}$ of $1.0 \times 10^{14}$ M$_{\odot}$ (SXDF21XGG; F10). 
Note that there is no $z_{spec}$ available for this structure in F10, 
and their redshift of $z=0.860$ is based on $z_{phot}$. 
Our redshift measurement suggests that this structure is made of two peaks 
at $z_{spec} = 0.840$ and 0.865.
We also find that these overdensities have an intermediate $F_{q}$ among 
galaxy cluster candidates at $0.85 < z < 0.9$, of 0.44 and 0.53.

\subsubsection{UDSOD3}

This overdensity shows the highest quiescent galaxy 
fraction among three target galaxy cluster candidates at $z \sim 0.87$. 
Eleven galaxies (out of 18 candidates) are spectroscopically 
confirmed to be the members of this overdensity from our Magellan 
observation.
In addition, there are five galaxies that are members of this 
cluster based on the spectroscopic redshifts from the literature 
\citep{sma08,sim12,aki15}. 
The spectroscopic redshift distribution of these galaxies 
is shown in the lower left panel of Figure~\ref{clspzdst}, 
and we determine its redshift to be $z=0.8731$.   

We find that $M_{*,total}$ of this overdensity is 
$1.0 \times 10^{12}$ M$_{\odot}$, and stellar-mass calibrated 
$M_{200}$ is $1.0 \times 10^{14}$ M$_{\odot}$. 
The X-ray based $M_{200}$ is $1.4 \times 10^{14}$ M$_{\odot}$ 
(SXDF46XGG; F10), in agreement with the 
$M_{200}$ values we derived.  
The halo mass estimates suggest that this overdensity can 
be classified as a low mass cluster.
Note that F10 determined $z_{spec}$ of this structure to be 
$z=0.875$ based on six $z_{spec}$ measurements, which agrees 
well with our redshift determination. 
The quiescent fraction of this overdensity is found to be 0.51. 

\subsubsection{UDSOD4}

\begin{figure*}
\centering
 \includegraphics[scale=0.35]{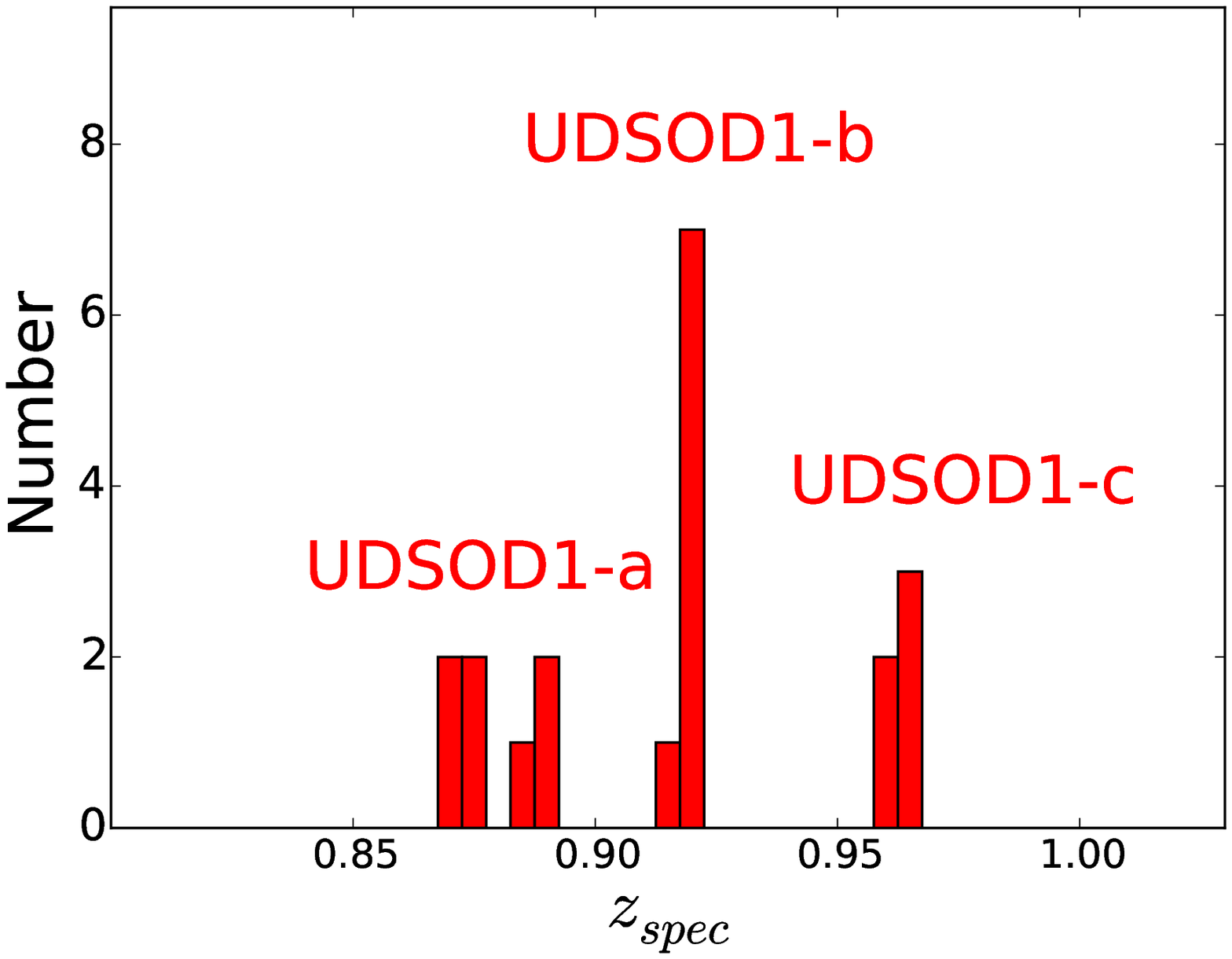}
\includegraphics[scale=0.35]{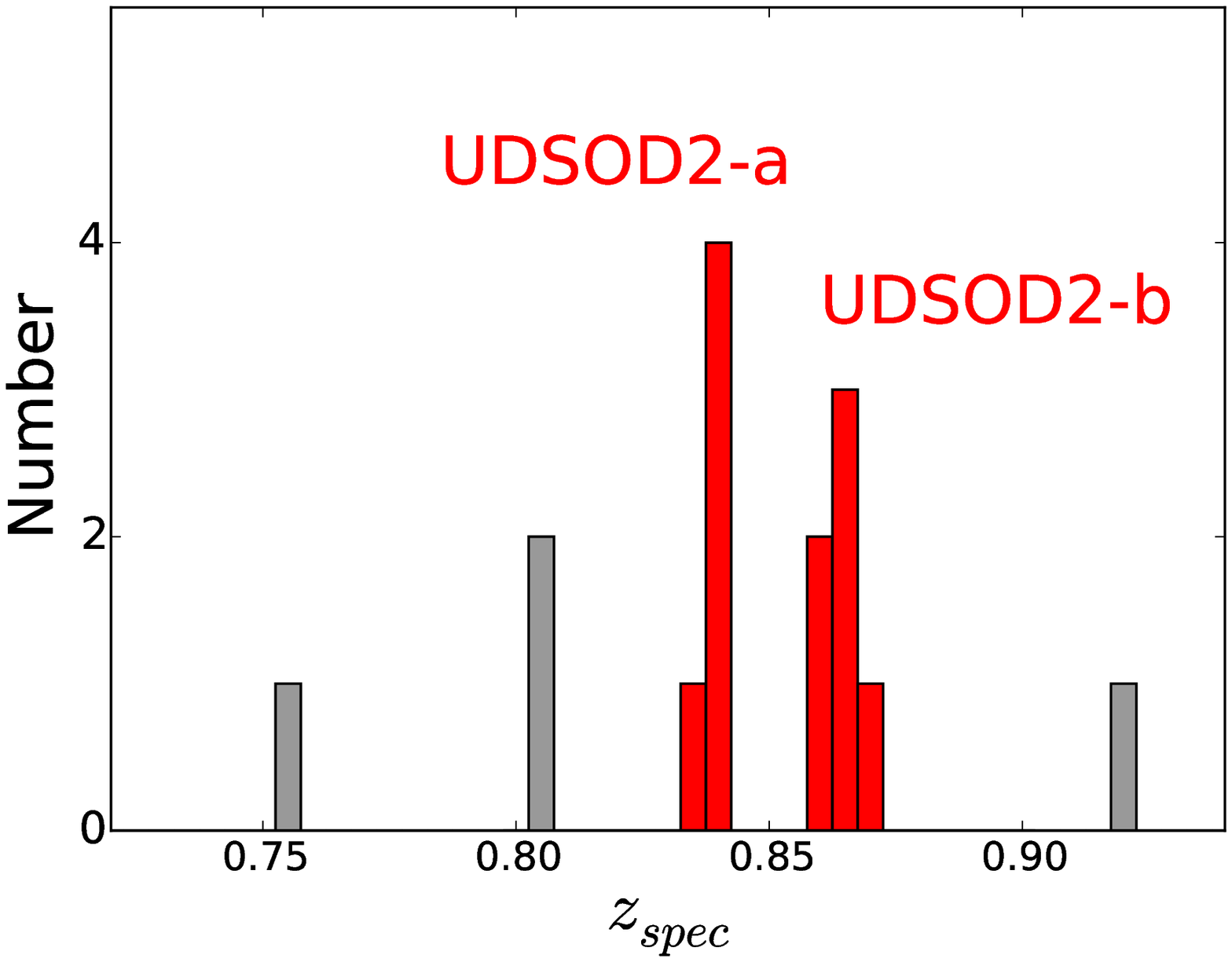}\\
\includegraphics[scale=0.35]{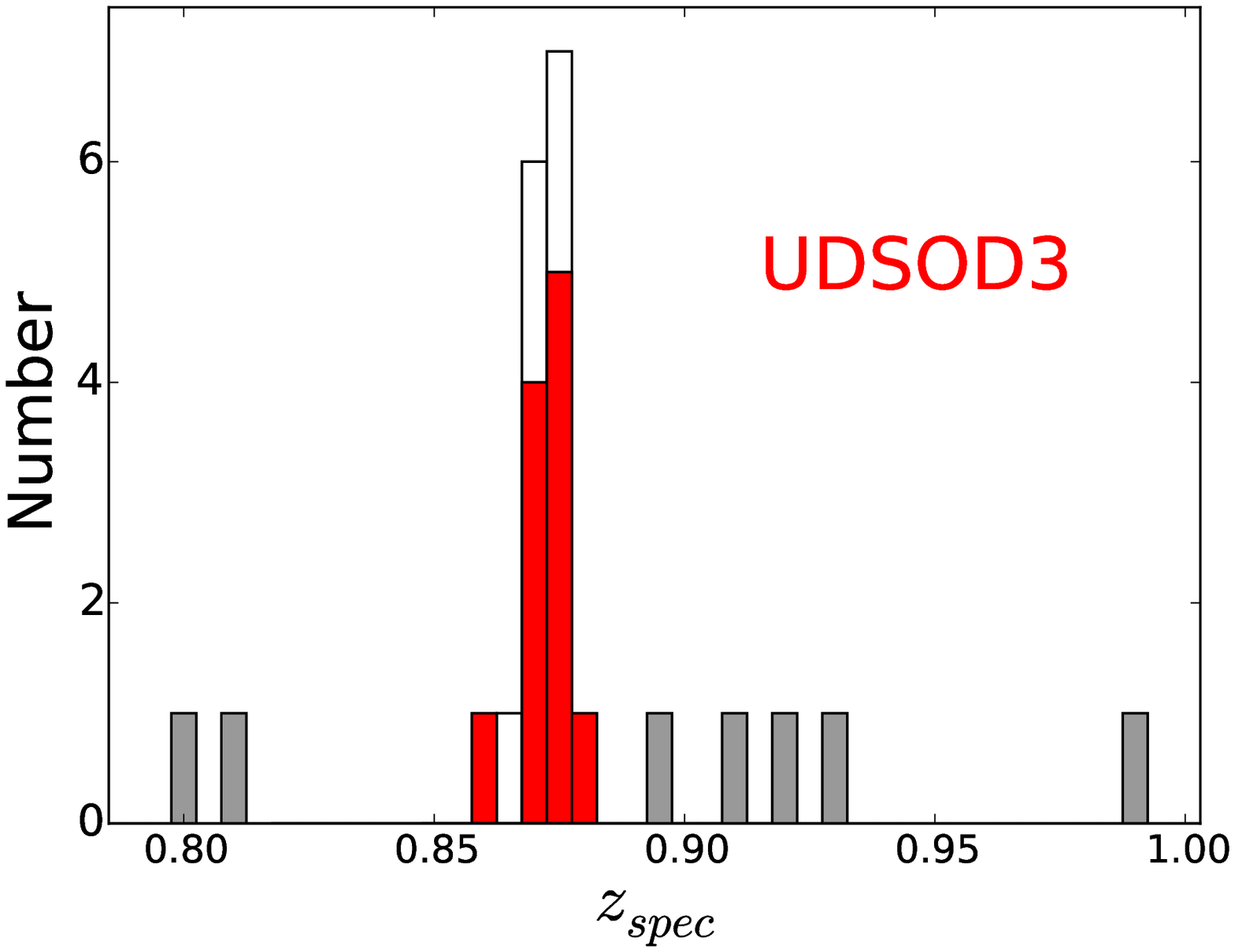}
\includegraphics[scale=0.35]{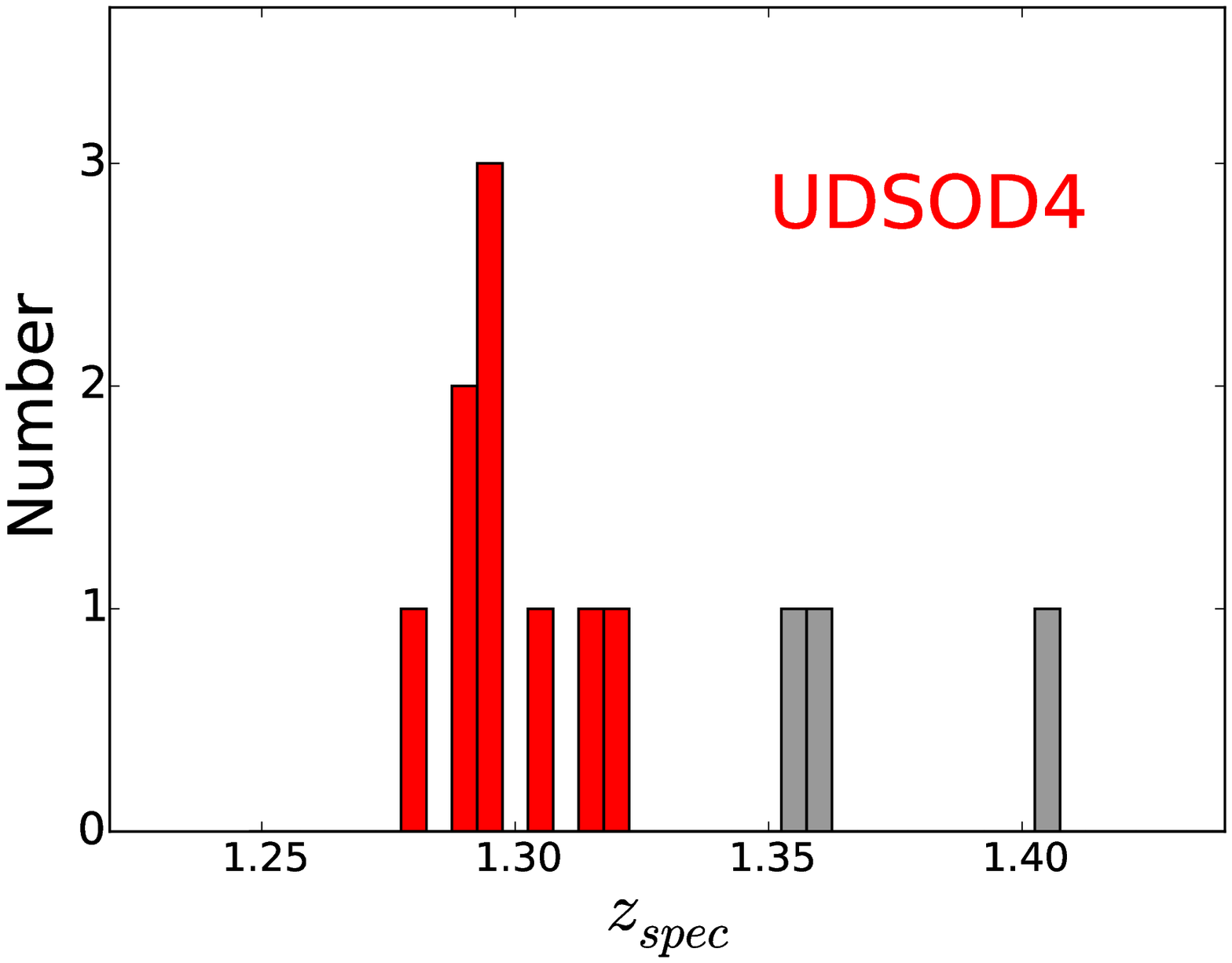}\\
\caption{The spectroscopic redshift distribution of candidate 
member galaxies in the confirmed overdensities. 
The red bars are for the confirmed members of each overdensity, 
while the gray bars show galaxies that turned out to be 
outliers. 
The open bars show galaxies with a spectroscopic 
redshift from the literature.}
 \label{clspzdst}
\end{figure*}

\begin{figure*}
\centering
\includegraphics[scale=0.35]{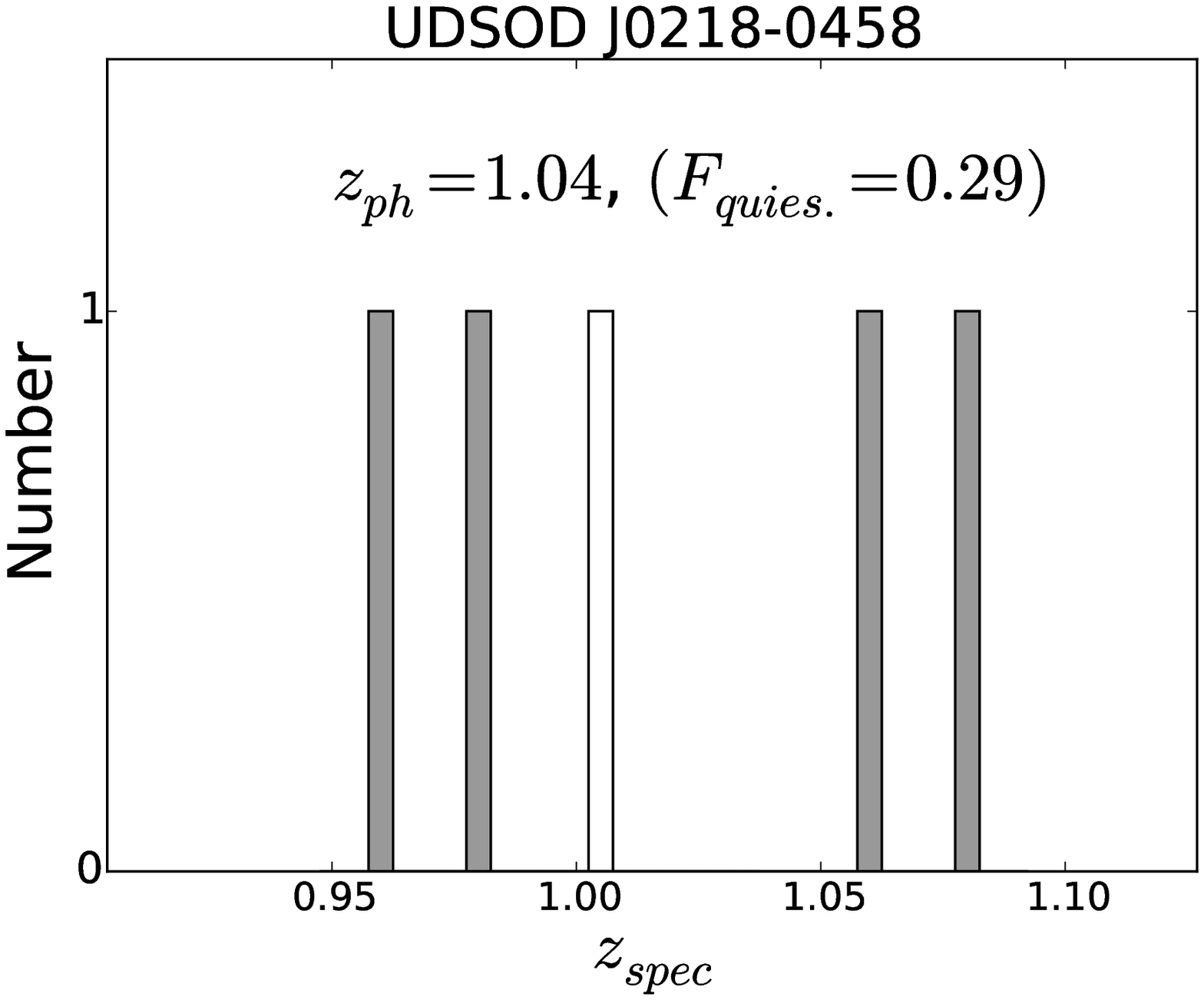}
\includegraphics[scale=0.35]{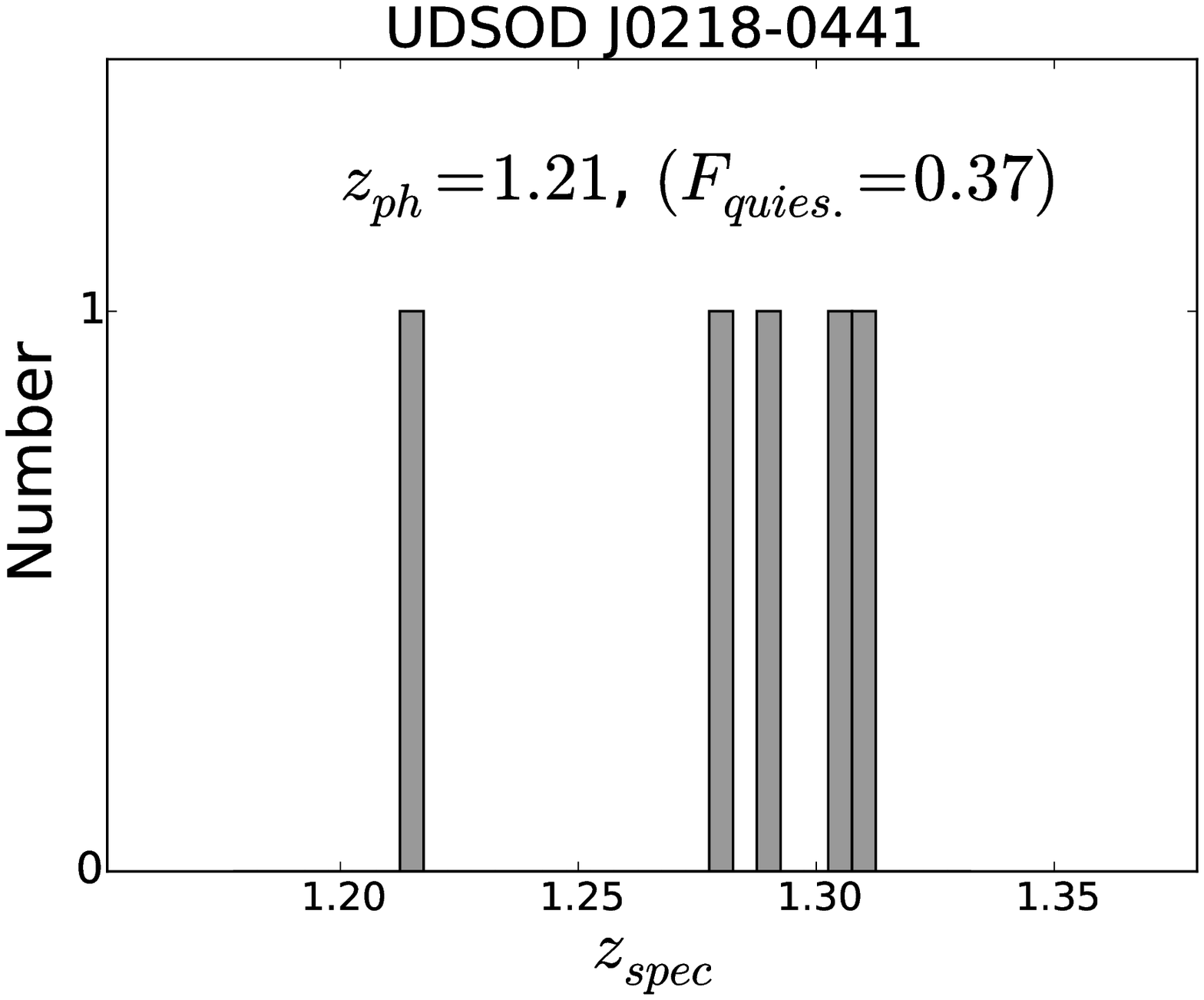}\\
\includegraphics[scale=0.35]{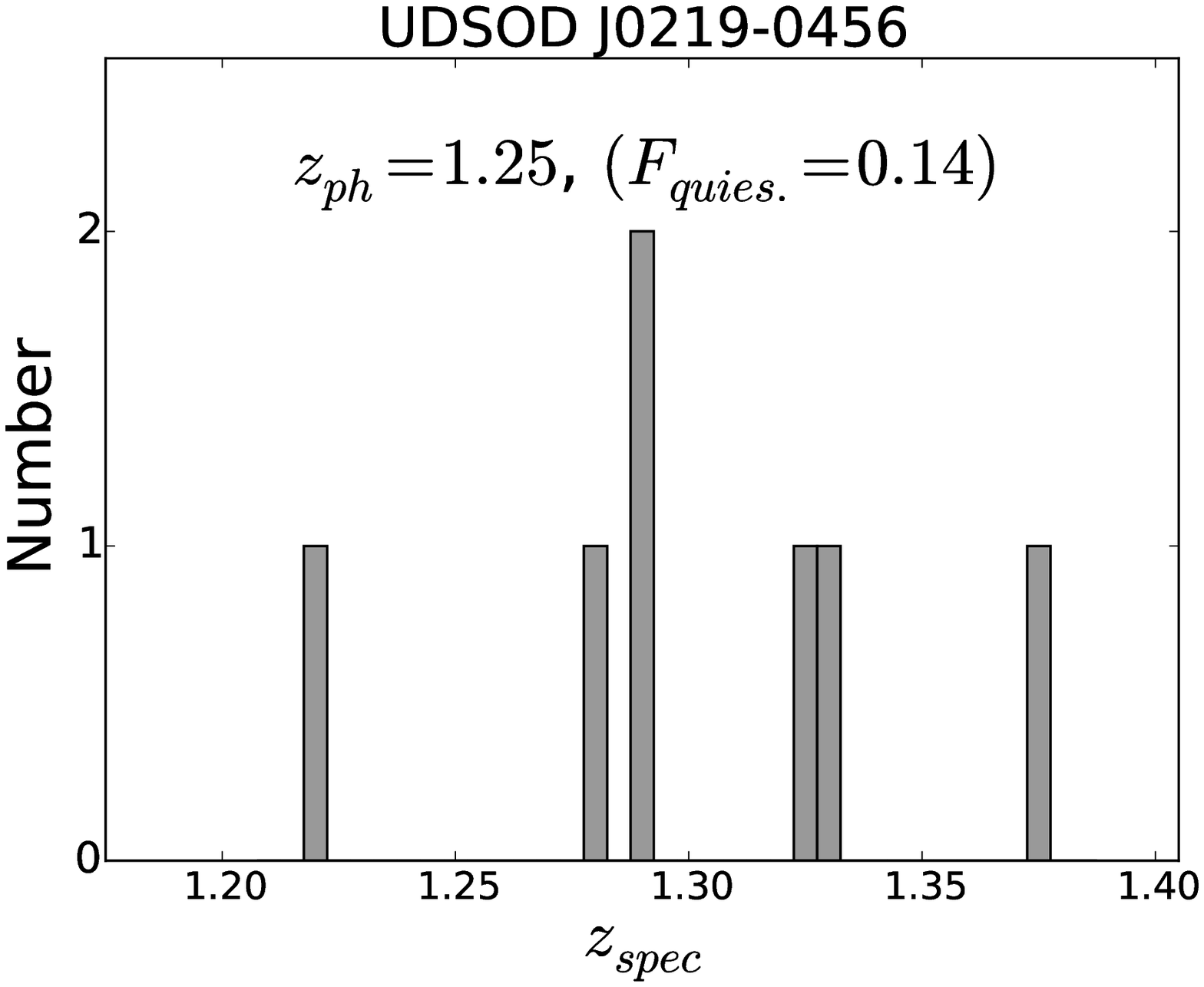}
\includegraphics[scale=0.35]{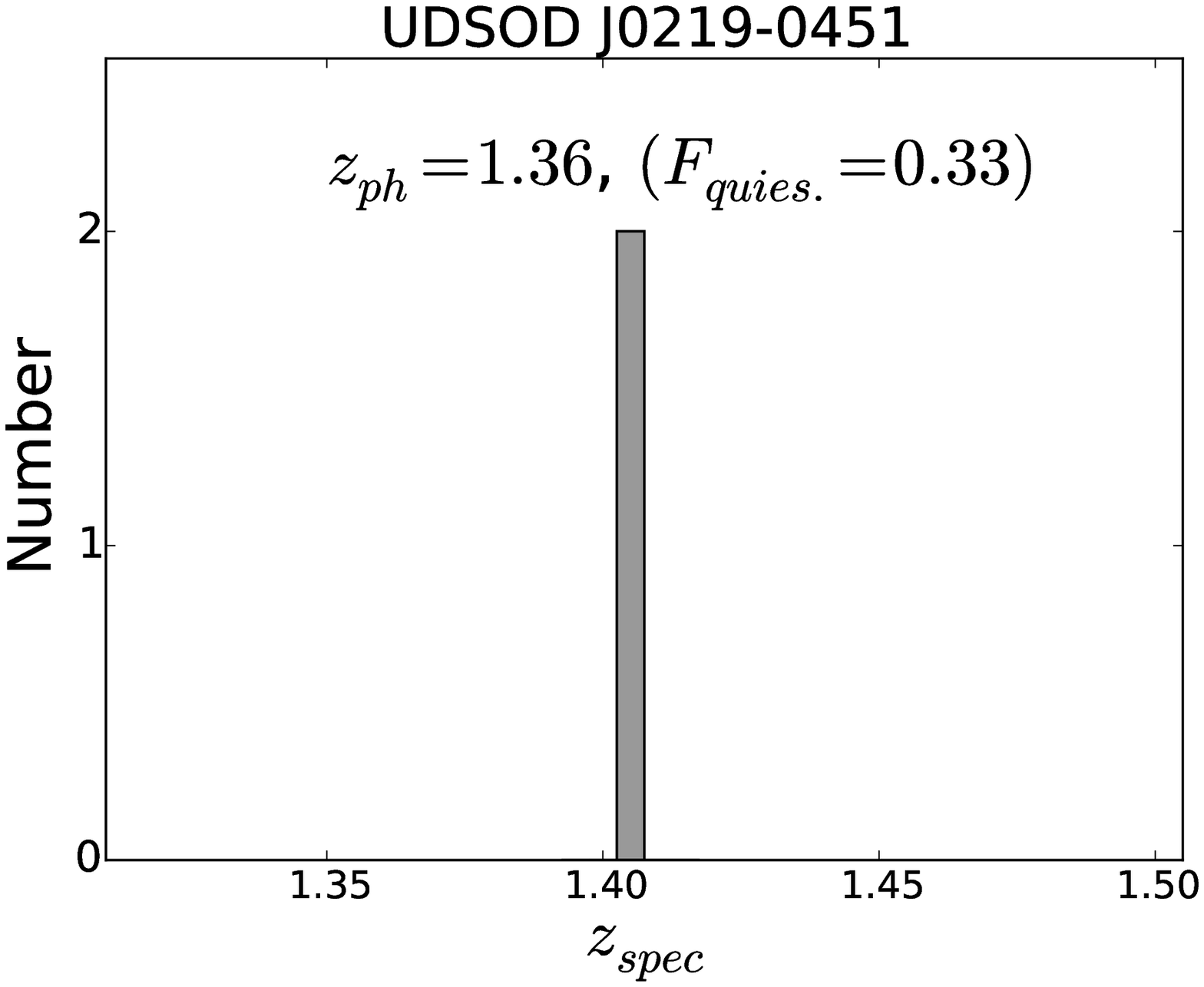}\\
\includegraphics[scale=0.35]{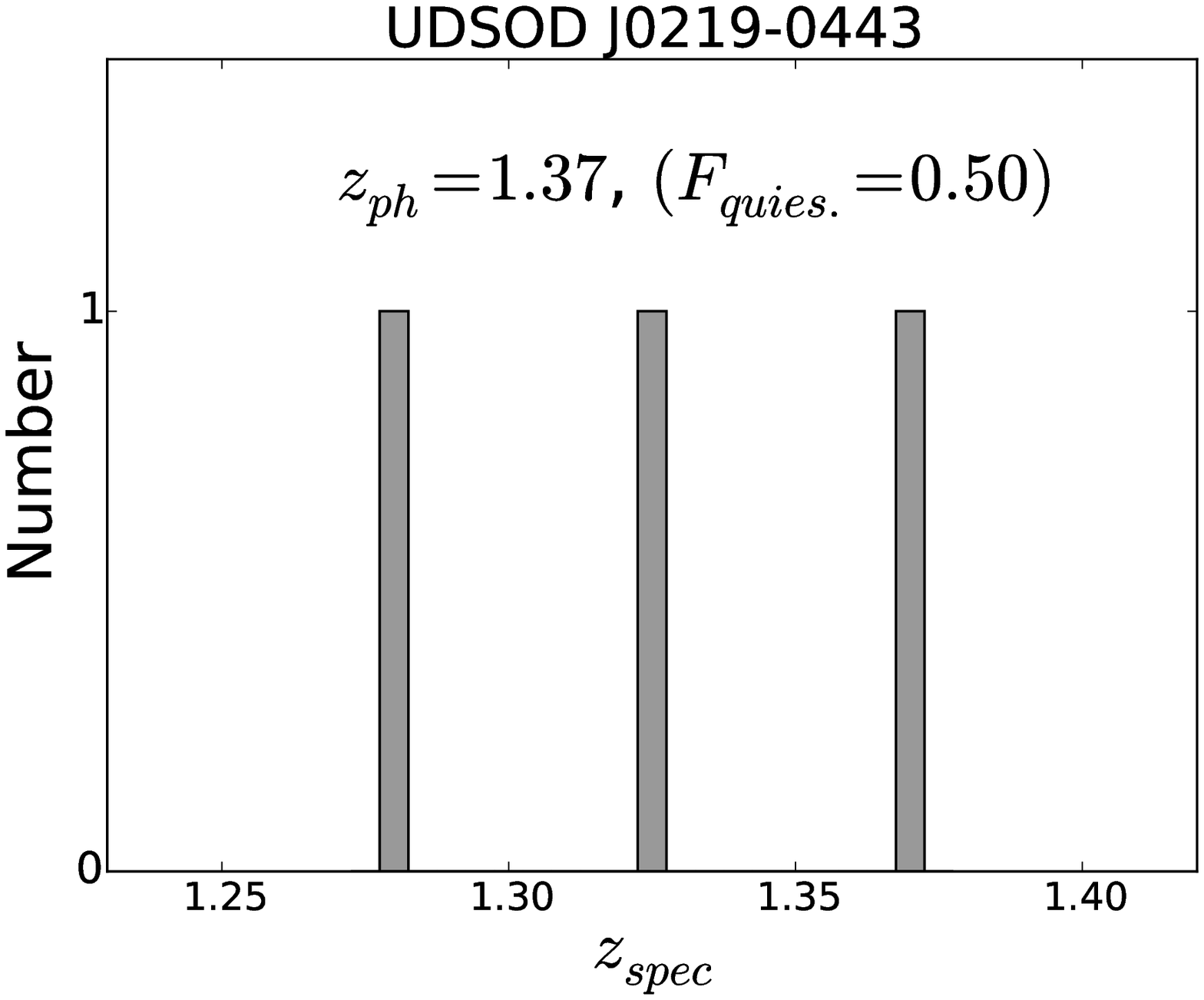}
\caption{Distribution of spectroscopic redshift of unconfirmed cluster candidates.
The open bar indicates galaxies with spectroscopic 
redshift from literature. 
We cannot confirm these candidates because of the small 
numbers of available redshifts.} 
 \label{uncandspzdst}
\end{figure*}

\begin{figure*}
\centering
 \includegraphics[scale=0.75]{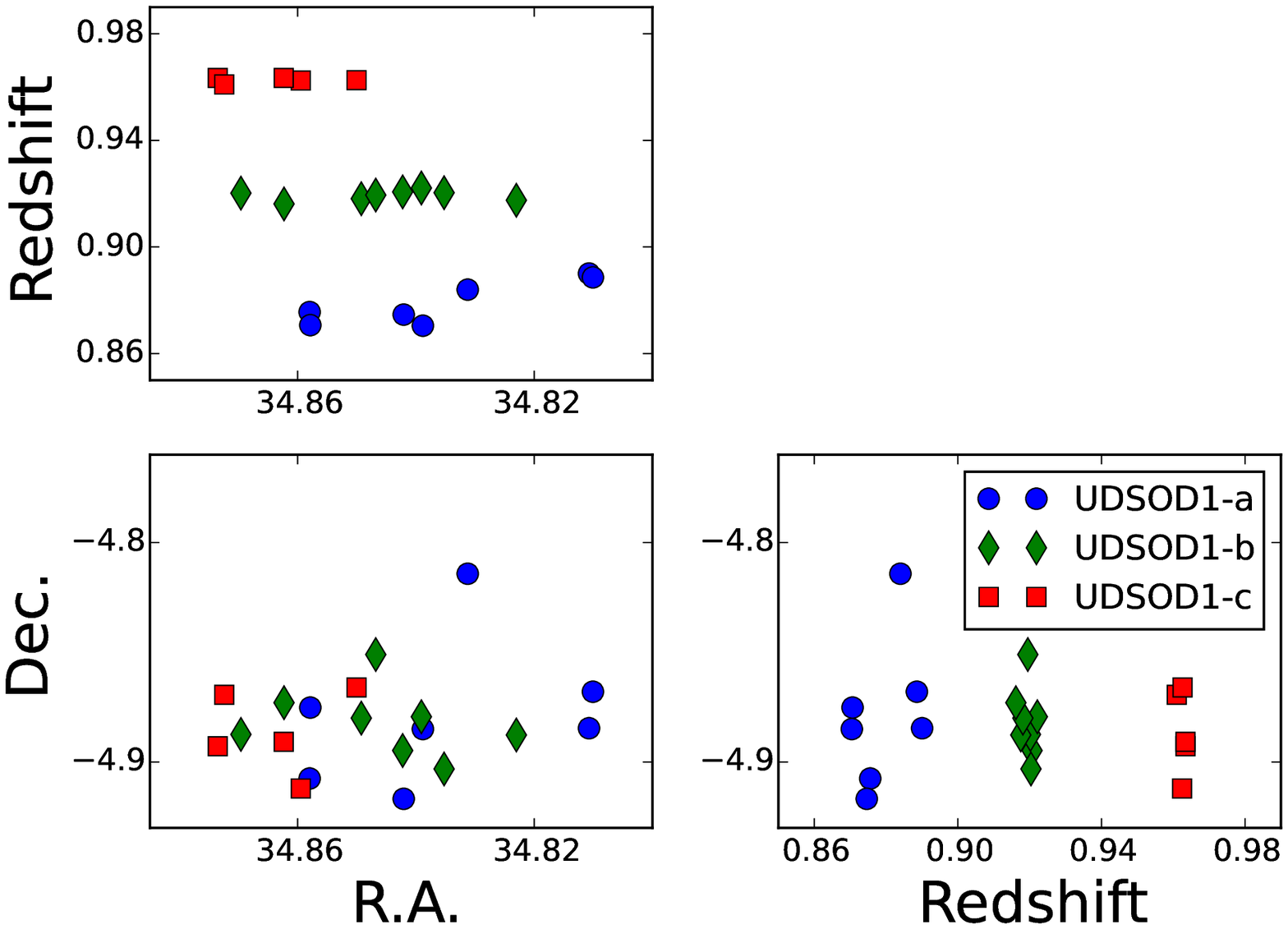}
\caption{The three-dimensional distribution of three 
overdense structures belonging to UDSOD1. 
The blue circles, the green diamonds, and the red squares 
are the spectroscopically confirmed members of UDSOD1-a, 
UDSOD1-b, and UDSOD1-c, respectively. 
Among these three structures, UDSOD1-c shows a slight offset 
in its central position compared to UDSOD1-a and UDSOD1-b.}
 \label{3dudsod1}
\end{figure*}

\begin{figure*}
\centering
 \includegraphics[scale=0.75]{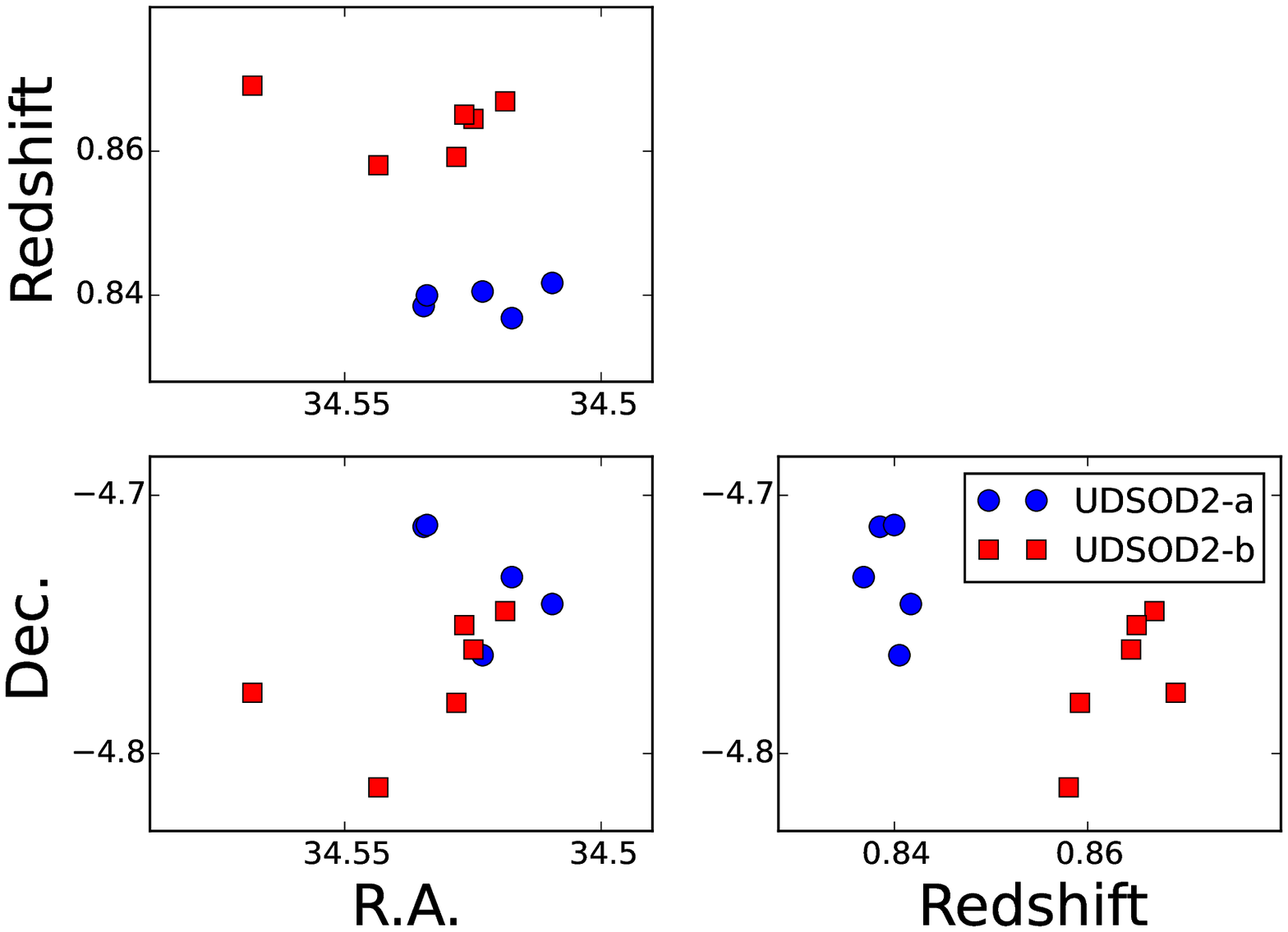}
\caption{The three dimensional distribution of two 
overdense structures belonging to UDSOD2. 
The blue circles and the red squares 
are the spectroscopically confirmed members of UDSOD2-a 
and UDSOD2-b, respectively.}
 \label{3dudsod2}
\end{figure*}

This overdensity is at a higher redshift, $z = 1.294$, than 
the above three overdensities. 
Within a 2 Mpc (1 Mpc) projected radius, there are nine (seven) galaxies 
whose redshifts are measured and confirmed as the overdensity members. 
The redshift distribution is shown in the lower right panel 
of Figure~\ref{clspzdst}. 

As can be seen in Figure~\ref{clspzdst}, the redshift 
distribution of this cluster has a red wing. 
The two galaxies at $z\sim 1.32$ are separated 
by $\sim 1300$ km s$^{-1}$ from the other seven members. 
If we apply a tighter velocity gap of 1200 km s$^{-1}$ 
to identify cluster/group members, these two galaxies will be excluded. 
The total stellar mass and stellar-mass 
calibrated $M_{200}$ are $1.0 \times 10^{12}$ M$_{\odot}$ and 
$1.0 \times 10^{14}$ M$_{\odot}$. 
We suggest that this overdensity is either a cluster or group 
with M$_{halo} \sim 10^{14}$ M$_{\odot}$, although no 
X-ray counterpart to this structure was identified. 
The quiescent galaxy fraction is 0.40, which is intermediate 
among three $z \sim 1.3$ targets.

\section{Extended Overdense Structures}\label{ESS}

Besides identifying galaxy clusters/groups, we have also 
searched for LSSs around the spectroscopically identified 
clusters/groups, combining 
the spectroscopic redshifts from our Magellan observation and 
those from the literature \citep{sma08,sim12,aki15}.
Through this search, we find extended LSSs, 
with sizes (physical scales) of up to $\sim 10$ Mpc, 
around three overdensities (UDSOD1, UDSOD2, and UDSOD4), 
while no significant extended LSS is found near UDSOD3.  

\subsection{Large-scale Structure near UDSOD1-a}

\begin{figure*}
\centering
 \includegraphics[scale=0.23]{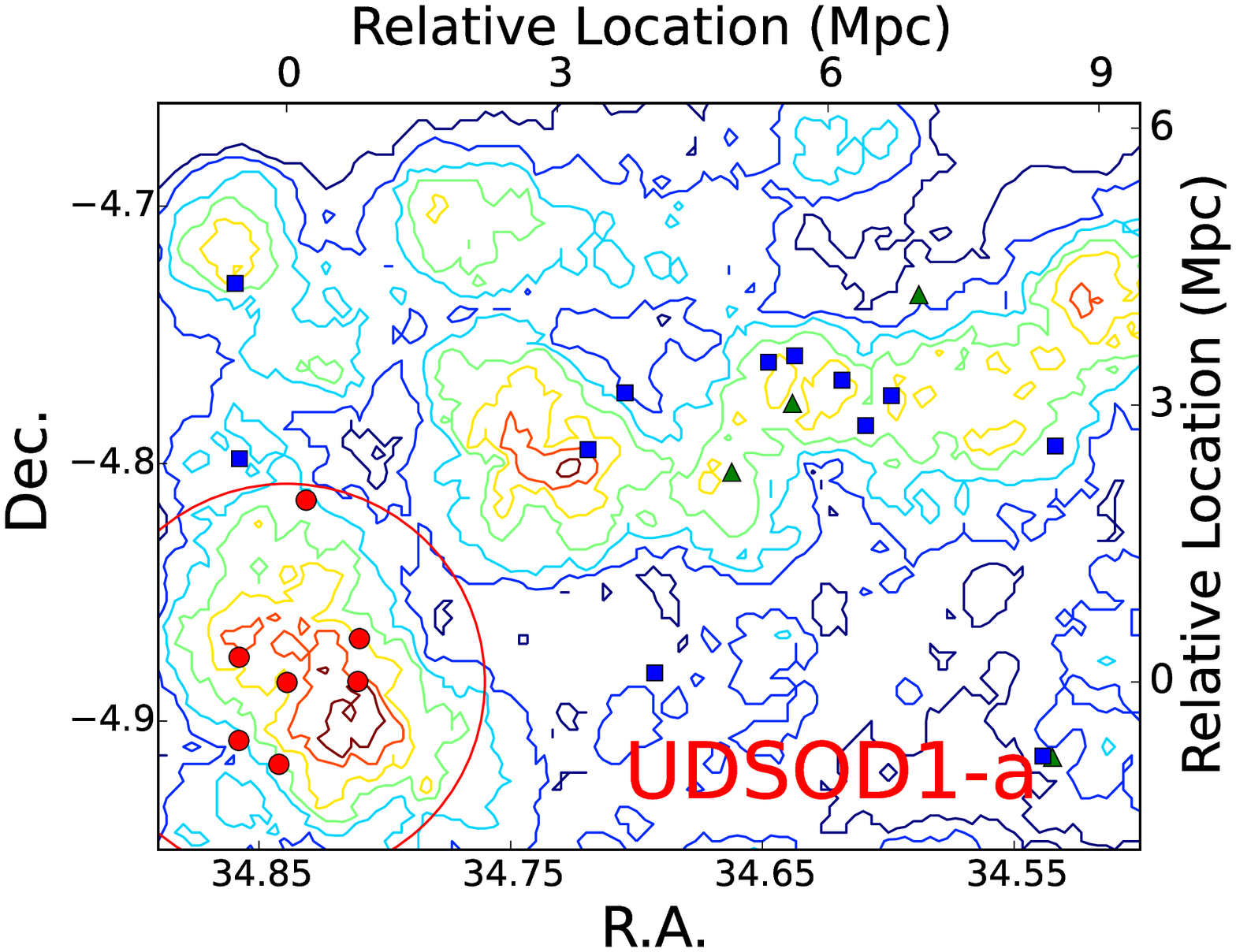}
 \includegraphics[scale=0.21]{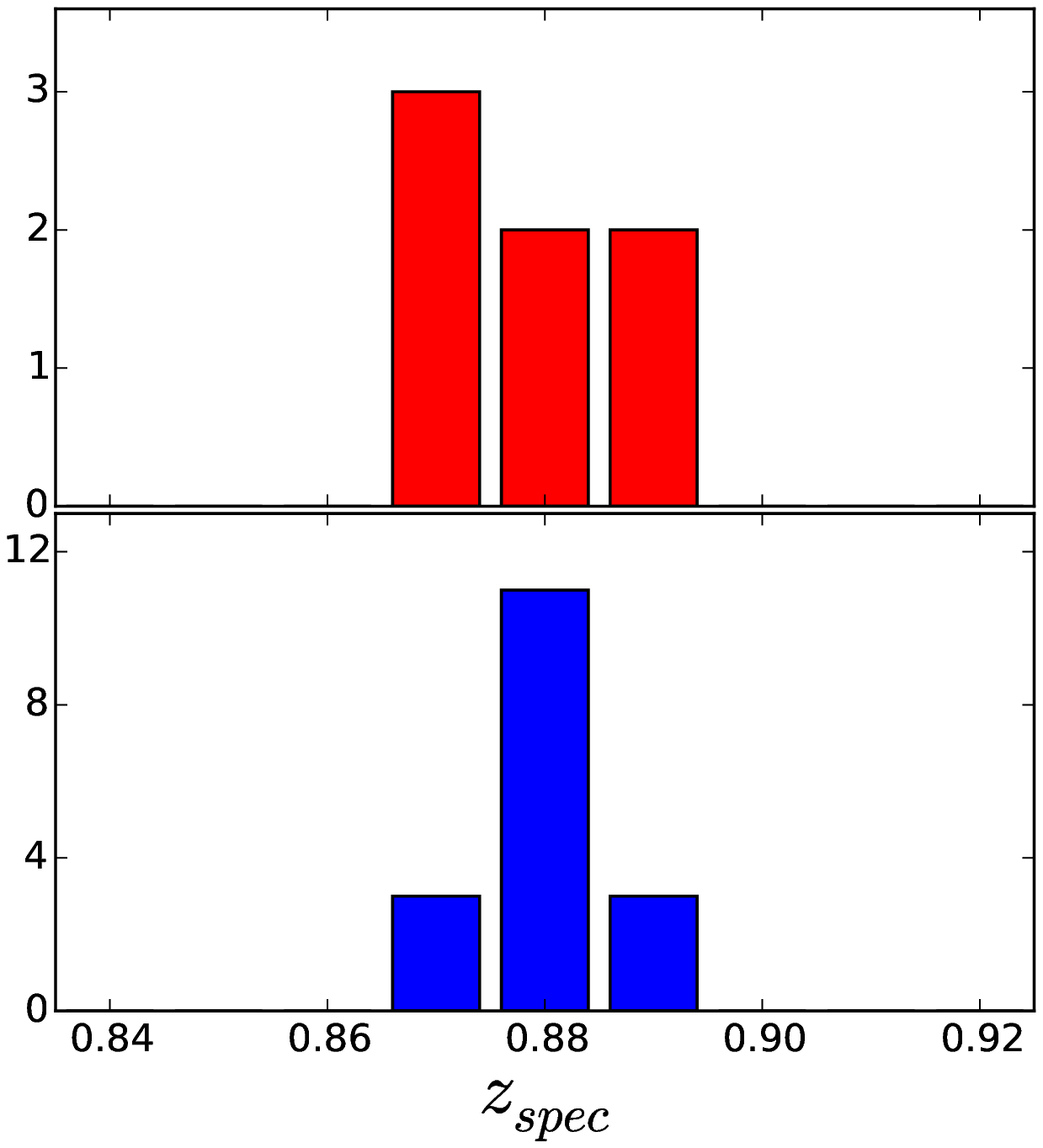}
\includegraphics[scale=0.23]{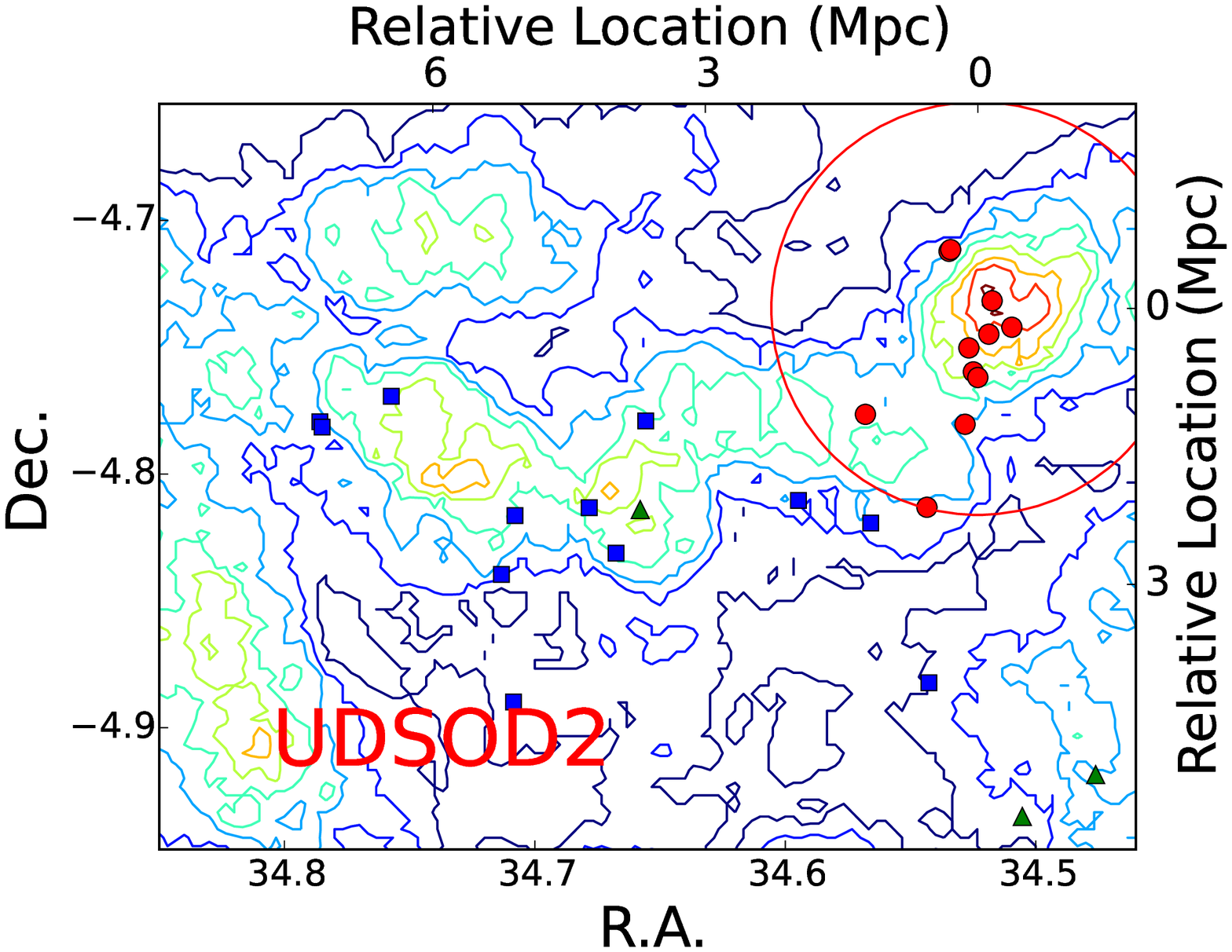}
\includegraphics[scale=0.21]{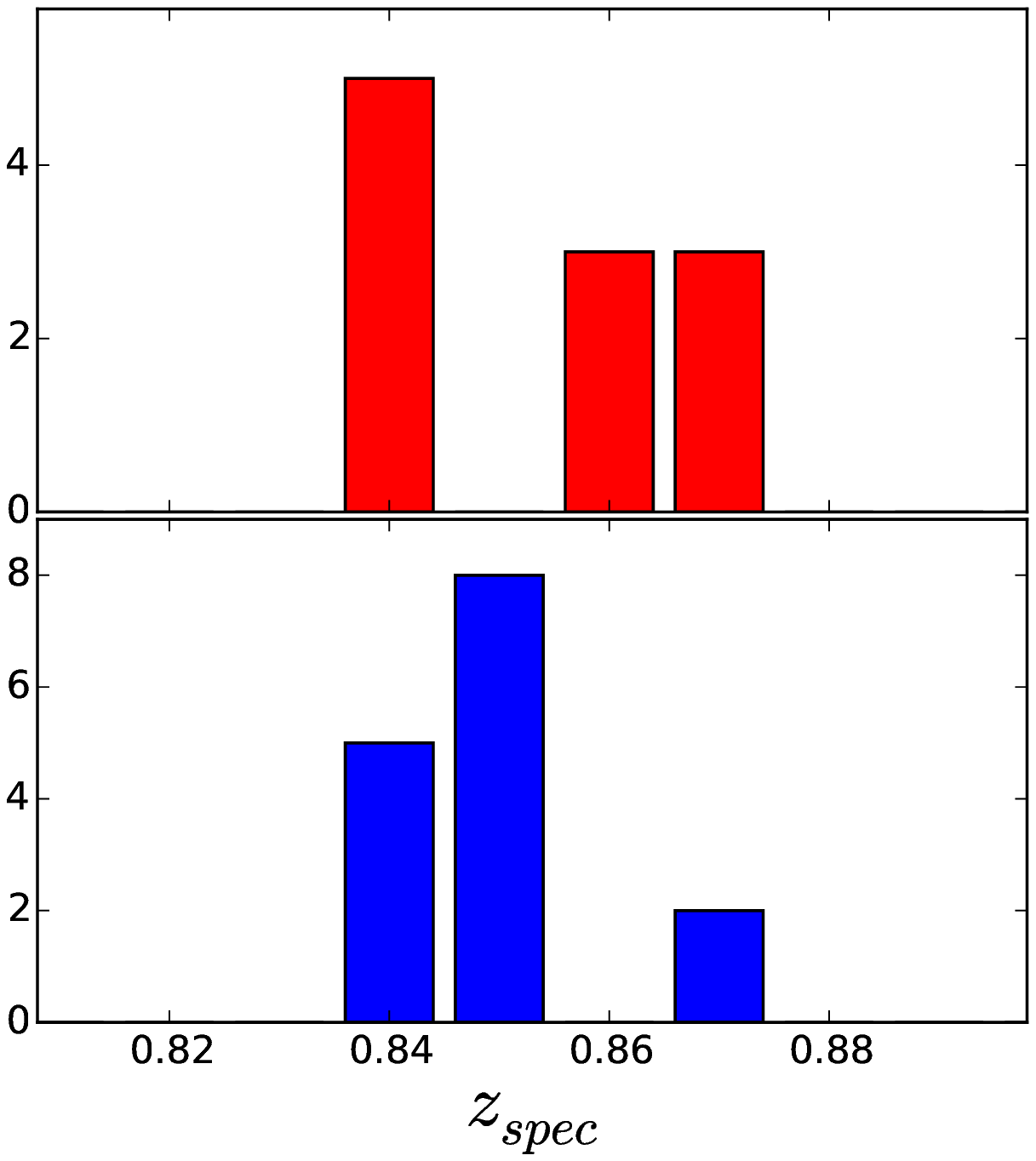}\\
\includegraphics[scale=0.23]{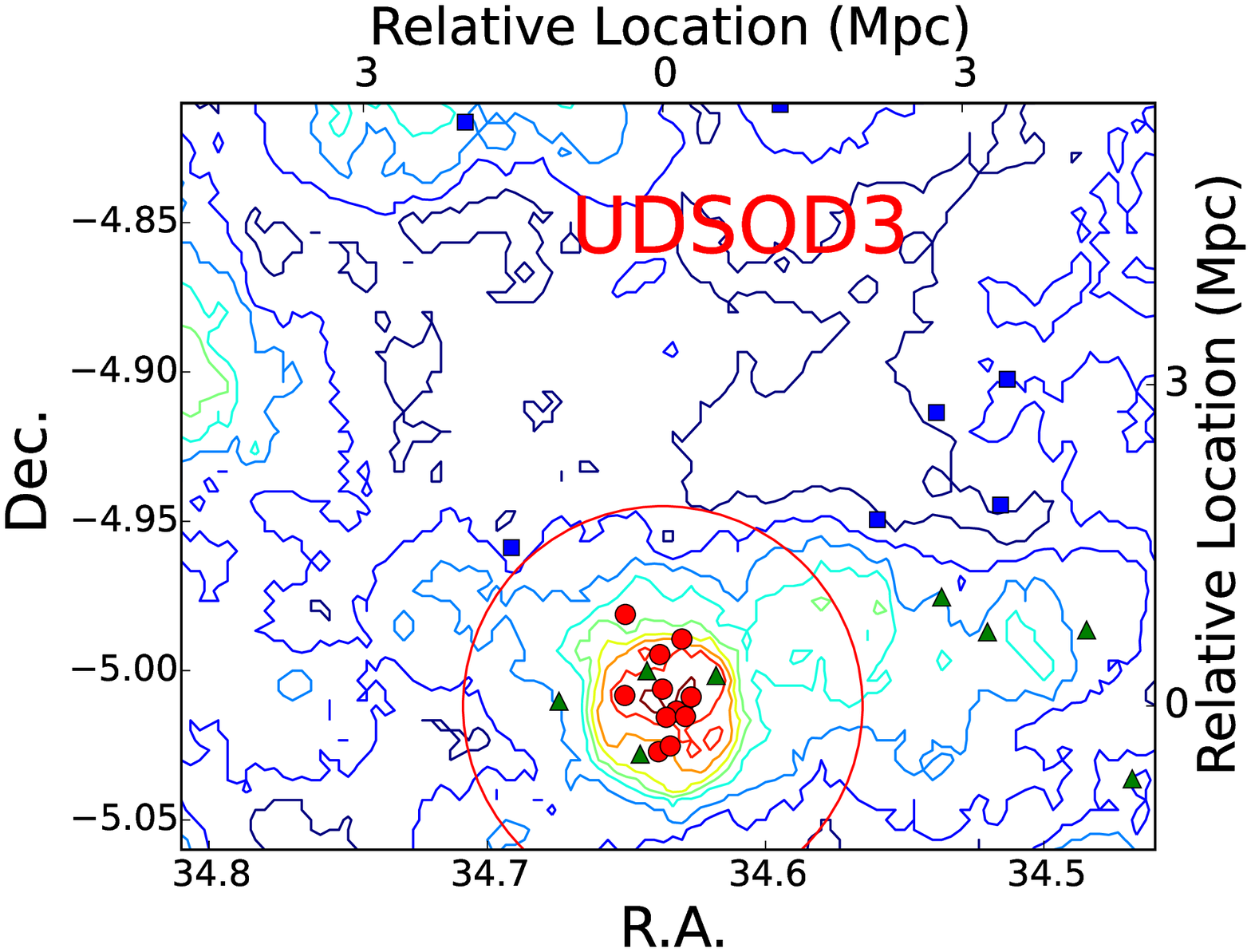}
\includegraphics[scale=0.21]{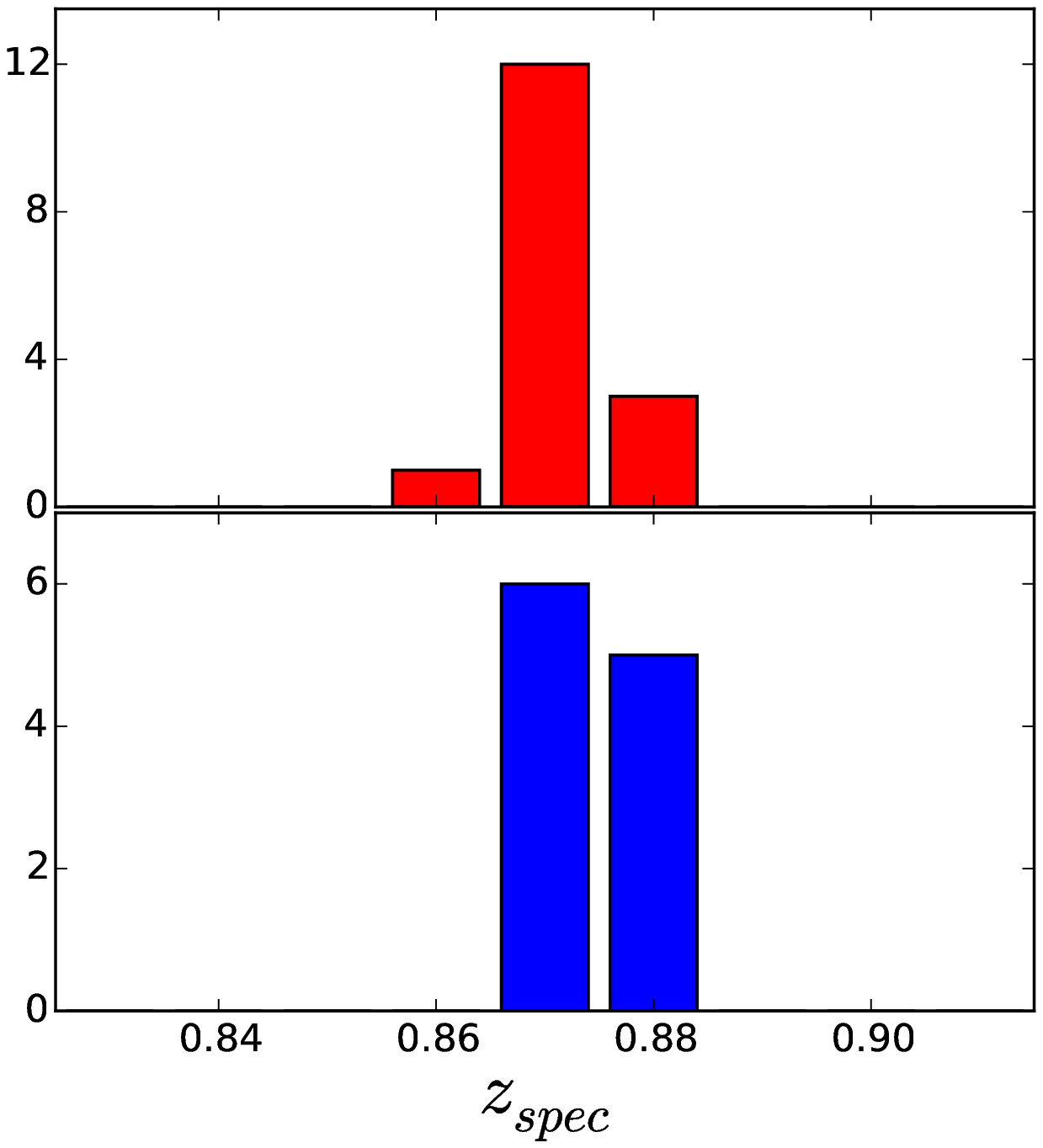}
\includegraphics[scale=0.23]{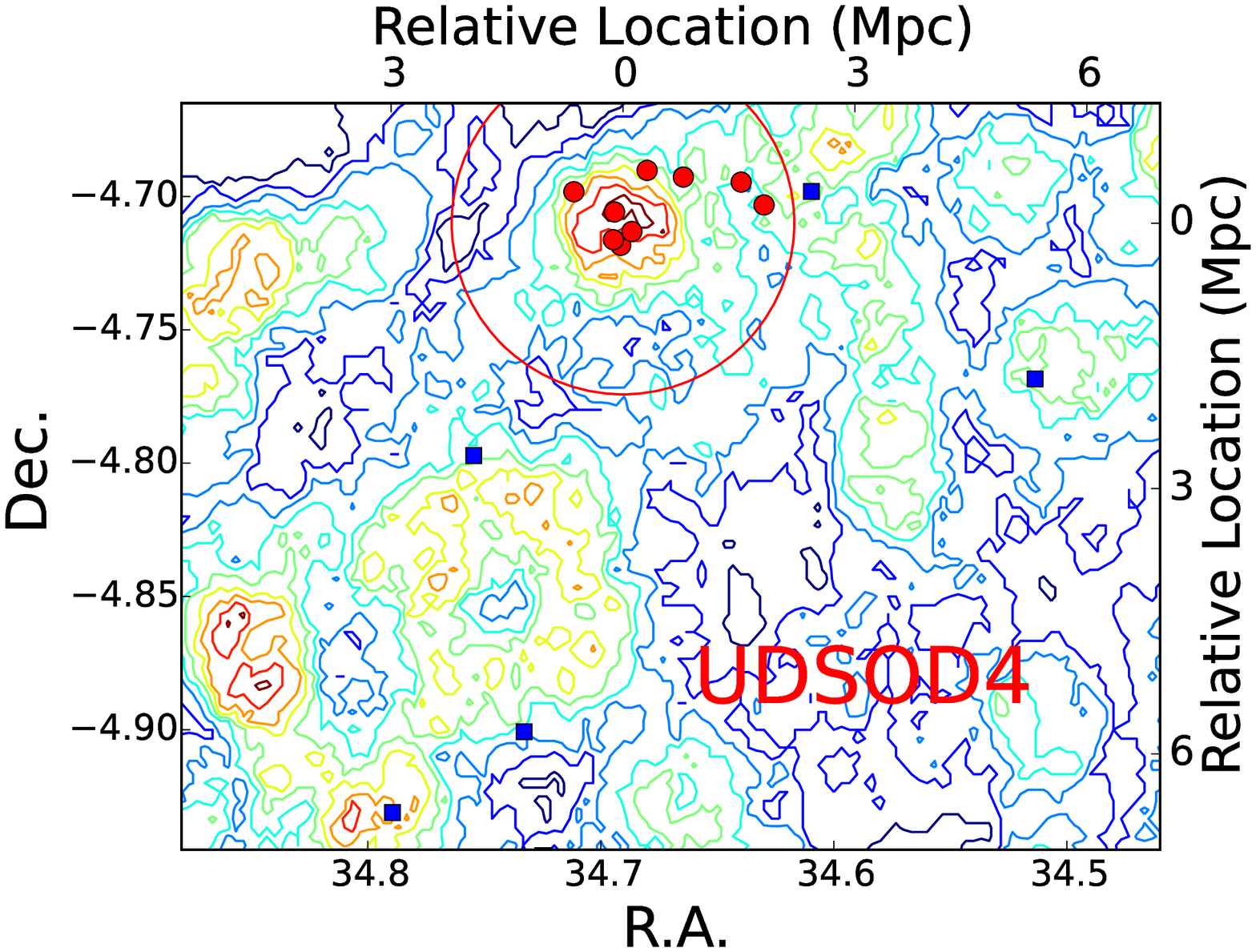}
\includegraphics[scale=0.21]{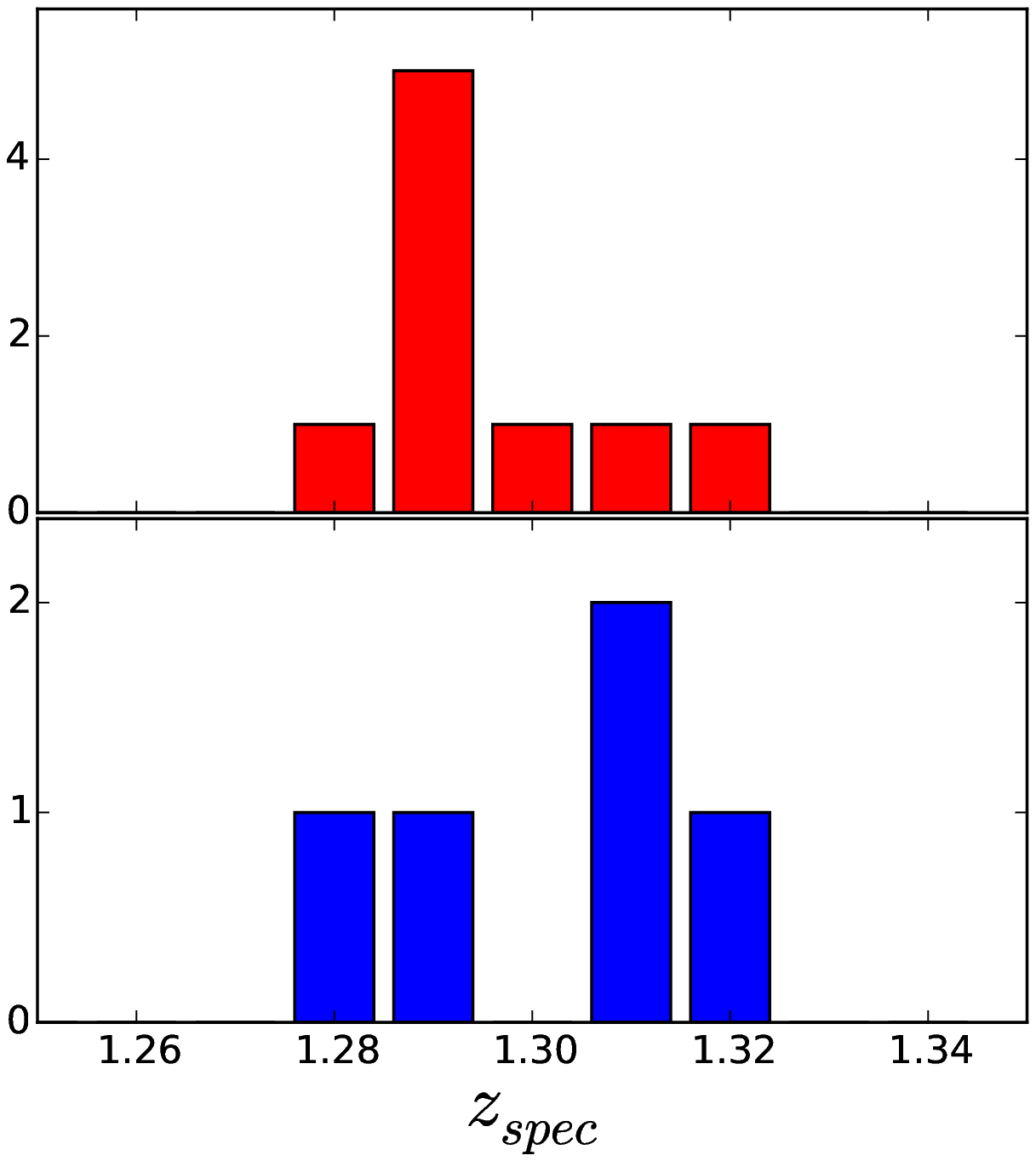}
\caption{Galaxy density map in the regions near the galaxy clusters/groups: 
{\bf(Upper left)} near UDSOD1-a, {\bf(Upper right)} near UDSOD2, 
{\bf(Lower left)} near UDSOD3, and {\bf(Lower right)} near UDSOD4.  
Galaxies with only photometric redshift are also included 
in the galaxy number density measurement shown as contours. 
The red filled circles and blue squares show galaxies whose 
spectroscopic redshifts are measured from our Magellan observation. 
The red filled circles are the members of each cluster whose 
location is shown as the red large open circle in each figure.
The blue squares are galaxies whose spectroscopic redshifts 
are within the spectroscopic redshift range of each cluster. 
The green triangles indicate galaxies with spectroscopic 
redshifts taken from the literature. 
The large red circles enclose the overdensities 
within 2 Mpc projected radius. 
Relative locations (from cluster center) are given in physical scales.
The red and blue histograms in each panel show the spectroscopic redshift 
distributions of overdensity members and non-members, respectively.}
 \label{lssdens}
\end{figure*}

Contours in the upper left panel of Figure~\ref{lssdens} show 
the galaxy surface number density near UDSOD1-a within a narrow 
redshift bin. 
The galaxy surface number density is measured by combining 
spectroscopic as well as photometric redshifts. 
For galaxies with photometric redshift only, we 
include galaxies whose redshifts are within the range, 
$|\Delta z| / (1+z_{cl}) \leq 0.023$, where 
$z_{cl}$ is the redshift of UDSOD1-a. 
For galaxies with spectroscopic redshift, we include 
galaxies whose spectroscopic redshifts are within the 
redshift range of UDSOD1-a, of 
$0.87 \leq z \leq 0.89$.
The UDSOD1-a is at the bottom left region of this 
density map.

Interestingly, we find an extended LSS 
in the western side of UDSOD1-a with a 
scale of $\sim 8$ Mpc.
There are also several dense, group-like, structures along 
this LSS, including the ones 
at ($\alpha, \delta$) $\sim$ (34$\,^{\circ}$.74, $-4\,^{\circ}$.80) 
and (34$\,^{\circ}$.62, $-4\,^{\circ}$.77). 
Although the density contours are constructed using mostly 
galaxies with photometric redshifts, galaxies with spectroscopic 
redshifts follow the density contour closely, 
lending credibility to the association of the LSS to UDSOD1-a. 
We note that UDSOD1-b and UDSOD1-c show similar LSSs around them, 
and we suggest that the UDSOD1-a through c are all a part of 
a single LSS.  

\subsection{Large-scale Structure near UDSOD2}

We also find an extended LSS near UDSOD2, as 
shown in the upper right panel of 
Figure~\ref{lssdens}, which extends up to $\sim 10$ Mpc. 
The UDSOD2 is in the top right region.
In this figure, we show the spectroscopically confirmed 
member galaxies of UDSOD2-a and UDSOD2-b together, 
as red circles, since the velocity interval between the 
redshifts of UDSOD2-a and UDSOD2-b is about 4000 km s$^{-1}$, 
or 30 Mpc on the physical scale, indicating 
these two structures may belong to a single LSS.
For galaxies with only a photometric redshift, we include 
those within $|\Delta z| / (1+z_{cl,mean}) \leq 0.023$ to 
construct the density contour.
Here, $z_{cl,mean}$ is the mean redshift of UDSOD2-a and 
UDSOD2-b. 
Over 20 galaxies with spectroscopic redshifts in a similar 
range as UDSOD2 are distributed in the filamentary 
structure toward the east side of UDSOD2. 

\subsection{Large-scale Structure near UDSOD3}

UDSOD3 is 
relatively isolated without no significant LSSs 
near it, 
as can be seen in the lower left panel of 
Figure~\ref{lssdens}. 
This shows a sharp contrast to the other overdensities 
at a similar redshift. 

\subsection{Large-scale Structure near UDSOD4}

In the lower right panel of Figure~\ref{lssdens}, we 
show the LSS around UDSOD4, which is 
at $z \sim 1.3$. 
Not many spectroscopic redshifts are available around 
UDSOD4, but the density map from the photometric redshift 
sample suggests an association of UDSOD4 with 
extended intermediate density regions over 3-5 Mpc 
from cluster UDSOD4. 

\section{Connection between SF Activities in Clusters and LSSs}\label{cause}

In Figure~\ref{qfrac}, we have shown that there is a large variation in 
the quiescent galaxy fraction among individual galaxy cluster candidates  
even though they reside at a similar redshift. 
Here, we investigate if the cluster SF activities are linked to LSSs. 
We concentrate our analysis on galaxy clusters/groups with 
log($M_{200}/M_{\odot}$) $\geq 13.4$. 
The overdensities that satisfy this condition are UDSOD1-a, b, UDSOD2-a, b, 
UDSOD3 and UDSOD4.   

\subsection{Enhancement of SF Activity in Galaxy Overdensities near LSSs}

In Section~\ref{ESS}, we presented the extended LSSs around 
galaxy overdensities, UDSOD1-a, UDSOD2, and UDSOD4, 
with scales as large as $\sim 10$ Mpc. 
On the other hand, there is no significant extended LSS 
around UDSOD3 --- i.e., this overdensity is relatively isolated.
Interestingly, galaxy overdensities that are embedded in the 
extended LSSs appear to show a higher fraction of SF galaxies 
among members, compared to the relatively isolated galaxy overdensity.

To quantify this large-scale environmental trend, we measure 
the fraction of the area covered by the extended LSSs within 10 Mpc 
around each galaxy cluster using the friends-of-friends (FOF) 
method.
We study the LSSs within 10 Mpc, based on our finding of LSSs 
with sizes of $\sim 10$ Mpc as presented in Section~\ref{ESS}.
First, we select the sky area around each galaxy cluster within 
a projected radius of 10 Mpc and within the survey 
boundary and divide the area into cells with a size of 
$12 \arcsec \times 12 \arcsec$. 
Next, we measure galaxy number counts in each cell  
within the redshift range of $|\Delta z| / (1+z_{cl}) \leq 0.023$. 
Then, we find overdense cells with a number count greater 
than 2-$\sigma$ from the mean background count. 
Finally, we mark the connected overdense cells with a linking 
length of 2 Mpc using the FOF algorithm. 

\begin{figure*}
 \centering
\plotone{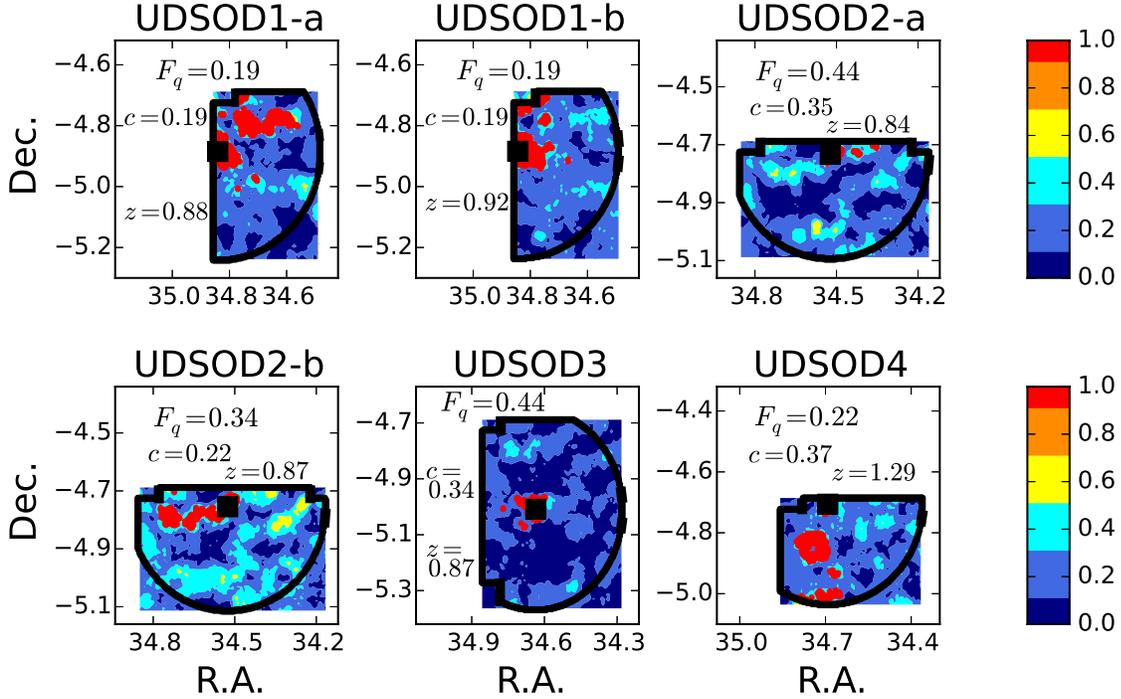}
\caption{Distribution of overdense ($ \geq 2 \sigma$) 
LSSs near confirmed galaxy overdensities 
(shown as red region). Thick black line in each panel shows the boundary 
which encloses the total area. The boundary is set by either 10 Mpc projected radius or 
the survey boundary. These connected structures are 
found using the friend-of-friend (FOF) algorithm. 
Coloured contours show the normalized galaxy surface number density levels.
Here, similarly as in density maps (Figure~\ref{lssdens}), 
we can see that UDSOD3 is relatively isolated, unlike the other 
confirmed overdensities at a similar redshift, such as UDSOD1-a and 
UDSOD2-b. In each panel, we show the quiescent fraction ($F_{q}$), redshift, 
and the concentration parameter ($c$) of each cluster.}
 \label{fofmap}
\end{figure*}

\begin{figure*}
\centering
 \includegraphics[scale=0.65]{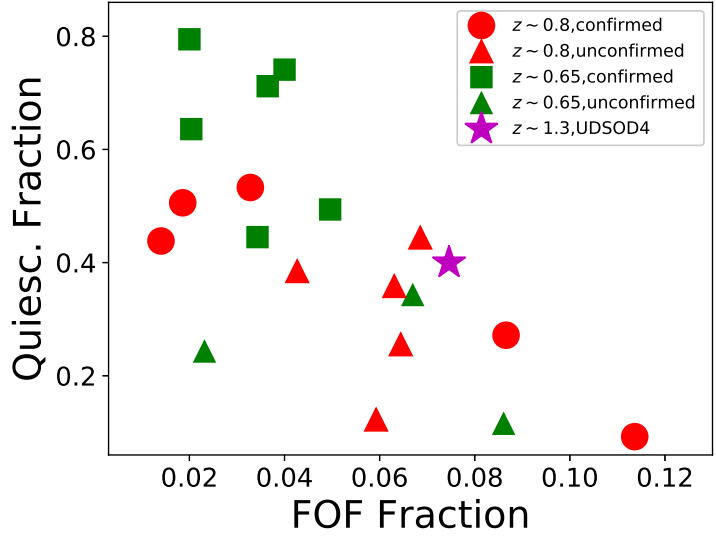}
\caption{Correlation between the fraction of the area covered by 
the overdense region connected with cluster ($x$-axis) and 
the quiescent galaxy fraction ($y$-axis).  
The red circles show the values for spectroscopically 
studied overdensities at $z \sim 0.8$ from this work, 
and the red triangles are for cluster 
candidates at a similar redshift. The green triangles and 
green squares show the values of cluster candidates at a lower 
redshift ($z \sim 0.65$). Among these, green squares 
are for $z \sim 0.65$ candidates confirmed in other works \citep{van06,gal18}.
The magenta star shows the value of UDSOD4, which is at $z=1.29$.
The FOF fraction and the quiescent fraction show anti-correlation. 
We do not include UDSOD1-c because of its very small mass 
($\sim 10^{12}$ M$_{\odot}$).}
 \label{qffof}
\end{figure*}

The results are shown in Figure~\ref{fofmap} 
for the confirmed overdensities 
with log(M$_{200}$/M$_{\odot}) > 13.4$.
In this figure, the thick black line in each panel shows 
the boundary which encloses the total area. This boundary is set 
by either 10 Mpc projected radius from each overdensity (whose 
center is shown with the black square) or by the survey boundary.
The red-coloured region shows the connected LSS found. 
Here, we can see that UDSOD2-a and UDSOD3 are relatively isolated 
while the other overdensities are surrounded with extended LSSs.
The fraction of the area covered by the extended LSS (red region) 
within the total area (gray area) --- `FOF fraction' 
--- is 0.11, 0.09, 0.01, 0.03, 0.02, and 0.08 for UDSOD1-a, 
UDSOD1-b, UDSOD2-a, UDSOD2-b, UDSOD3, and UDSOD4, respectively. 
Note that we did not make a correction for the region where no 
data are available. 
As shown with the red circles and the purple star in Figure~\ref{qffof}, 
there is a strong anti-correlation between the 
FOF fraction and the $F_{q}$, 
with the Pearson correlation coefficient of -0.89. 

We test if this FOF fraction is a reliable tracer of actual LSSs 
by checking if the FOF fractions from random distribution of galaxies 
have low values. 
We find that the median number of galaxies in a photometric redshift 
slice is about 500 within 10 Mpc projected radius circle for 
our sample galaxies.  
Therefore, we distribute 500 points randomly within 10 Mpc 
projected radius circle, 
then calculate FOF fraction following the same procedure explained 
in the above paragraph. 
We repeat this procedure 100 times, and find that FOF fractions are 
smaller than 0.01 in all cases.

\begin{figure*}
\plotone{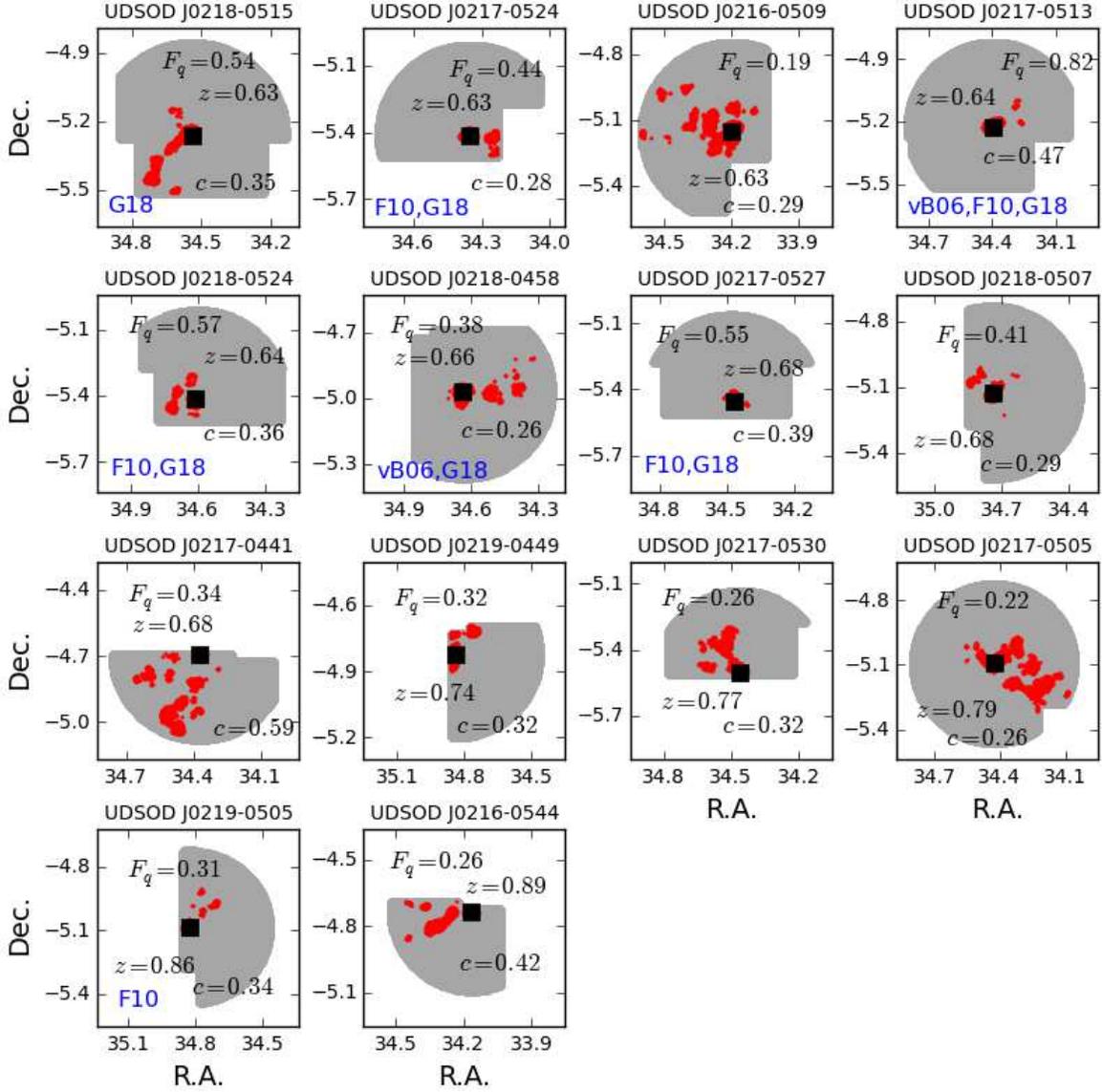}
\caption{Distribution of the overdense ($ \geq 2 \sigma$) 
large scale structures (shown as red regions) 
near galaxy cluster candidates at $0.55 < z < 0.9$. 
Gray region shows the total area whose 
boundary is set by either 10 Mpc projected radius or the survey 
boundary. These connected structures are found using the friend-of-friend 
(FOF) algorithm. 
In each panel, we show the quiescent fraction ($F_{q}$), redshift, 
and the concentration parameter ($c$) of each cluster.
Some of our candidates are detected are detected in 
X-ray and their halo mass estimated with the X-ray data (F10) 
or spectroscopically confirmed in other works 
\citep[][vB06; G18]{van06,gal18}, and the references are shown in 
each panel.}
 \label{fofmapphz}
\end{figure*}

Encouraged by this trend, we also analyzed the LSSs 
around 16 other cluster candidates with log $(M_{200}/M_{\odot}) > 13.4$ 
at $0.55 < z < 0.9$ in the UDS field from L15. 
Note that 6 out of these 16 candidates have been identified 
as clusters or groups either spectroscopically or through 
X-ray detection. 
These cluster candidates are listed in Table~\ref{clcndprop} 
along with their associated properties.
We choose this redshift range since the photometric 
redshift uncertainty is relatively small 
($|\Delta z|$/$(1+z)$ = 0.017), 
and the overdensity identification has been proven to be 
secure even with photometric redshifts only 
\citep[this work and others, e.g.,][]{lai16} in this redshift 
range. 
We show the large-scale environment around these cluster 
candidates in Figure~\ref{fofmapphz}.
The $F_{q}$ as a function of the FOF fraction is shown 
in Figure~\ref{qffof}, where the red triangles and 
the green symbols 
are the cluster candidates at $0.7 < z < 0.9$ and 
at $0.55 < z < 0.7$, respectively. 
With this larger set of data, we confirm the 
anti-correlation between the quiescent galaxy fraction 
of galaxy clusters and the FOF fraction. 

We test the significance of this correlation using the Pearson 
test. The Pearson correlation coefficient, $r$ is $-0.76$ for 
$z \sim 0.85$ samples of clusters and cluster candidates, 
indicating strong anti-correlation. For $z \sim 0.65$ 
cluster candidates, $r = -0.63$.
This trend remains even if the samples from the two redshift 
ranges are combined, although in general, the lower 
redshift sample tends to have higher $F_{q}$ 
for a given FOF-fraction. 

Note that the use of the photometric resdhift cut could include 
foreground and background structures which are not actually at 
the same distance with the overdensity. 
Depending on whether these foreground or background structures 
are connected with the central overdensity (in the photometric 
redshift slice), this could overestimate or underestimate the FOF fractions. 
Therefore, we check this using mock galaxy catalog from \citet{mer13}, 
by comparing the FOF fractions computed from two methods: 
one using galaxies in a narrow physical distance ($\pm$ 20 Mpc) from the 
cluster center, and another using galaxies from the photometric redshift 
cut ($\pm$ 0.023$\times (1+z)$) as we did for the simulation 
(see APPENDIX A) that would include galaxies out to $\pm$ 100 Mpc 
from the cluster center in comoving distance.

\begin{figure*}
\centering
 \includegraphics[scale=0.55]{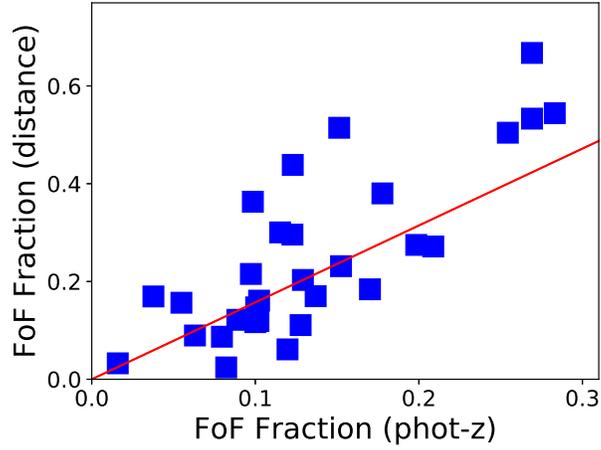}
\caption{Correlation between FOF fraction of mock clusters 
measured based on photometric redshift ($x$-axis) and 
physical radial distance ($y$-axis).  
These two quantities show a good positive correlation inferring 
the reliability of measuring FOF fraction using photometric redshift.
The red line shows the line where 
[FoF Fraction (photo-z)]/[FoF Fraction (distance)] = 0.64, which is 
the median value of this ratio.}
 \label{fofphsp}
\end{figure*}

Figure~\ref{fofphsp} shows the comparison between FOF fractions of 
mock clusters measured using the sample from the `photometric redshift' 
cut and the sample with the `physical distance' cut. 
This figure shows the FOF fractions from the two methods 
show a positive correlation with the rms scatter of 0.60, but 
the photometric redshift based FOF fraction being somewhat underestimated by a 
factor of $\sim 0.64$.
The underestimate in the FOF fraction for the photometric redshift cut 
sample can be understood as the following. 
More galaxies are included in the photometric redshift cut sample, and 
this increases the number density fluctuation, $\sigma$. 
Due to this increase in $\sigma$, the area above the 2-$\sigma$ fluctuation 
threshold decreases, leading to the underestimation of the FOF fraction.

\subsection{Dependence of SF Activity of Galaxy Clusters on Halo Concentration}

Simulations show the growth of overdense areas in 
connection to LSSs of filaments \citep[e.g.,][]{ebe04}.
Under this kind of scenario, isolated galaxy clusters like 
UDSOD3 are the ones that get disconnected from the LSS and 
finish its dynamical evolution early. 
Then, we can expect that clusters like them are virialized 
and more concentrated 
as they evolve further in an isolated environment.

We investigate the correlation between the quiescent 
galaxy fraction and the concentration of galaxy distribution
of the confirmed overdensities and cluster candidates 
with log (M$_{200}$/M$_{\odot}) \geq 13.4$.
For this, we define the concentration parameter of the 
galaxy overdensity as `$-$log ($A_{0.3}/A_{0.7}$)', where 
$A_{0.3}$ and $A_{0.7}$ are the areas that contain 
$30 \%$ and $70 \%$ of the total member galaxies. 

\begin{figure*}
\centering
 \includegraphics[scale=0.65]{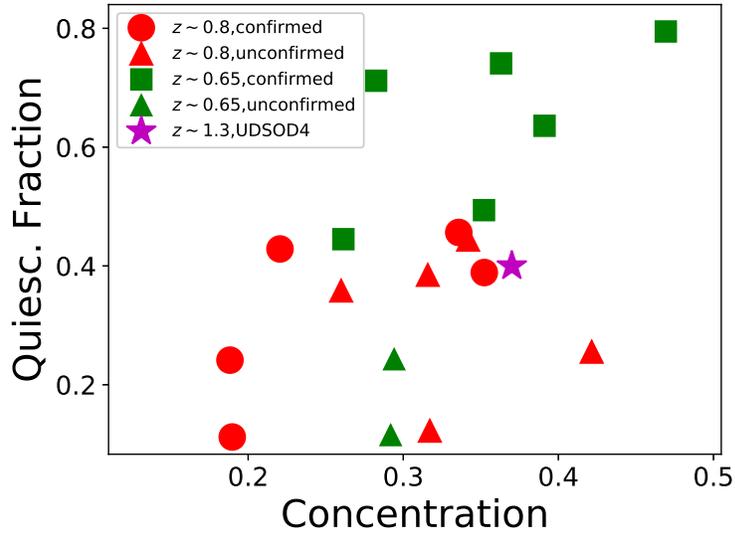}
\caption{Correlation between concentration of galaxy clusters 
($x$-axis) and quiescent galaxy fraction ($y$-axis).  
The symbol assignments are the same as Figure \ref{qffof}.}
 \label{qfconc}
\end{figure*}

Figure~\ref{qfconc} shows the correlation 
between the concentration parameter 
and the quiescent fraction of various overdensities. 
The correlation between $F_{q}$ and concentration is weak 
with the Pearson coefficient of 0.61, mainly due to three 
outliers with a low $F_{q}$ and a high concentration at 
$z \sim 0.8$ and $z \sim 1.3$. 
However, if we remove these three points, the correlation 
becomes stronger (Pearson coefficient $\sim 0.70$). 
The same is true if we focus on the four galaxy overdensities 
at $z \sim 0.8$. 
We conclude that at a given redshift, there seems to be a 
correlation between $F_{q}$ and concentration of 
galaxy overdensities, although a larger sample of 
spectroscopically confirmed samples is needed to reach a 
firm conclusion.

\subsection{Web-feeding Model of SF in Overdensities at $z \sim 1$}
 
Based on the correlation of the quiescent galaxy fraction with two parameters, 
the FOF fraction of the surrounding LSS and the concentration parameter of 
the overdensity, we suggest the ``Web-feeding model" to explain the variety 
of SF activities in clusters at $z \sim 1$. 
In the model, we attribute the enhanced SF activities in overdensities 
are due to the inflow of gas and SF galaxies to 
localized overdense areas \citep{fad08,mah12,pin13,dar14,hay17,kle17}. 
The concentration parameter of overdensities stays low in such regions due 
to new SF galaxies filling up the outskirt of galaxy overdensities even 
if overdensity members in-fall to the overdensity center. 
However, once the supply of gas and galaxies stops, i.e., LSS around the 
overdensities disappears, SF activity of the overdensity stops and 
the overdensities become concentrated with continued in-fall of member 
galaxies toward their centers. 
This picture can explain the observed association of the LSS with 
low $F_{q}$ overdensities and the possible high concentration of 
overdensities with high $F_{q}$.
This scenario is consistent with the CWD model \citep{ara16}.

The biased galaxy formation, where galaxies formed earlier in higher 
density regions connected to LSSs, could work against our finding. 
Despite of this opposite effect of the biased galaxy formation working 
on the LSS-$F_{q}$ connection, the fact that we find the anti-correlation 
between the FOF fraction and $F_{q}$ suggests that the Web-feeding effect 
could be stronger than what is found here. 

\section{Summary and Conclusion}\label{summa}

In this paper, we present the results of the MOS observation of high 
redshift ($0.8 < z < 1.3$) galaxy cluster candidates in the UDS field.
Based on the spectroscopic observation and the photometric redshifts, 
we investigate whether SF galaxy fraction in overdense areas is 
enhanced due to the presence of adjacent LSSs. 
The results are summarized below.

\begin{enumerate}

\item {We identify four high-redshift ($0.8<z<1.3$) galaxy 
overdensities, containing 7 to 8 clusters/groups of galaxies.  
The spectroscopically identified clusters/groups are found to have 
halo masses in the range of 
$13 \lesssim$ log ($M_{200}/M_{\odot}$) $\lesssim 14$.}

\item {We also find elongated LSSs near three confirmed 
galaxy overdensities, 
which extend as large as 10 Mpc. 
The LSSs are found from the photometric redshift sample, 
but spectroscopically confirmed galaxies also follow the 
distribution of LSSs.}

\item {There is an anti-correlation between the quiescent 
galaxy fraction in clusters/groups and the significance of the 
surrounding LSS. The significance of LSSs is quantified 
using the FOF algorithm, expressed as the `FOF fraction'. 
This finding indicates that SF activity in galaxy 
overdensities is enhanced when the overdensities at 
$z \sim 1$ are surrounded by LSSs.}

\item {We find that more concentrated 
overdensities tend to have 
a higher quiescent galaxy fraction, with some exceptions. 
However this trend needs to be confirmed with a larger set of 
data in the future.} 

\item {Based on the results 3 and 4, we put forward the 
``Web-feeding model" 
for SF activities in overdensities at $z \sim 1$. 
In this model, the low $F_{q}$ of overdensities surrounded by LSS 
can be explained with inflow of gas and SF galaxies from the 
surrounding LSSs into the overdensities.
Once the supply of the gas and galaxies stops and the surrounding LSS disappears, 
galaxy overdensities become filled with quiescent galaxies that infall to 
cluster center producing the high concentration of overdensities. }

\end{enumerate}

Our results show that SF activities in galaxy overdensities at $0.55 < z < 1.3$ 
with $13.4 \lesssim$ log ($M_{200}/M_{\odot}$) $\lesssim 14$ are closely 
connected with the presence of LSSs. Future spectroscopy study of dense 
sampling of galaxies over a much larger area should reveal much clearer 
picture about the proposed connection between SF activity in clusters/groups 
and LSSs surrounding them.

\acknowledgments

This work was supported by the National Research Foundation of Korea (NRF) grant, 
No. 2017R1A3A3001362, funded by the Korea government (MSIP). 
M.H. acknowledges the support from Global PH.D Fellowship Program through the National Research Foundation of Korea (NRF) funded by the Ministry of Education (NRF-2013H1A2A1033110).
This paper includes data gathered with the 6.5 meter Magellan Telescopes located 
at Las Campanas Observatory, Chile, and the United Kingdom Infrared Telescope (UKIRT) 
which is supported by NASA and operated under an agreement among the University of 
Hawaii, the University of Arizona, and Lockheed Martin Advanced Technology 
Center; operations are enabled through the cooperation of the Joint Astronomy 
Centre of the Science and Technology Facilities Council of the U.K.

{\it Facilities:} \facility{Magellan/IMACS}.



\appendix

\section{Selection of Massive Structures based on Photometric Data}

There are several methods used to find massive structures of galaxies 
(e.g., galaxy clusters) at high redshift.
Among these, the selection method based on photometric redshift is 
an efficient way, especially considering increasing number of 
panchromatic photometric surveys 
\citep[e.g.,][]{tre07,eis08,kan15,lee15,kim16}.
Here, we test if this kind of photometric selection of massive 
structures is reliable --- i.e., how many photometrically-found 
overdensities are actually clusters or groups.

For this, we use the mock galaxy catalog of \citet{mer13}. 
From the mock catalog, we select four sky areas, each of which 
has a similar sky area with the UDS field.
Then we add noise to the spectroscopic redshift of each galaxy with 
the amount of the photometric redshift error from the observed data 
(see Figure~\ref{szpz}) assuming Gaussian distribution. 
From this mock `photometric redshift' catalog, we select 
overdensities following the same procedure as the one 
explained in Section~\ref{sample} and also in L15.

After identifying overdensities, we check the spectroscopic 
redshift distribution of each overdensity, finding $95 \%$ of 
the photometrically selected overdensities contain 
massive structures like galaxy clusters or groups. 

\begin{figure}
\centering
\includegraphics[scale=0.5]{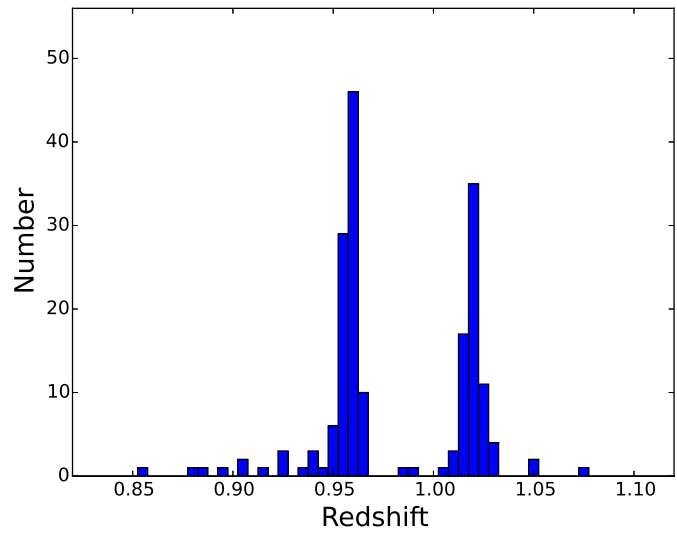}
\caption{An example of the spectroscopic redshift distribution of a 
photometrically selected mock overdensity from the mock 
photometric redshift catalog. $49 \%$ of photometrically selected 
model overdensities contain multiple peaks in their spectroscopic 
redshift distribution.}
\label{mpmodex}
\end{figure}

From this test, we also find that about half ($\sim 49 \%$) 
of photometrically found overdensities contain multiple peaks 
in their spectroscopic redshift distribution as shown 
in Figure ~\ref{mpmodex}. 
The existence of spatially overlapped multiple structures 
within a single photometrically selected overdensity is 
found in our sample as well (with a similar 
occurrence fraction), and also 
often found in other studies as well \citep[e.g.][]{van07,van09}.

\section{Simulation Test of Correlation between Halo Mass and Total Stellar Mass}
\label{msmhmodel}

Studies show that there exists a tight correlation between the mass of 
dark matter halos of clusters or groups and the total stellar mass of 
cluster/group member galaxies \citep[e.g.,][]{kim15,pat15,lin17,kra18}. 
Here, we check this correlation in model halos in the similar ways as 
used for our observationally selected clusters/groups.
The mock catalogs in APPENDIX A are used here.
For each model halo, we first select galaxies within the photometric 
redshift error range. Then, among these galaxies, fifteen galaxies are 
randomly selected, and we check if these galaxies are cluster members or not. 
For these fifteen galaxies, we derive the weight, $w_{i}$, 
to be the number of member galaxies divided by fifteen. Then we calculate 
the total stellar mass applying this weight. We repeat this procedure 
1000 times, and use the median stellar mass from these 1000 run in the 
following analysis.

\begin{figure}
\centering
 \includegraphics[scale=0.7]{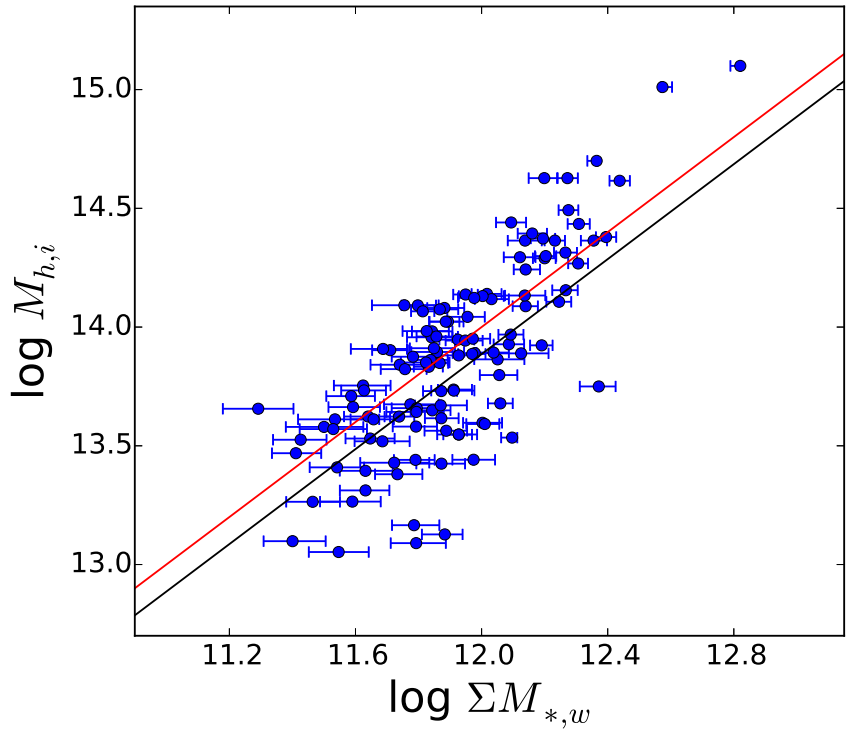}\\
\caption{Correlation between halo mass and 
the total stellar mass of overdensities found from the mock catalog. 
The total stellar mass is calculated using the same procedure done 
for analyzing observed overdensities (Section~\ref{odprop}). 
The red line is the line with log ($M_{halo}/M_{*,total}$)=2.0, 
while the black line shows the correlation between halo mass and 
total stellar mass from L15.}
 \label{smhmmck}
\end{figure}

The Figure~\ref{smhmmck} shows that the total 
stellar mass is proportional to the halo mass. 
The error bars in the figure show the scatter (the first 
and third quartiles) in the measured total stellar mass for each halo from 1000 
Monte Carlo runs. 
The median value of these scatters is 0.06 (same for both upper and lower scatter) 
in logarithmic scale. 
The median and the rms scatter of 
log ($M_{halo}/M_{*,total}$) are 2.00 and 0.25. The scatter of the median value 
of log ($\Sigma M_{*,i}$) is much smaller than the rms scatter in log ($M_{halo}/M_{*,total}$), 
and can be considered negligible when converting $M_{*,total}$ to $M_{halo}$.

The scatter in the $M_{halo}-M_{*,total}$ correlation is modest, which is 
well consistent with other studies \citep{pat15,kra18}.
The black line in this figure shows $M_{*,total}$ to $M_{200}$ ratio 
of 0.013 found in L15.

This correlation between $M_{*,total}$ and $M_{200}$ 
of clusters/groups indicates that $M_{*,total}$ 
can be a good proxy for cluster/group mass \citep{and12}. 
Therefore, in Section~\ref{odprop}, we provide $M_{200}$ values converted from the total 
stellar mass, applying the above mentioned best-fit correlation.
The dispersion in the correlation and scatter in $M_{*,total}$ 
from the Monte Carlo runs are also included (as a quadrature) in the error budget 
of $M_{200}$ (column (9) in Table 3).

\clearpage

\clearpage

\clearpage

\begin{deluxetable}{ccc}
\tablecolumns{3} \tablewidth{0pc} \tablecaption{Overdensity targets \label{cltarget}}
\tablehead{ \colhead{Name} & \colhead{$z_{phot}$} &
\colhead{$z_{spec}$ measured/targets$^{a}$}}

\startdata
   UDSOD1$^{b,d}$ & 0.89 & 20/24 \\      
   UDSOD2$^{b,d}$ & 0.89 & 15/16 \\
   UDSOD3$^{b,d}$ & 0.87 & 18/22 \\
   UDSOD4$^{b,d}$ & 1.24 & 12/15 \\
   UDSOD J0218-0441$^{b,e}$ & 1.21 & 5/6 \\ 
   UDSOD J0219-0456$^{b,e}$ & 1.25 & 7/9 \\
   UDSOD J0218-0458$^{c,e}$ & 1.04 & 4/12 \\
   UDSOD J0219-0451$^{c,e}$ & 1.36 & 2/6 \\
   UDSOD J0219-0443$^{c,e}$ & 1.37 & 3/7 \\
\enddata

\tablenotetext{a}{The number of spectroscopically observed targets with reliable $z_{spec}$ over the number of member galaxy candidates observed with IMACS. The objects with $z_{spec}$ are not necessarily the overdensity members.}
\tablenotetext{b}{Primary targets (red circles in Figure~\ref{qfrac})}
\tablenotetext{c}{Secondary targets (green diamonds in Figure~\ref{qfrac})}
\tablenotetext{d}{Confirmed overdensities}
\tablenotetext{e}{Overdensity candidates which are not confirmed, mostly due to the small number of galaxies with $z_{spec}$ identification.}

\end{deluxetable}

\clearpage

\begin{deluxetable}{cccccccc}
\tablecolumns{8} \tablewidth{0pc} \tablecaption{Target UDS Galaxies 
for our Magellan spectroscopic observation \label{tab1}}
\tablehead{ \colhead{ID} & \colhead{R.A. (J2000) }   &
\colhead{dec. (J2000)}   &
\colhead{$z_{spec}$}   &
\colhead{$flag$}   &
\colhead{$z_{phot}$}   &
\colhead{$R_{AB}$}  & 
\colhead{Cluster ID} \\
\colhead{(1)} & \colhead{(2)} & \colhead{(3)} & \colhead{(4)} & 
\colhead{(5)} & \colhead{(6)} & \colhead{(7)} & \colhead{(8)}}

\startdata
UDS-228531 &  34.64690 & -5.04229 & --- & --- & 0.865  &   24.49 &  ---\\
UDS-229694 & 34.64075 & -5.03853 &  0.8116 &  b & 0.918  &    24.47 &  --- \\
UDS-231087 &  34.64948 & -5.03574 & --- & --- & 0.848  &	23.97 &  ---\\
UDS-234710 & 34.63850 & -5.02716 &  0.8743 &  b & 0.832  &    23.34 & 3 \\
UDS-235482 & 34.63426 & -5.02524 &  0.8753 &  b & 0.855  &    23.73 & 3 \\  
UDS-235622 & 34.63008 & -5.02307 &  0.8972 &  b & 0.833  &    24.49 &  --- \\  
UDS-236426 & 34.63672 & -5.02090 &  0.7986 &  a & 0.819  &    23.27 &  --- \\  
UDS-236747 & 34.60576 & -5.01930 &  0.9114 &  b & 0.871  &    23.90 &  --- \\  
UDS-238512 & 34.63570 & -5.01564 &  0.8723 &  b & 0.884  &    23.06 & 3 \\
UDS-238531 & 34.62882 & -5.01538 &  0.8772 &  b & 0.910  &    23.38 & 3 \\
UDS-239219 &  34.64142 & -5.01153 & --- & --- & 0.826  &	24.47 &  ---\\
UDS-239439 & 34.63197 & -5.01353 &  0.8750 &  a & 0.881  &    22.79 & 3 \\  
UDS-240385 & 34.65060 & -5.00833 &  0.8779 &  b & 0.857  &    24.25 & 3 \\  
UDS-240908 & 34.62673 & -5.00883 &  0.8693 &  b & 0.853  &    23.22 & 3 \\  
UDS-241334 & 34.63706 & -5.00619 &  0.8745 &  b & 0.830  &    24.42 & 3 \\  
UDS-243040 & 34.66692 & -5.00251 &  0.9298 &  b & 0.882  &    23.11 &  --- \\   
UDS-243163 &  34.60729 & -5.00077 & --- & --- & 0.911  &   23.17 &  ---\\
UDS-244398 &  34.60018 & -4.99799 & --- & --- & 1.004  &   23.80 &  ---\\
UDS-245018 & 34.63802 & -4.99461 &  0.8676 &  a & 0.836  &    23.37 & 3 \\
UDS-246462 & 34.64983 & -4.99181 &  0.9884 &  b & 0.875  &    24.00 &  --- \\  
UDS-247109 & 34.63007 & -4.98942 &  0.8710 &  b & 0.839  &    23.68 & 3 \\  
UDS-249003 &  34.58368 & -4.98388 & --- & --- & 1.056  &   24.13 & --- \\ 
UDS-249146 & 34.64603 & -4.98557 &  0.9176 &  a & 0.866  &    23.35 & ---  \\  
UDS-249702 & 34.65040 & -4.98120 &  0.8604 &  b & 0.887  &    24.36 & 3 \\  
UDS-249779 &  34.58160 & -4.98399 & --- & --- & 1.061  &   23.30 &  ---\\  
UDS-251072 &  34.59115 & -4.98098 & --- & --- & 1.055  &   22.86 &  ---\\
UDS-253772 &  34.59892 & -4.97094 & --- & --- & 1.069  &   24.48 &  ---\\
UDS-254106 & 34.75435 & -4.96990 &  1.2354 &  b & 1.199  &    23.36 &  --- \\  
UDS-254869 & 34.74907 & -4.96749 &  1.2904 &  b & 1.239  &    23.33 &  --- \\  
UDS-254902 & 34.59714 & -4.96754 &  1.0806 &  a & 1.018  &    23.32 &  --- \\  
UDS-254980 &  34.59380 & -4.96865 & --- & --- & 1.012  &   23.19 &  ---\\
UDS-255098 &  34.61484 & -4.96656 & --- & --- & 1.016  &   24.00 &  ---\\
UDS-256270 &  34.78955 & -4.96476 & --- & --- & 1.286  &   23.48 &  --- \\ 
UDS-256378 &  34.61165 & -4.96365 & --- & --- & 1.020  &   24.37 & ---  \\
UDS-256643 &  34.59917 & -4.96423 & --- & --- & 1.016  &   22.18 & ---  \\
UDS-257227 &  34.57320 & -4.96207 & --- & --- & 1.015  &   23.94 & ---  \\
UDS-258072 & 34.69135 & -4.95880 &  0.8733 &  a & 0.851  &    23.28 &   ---  \\  
UDS-260259 & 34.77536 & -4.95349 &  0.7928 &  b & 0.821  &    22.68 &  ---  \\  
UDS-260634 & 34.57821 & -4.95170 &  0.9808 &  b & 0.998  &    24.41 &  ---  \\  
UDS-261211 &  34.80573 & -4.94992 & --- & --- & 1.297  &   23.98 &  --- \\ 
UDS-261778 & 34.55985 & -4.94946 &  0.8751 &  b & 0.834  &    22.67  &  --- \\  
UDS-262504 & 34.72490 & -4.94716 &  1.2915 &  b & 1.182  &    22.32  &  --- \\  
UDS-263004 &  34.80919 & -4.94550 & --- & --- & 1.253  &   24.02 &  --- \\
UDS-263423 & 34.51558 & -4.94456 &  0.8735 &  b & 0.875  &    22.82 &  ---  \\  
UDS-264476 &  34.83005 & -4.94270 & --- & --- & 1.214  &   23.20 &  --- \\
UDS-265946 & 34.56188 & -4.93772 &  0.9607 &  b & 0.913  &    23.3 & ---  \\  
UDS-266173 & 34.78123 & -4.93566 &  0.9268 &  b & 0.932  &    23.38 &   --- \\   
UDS-267294 & 34.75909 & -4.93306 &  1.3241 &  b & 1.242  &    23.33 &  ---  \\  
UDS-267966 & 34.78945 & -4.93104 &  1.2904 &  b & 1.225  &    23.58 & ---   \\  
UDS-268909 &  34.82768 & -4.92870 & --- & --- & 1.273  &   23.69 & ---  \\
UDS-269070 & 34.54436 & -4.92888 &  1.0589 &  b & 1.024  &    22.84  & ---  \\  
UDS-270063 & 34.83935 & -4.92447 &  1.2794 &  b & 1.221  &    24.17 &  ---  \\  
UDS-270445 & 34.76143 & -4.92622 &  0.7905 &  a & 0.832  &    21.85 &  ---  \\  
UDS-271576 & 34.82034 & -4.92197 &  1.2185 &  b & 1.252  &    23.80 &  ---  \\  
UDS-271680 & 34.53366 & -4.92030 &  1.1026 &  b & 1.021  &    23.48 &  ---  \\  
UDS-272156 & 34.69100 & -4.91928 &  1.2817 &  b & 1.213  &    23.31  & ---  \\  
UDS-272637 &  34.52046 & -4.91798 & --- & --- & 1.233  &   23.21 & ---  \\
UDS-272659 & 34.82060 & -4.91882 &  1.3764 &  b & 1.231  &    23.56  & ---  \\  
UDS-272708 & 34.84207 & -4.91679 &  0.8746 &  a & 0.873  &    24.11  & 1-a \\  
UDS-273396 & 34.78667 & -4.91462 &  1.3325 &  b & 1.305  &    24.37  &  ---\\   
UDS-274407 & 34.85949 & -4.91211 &  0.9624 &  b & 0.928  &    23.59  & 1-c \\  
UDS-274524 & 34.53860 & -4.91359 &  0.8754 &  a & 0.877  &    22.98  & --- \\  
UDS-275441 & 34.51866 & -4.90938 &  0.9622 &  b & 0.899  &    22.87  & --- \\  
UDS-275596 & 34.71127 & -4.90992 &  0.9299 &  b & 0.863  &    23.12  & --- \\  
UDS-276058 & 34.85797 & -4.90754 &  0.8756 &  b & 0.894  &    23.35  & 1-a \\  
UDS-276882 & 34.53532 & -4.90534 &  1.2711 &  b & 1.199  &    23.42  & --- \\  
UDS-277386 & 34.68735 & -4.90490 &  0.9217 &  b & 0.923  &    22.62  & --- \\  
UDS-277731 & 34.73405 & -4.90280 &  1.2312 &  b & 1.181  &    23.26  & --- \\  
UDS-277735 & 34.83523 & -4.90318 &  0.9204 &  b & 0.910  &    24.17  & 1-b \\   
UDS-277802 & 34.51316 & -4.90243 &  0.8698 &  b & 0.842  &    23.27  & --- \\  
UDS-278440 & 34.73294 & -4.90075 &  1.2842 &  b & 1.216  &    23.34  & --- \\  
UDS-279317 & 34.85374 & -4.89782 &  1.4026 &  b & 1.331  &    23.48  & --- \\  
UDS-279506 & 34.71846 & -4.89810 &  1.0416 &  b & 1.020  &    23.06  & --- \\  
UDS-280301 & 34.84224 & -4.89478 &  0.9207 &  b & 0.886  &    23.65  & 1-b \\  
UDS-281576 & 34.87351 & -4.89281 &  0.9634 &  b & 0.940  &    23.30  & 1-c \\  
UDS-281951 & 34.86235 & -4.89081 &  0.9634 &  b & 0.923  &    22.53  & 1-c \\  
UDS-282637 & 34.82302 & -4.88770 &  0.9175 &  b & 0.908  &    23.69  & 1-b \\  
UDS-282670 & 34.52402 & -4.88735 &  1.2808 &  b & 1.169  &    23.37  &  ---\\  
UDS-282683 & 34.70849 & -4.88985 &  0.8510 &  b & 0.994  &    22.96  & --- \\  
UDS-282725 &  34.86685 & -4.89073 & --- & --- & 0.924  &   23.37 & --- \\
UDS-283427 & 34.86960 & -4.88744 &  0.9202 &  b & 0.923  &    23.38  & 1-b \\  
UDS-283892 & 34.83882 & -4.88500 &  0.8704 &  a & 0.839  &    22.13  & 1-a \\  
UDS-284432 &  34.86063 & -4.88256 & --- & --- & 0.869  &   22.97  & --- \\ 
UDS-284464 & 34.81072 & -4.88463 &  0.8900 &  a & 0.901  &    22.84  & 1-a \\  
UDS-284743 & 34.54264 & -4.88225 &  0.8488 &  a & 0.889  &    23.17  & --- \\  
UDS-285335 & 34.69280 & -4.88131 &  0.8866 &  b & 0.814  &    23.13  & --- \\  
UDS-285442 & 34.84924 & -4.88002 &  0.9181 &  b & 0.865  &    23.41  & 1-b \\  
UDS-285528 &  34.82124 & -4.88199 & --- & --- & 0.923  &   23.87 & --- \\
UDS-286367 & 34.84731 & -4.87751 &  1.4033 &  b & 1.378  &    23.62  &  ---\\  
UDS-286518 & 34.83908 & -4.87939 &  0.9221 &  b & 0.879  &    22.41  & 1-b \\  
UDS-286588 & 34.73166 & -4.87695 &  0.9712 &  b & 0.927  &    23.12  &  ---\\  
UDS-287242 & 34.85786 & -4.87521 &  0.8707 &  b & 0.845  &    23.62  & 1-a \\  
UDS-287852 & 34.86231 & -4.87299 &  0.9161 &  b & 0.873  &    23.19  & 1-b \\  
UDS-287888 &  34.84760 & -4.87508 & --- & --- & 1.331  &   23.51 & --- \\
UDS-288765 &  34.71830 & -4.87180 & --- & --- & 0.928  &   22.93 & --- \\
UDS-288805 &  34.86459 & -4.87180 & --- & --- & 0.854  &   23.07 & --- \\
UDS-289121 & 34.87242 & -4.86937 &  0.9610 &  b & 0.937  &    22.23  & 1-c \\  
UDS-289271 & 34.81006 & -4.86798 &  0.8886 &  b & 0.887  &    24.03  & 1-a \\  
UDS-290427 & 34.71991 & -4.86649 &  0.9266 &  b & 0.874  &    23.15  &  ---\\  
UDS-290579 & 34.85003 & -4.86603 &  0.9626 &  a & 0.930  &    22.49  & 1-c \\  
UDS-291394 &  34.86009 & -4.86516 & --- & --- & 0.914  &   22.97 & --- \\
UDS-291643 & 34.70777 & -4.86344 &  1.3939 &  b & 1.374  &    21.90  &  ---\\  
UDS-292029 &  34.84612 & -4.86106 & --- & --- & 1.301  &   23.93 & --- \\
UDS-292229 &  34.85287 & -4.86294 & --- & --- & 1.349  &   23.75 & --- \\
UDS-292371 & 34.74278 & -4.86118 &  1.0560 &  a & 1.034  &    23.25  & --- \\  
UDS-295775 & 34.84679 & -4.85103 &  0.9195 &  b & 0.894  &    23.98  & 1-b \\  
UDS-295803 & 34.75636 & -4.85083 &  0.9310 &  b & 0.925  &    23.22  & ---  \\  
UDS-296910 & 34.67932 & -4.84771 &  1.0094 &  b & 0.901  &    23.28  & ---  \\  
UDS-298584 &  34.86310 & -4.84316 & --- & --- & 1.392  &   23.97  & --- \\
UDS-299358 &  34.64368 & -4.84205 & --- & --- & 0.842  &   23.38  & --- \\
UDS-300797 & 34.71334 & -4.83955 &  0.8411 &  b & 0.862  &    22.50  & ---  \\    
UDS-301305 &  34.70603 & -4.83559 & --- & --- & 1.082  &   23.47  & --- \\
UDS-303047 & 34.66753 & -4.83115 &  0.8478 &  a & 0.846  &    22.69  & ---  \\  
UDS-303667 & 34.72319 & -4.82904 &  1.0554 &  b & 1.021  &    22.75  & ---  \\  
UDS-303953 & 34.55543 & -4.82916 &  1.0770 &  a & 0.998  &    23.46  & ---  \\  
UDS-304202 & 34.62020 & -4.82734 &  1.1809 &  b & 1.064  &    23.05  & ---  \\  
UDS-304566 & 34.51992 & -4.82607 &  1.1561 &  b & 1.024  &    22.65  & ---  \\   
UDS-304585 &  34.71726 & -4.82676 & --- & --- & 1.229  &   22.84 & ---  \\
UDS-305878 & 34.58210 & -4.82342 &  0.7772 &  a & 0.830  &    23.24   & --- \\   
UDS-306056 & 34.70878 & -4.82250 &  1.0189 &  b & 1.008  &    23.08   & --- \\  
UDS-306573 & 34.70792 & -4.81634 &  0.8679 &  b & 0.835  &    21.70  & ---  \\  
UDS-307198 & 34.56587 & -4.81928 &  0.8401 &  b & 0.790  &    22.88  & ---  \\  
UDS-307542 & 34.71526 & -4.81900 &  1.3859 &  b & 1.365  &    23.37  & ---  \\  
UDS-308248 & 34.83765 & -4.81644 &  0.9634 &  b & 0.937  &    23.33  & ---  \\  
UDS-308415 & 34.57049 & -4.81708 &  1.0523 &  b & 0.999  &    23.25   & ---  \\   
UDS-309250 & 34.54344 & -4.81307 &  0.8580 &  b & 0.838  &    22.97 & 2-b \\  
UDS-309800 & 34.83124 & -4.81421 &  0.8840 &  b & 0.925  &    23.49  & 1-a \\  
UDS-309849 & 34.82151 & -4.81154 &  0.8251 &  a & 0.842  &    22.63   &   ---\\  
UDS-310171 & 34.67828 & -4.81322 &  0.8456 &  b & 0.813  &    21.01  &  ---  \\  
UDS-310820 & 34.86150 & -4.80826 &  0.9019 &  b & 0.901  &    23.43  &  ---  \\  
UDS-310841 & 34.59476 & -4.81040 &  0.8677 &  b & 0.806  &    22.64   & ---  \\  
UDS-310952 &  34.65653 & -4.80873 & --- & --- & 0.833  &   23.20  & ---  \\
UDS-312133 & 34.54492 & -4.80747 &  0.8408 &  a & 0.813  &    22.66  & ---   \\  
UDS-313589 & 34.58816 & -4.80331 &  1.0986 &  b & 1.073  &    23.18  & ---   \\  
UDS-313974 & 34.87246 & -4.80101 &  0.8316 &  b & 0.880  &    23.14  &  ---  \\  
UDS-314141 & 34.59129 & -4.79898 &  0.8572 &  b & 0.869  &    23.47  & ---   \\  
UDS-314225 & 34.70618 & -4.80062 &  1.0504 &  b & 1.203  &    23.02   & ---  \\  
UDS-314832 & 34.85776 & -4.79807 &  0.8761 &  b & 0.884  &    23.10  & ---   \\  
UDS-314889 & 34.75458 & -4.79716 &  1.3178 &  b & 1.215  &    23.42  & ---   \\  
UDS-315323 & 34.64763 & -4.79703 &  1.0831 &  b & 1.068  &    23.48  & ---   \\  
UDS-316538 & 34.53368 & -4.79317 &  0.8825 &  a & 0.870  &    22.26  & ---   \\  
UDS-316584 & 34.72644 & -4.79254 &  0.8343 &  a & 0.881  &    23.15  & ---   \\  
UDS-316613 & 34.71928 & -4.79456 &  0.8719 &  b & 0.900  &    23.35  & ---   \\   
UDS-316643 &  34.87366 & -4.79501 & --- & --- & 0.883  &   23.37  & ---  \\
UDS-317085 & 34.83352 & -4.79248 &  1.1513 &  b & 1.179  &    23.34  &  ---  \\  
UDS-317860 & 34.76135 & -4.78958 &  1.4576 &  b & 1.369  &    23.26  &  ---  \\  
UDS-318479 & 34.62929 & -4.79089 &  1.0854 &  b & 1.256  &    23.11  &  ---  \\   
UDS-318699 & 34.66456 & -4.78748 &  1.2713 &  b & 1.176  &    22.63  &  ---  \\  
UDS-319198 &  34.87153 & -4.78745 & --- & --- & 1.014  &   22.67  &  --- \\ 
UDS-319682 & 34.60902 & -4.78522 &  0.8878 &  b & 0.842  &    22.11  &  ---  \\  
UDS-320687 & 34.78495 & -4.78144 &  0.8509 &  b & 0.867  &    23.25  &  ---  \\  
UDS-321054 &  34.62932 & -4.78077 & --- & --- & 0.813  &   23.48  &  --- \\
UDS-321110 & 34.52822 & -4.78036 &  0.8592 &  a & 0.794  &    22.34  &  2-b \\  
UDS-321367 & 34.65572 & -4.77906 &  0.8463 &  a & 0.864  &    22.96  &   --- \\  
UDS-322051 & 34.78581 & -4.77935 &  0.8503 &  b & 0.888  &    22.31  & ---   \\  
UDS-322278 & 34.67725 & -4.77674 &  1.0833 &  b & 1.011  &    23.30  & ---   \\  
UDS-322652 & 34.84634 & -4.77655 &  1.0497 &  a & 1.018  &    22.85  & ---   \\  
UDS-323047 & 34.56600 & -4.77644 &  0.8691 &  a & 0.798  &    22.23  &  2-b \\  
UDS-323500 & 34.70461 & -4.77254 &  0.8722 &  b & 0.862  &    23.50  & ---   \\  
UDS-324211 & 34.59881 & -4.77360 &  0.8804 &  a & 0.822  &    21.93  & ---   \\  
UDS-324306 & 34.84994 & -4.77155 &  1.0475 &  b & 0.936  &    23.26  & ---   \\  
UDS-324514 &  34.63706 & -4.76977 & --- & --- & 1.034  &   23.49  & ---  \\
UDS-324868 & 34.53780 & -4.77107 &  0.9555 &  a & 0.915  &    22.78  & ---   \\  
UDS-325191 & 34.75730 & -4.76932 &  0.8425 &  b & 0.801  &    22.81  & ---   \\   
UDS-325205 & 34.51362 & -4.76841 &  1.3110 &  b & 1.229  &    23.28  &  ---  \\  
UDS-325328 & 34.52161 & -4.76845 &  0.8049 &  b & 0.791  &    23.46  & ---   \\   
UDS-325804 & 34.61840 & -4.76758 &  0.8812 &  a & 0.862  &    22.99  & ---   \\  
UDS-325951 & 34.81144 & -4.76622 &  0.9325 &  a & 0.903  &    23.39  & ---   \\  
UDS-326679 & 34.72874 & -4.76341 &  1.1899 &  b & 1.197  &    23.31  & ---   \\  
UDS-326940 & 34.66651 & -4.76529 &  0.9734 &  b & 0.850  &    22.58   & ---  \\  
UDS-327832 & 34.80331 & -4.76028 &  0.8255 &  a & 0.814  &    22.53  & ---   \\  
UDS-328018 & 34.64759 & -4.76059 &  0.8805 &  a & 0.834  &    22.79  & ---  \\  
UDS-328084 & 34.52312 & -4.76191 &  0.8405 &  a & 0.872  &    22.79  &  2-a \\  
UDS-328399 & 34.52490 & -4.75969 &  0.8645 &  b & 0.851  &    22.72   &  2-b \\  
UDS-329293 & 34.63724 & -4.75814 &  0.8811 &  b & 0.850  &    22.77  & ---   \\  
UDS-330370 & 34.53937 & -4.75354 &  0.7551 &  a & 0.792  &    22.20   & ---  \\  
UDS-330393 &  34.56446 & -4.75353 & --- & --- & 1.007  &   23.25  & ---  \\
UDS-332031 &  34.54285 & -4.75128 & --- & --- & 0.875  &   23.46  & ---  \\
UDS-332415 & 34.52670 & -4.75025 &  0.8651 &  a & 0.809  &    22.32  &  2-b \\  
UDS-332893 & 34.82824 & -4.74734 &  1.2810 &  b & 1.350  &    24.49  & ---   \\  
UDS-333517 & 34.84335 & -4.74426 &  1.3698 &  b & 1.335  &    24.00  & ---   \\  
UDS-334546 & 34.50953 & -4.74210 &  0.8417 &  a & 0.869  &    23.66  &  2-a \\  
UDS-334552 & 34.51872 & -4.74485 &  0.8669 &  b & 0.885  &    22.54   &  2-b \\  
UDS-334823 &  34.82278 & -4.74166 & --- & --- & 1.393  &   24.47  &  --- \\  
UDS-334863 &  34.53096 & -4.74227 & --- & --- & 0.816  &   22.37  & ---  \\
UDS-338191 & 34.71060 & -4.73169 &  1.3557 &  b & 1.267  &    23.43   & ---  \\  
UDS-338315 & 34.51740 & -4.73168 &  0.8368 &  b & 0.825  &    23.34   &  2-a \\  
UDS-339117 & 34.69383 & -4.72919 &  1.4038 &  b & 1.260  &    23.21  & ---   \\  
UDS-339310 &  34.85504 & -4.72769 & --- & --- & 1.307  &   24.39  & ---  \\
UDS-339814 & 34.85943 & -4.72999 &  0.8893 &  b & 1.406  &    24.02  & ---   \\  
UDS-340102 &  34.52670 & -4.72659 & --- & --- & 0.809  &   24.26  &  --- \\
UDS-340421 & 34.55467 & -4.72521 &  0.8046 &  a & 0.797  &    23.08  & ---   \\   
UDS-341240 &  34.66593 & -4.72263 & --- & --- & 1.297  &   24.11  & ---  \\
UDS-341435 & 34.78862 & -4.72228 &  0.8246 &  a & 0.817  &    23.22  & ---   \\   
UDS-341894 &  34.52547 & -4.72222 & --- & --- & 0.793  &   23.21  & ---  \\
UDS-342357 & 34.70268 & -4.72055 &  1.3611 &  b & 1.180  &    23.48  & ---   \\  
UDS-343198 & 34.85323 & -4.71673 &  1.3258 &  b & 1.302  &    24.14  & ---   \\  
UDS-343250 & 34.69148 & -4.71830 &  1.2943 &  b & 1.190  &    23.34   &  4 \\  
UDS-343777 & 34.53057 & -4.71615 &  0.9203 &  b & 0.877  &    24.45  & ---   \\  
UDS-343860 & 34.69462 & -4.71620 &  1.2901 &  a & 1.182  &    22.63  & 4\\  
UDS-344737 &  34.70541 & -4.71390 & --- & --- & 1.213  &   23.67  &  --- \\
UDS-344878 & 34.68667 & -4.71312 &  1.2927 &  b & 1.225  &    22.61  &  4 \\  
UDS-345732 & 34.53461 & -4.71216 &  0.8385 &  b & 0.815  &    22.67  &  2-a \\  
UDS-345854 & 34.53396 & -4.71149 &  0.8400 &  b & 0.808  &    21.93  &  2-a \\  
UDS-346190 &  34.82927 & -4.71146 & --- & --- & 1.339  &   24.48  &  --- \\
UDS-346271 &  34.69253 & -4.71137 & --- & --- & 1.255  &   23.39  &  --- \\
UDS-346514 &  34.66734 & -4.71027 & --- & --- & 1.240  &   23.76  &  --- \\
UDS-346705 & 34.76465 & -4.70954 &  0.8204 &  b & 0.829  &    22.51  &   --- \\  
UDS-346750 &  34.65765 & -4.70804 & --- & --- & 1.184  &   23.52  &  --- \\
UDS-347025 & 34.79855 & -4.70749 &  0.9009 &  b & 0.864  &    22.60  &  ---  \\  
UDS-347036 & 34.69412 & -4.70580 &  1.3135 &  b & 1.285  &    23.92   &  4\\   
UDS-348003 & 34.62996 & -4.70317 &  1.2905 &  b & 1.233  &    23.80  &  4 \\  
UDS-348269 & 34.79899 & -4.70331 &  0.9604 &  b & 0.926  &    22.66   &  --- \\  
UDS-349053 & 34.78108 & -4.70123 &  0.9602 &  b & 1.193  &    23.18  &   --- \\  
UDS-349372 &  34.59794 & -4.70021 & --- & --- & 1.247  &   24.45  &  --- \\
UDS-349733 & 34.71157 & -4.69821 &  1.2940 &  b & 1.233  &    23.79  &  4 \\  
UDS-350345 & 34.60964 & -4.69812 &  1.3077 &  b & 1.237  &    24.29  &  ---  \\  
UDS-350773 & 34.75123 & -4.69593 &  0.9204 &  b & 0.890  &    23.42  &   --- \\  
UDS-351328 & 34.63979 & -4.69460 &  1.3034 &  b & 1.257  &    23.46  &  4 \\  
UDS-352105 & 34.66459 & -4.69271 &  1.3205 &  b & 1.237  &    23.17  &  4 \\  
UDS-352892 & 34.68014 & -4.69021 &  1.2823 &  b & 1.183  &    23.18   &  4\\  
UDS-354242 & 34.74165 & -4.68610 &  0.8992 &  a & 0.895  &    23.16  & --- \\  
UDS-354710 & 34.76780 & -4.68804 &  0.9003 &  a & 0.898  &    22.16  &  --- \\  
UDS-355005 & 34.61463 & -4.68410 &  1.2174 &  b & 1.236  &    24.22  &  ---  \\  
UDS-358074 & 34.61502 & -4.67741 &  1.2778 &  b & 1.216  &    23.84  &  --- \\
---    & 34.82014 & -4.81893 &  0.0000 & star & --- & 16.88$^{a}$  & ---  \\
---    & 34.65724 & -4.94467 &  0.0000 & star & --- & 15.78$^{a}$  &  --- \\ 			
\enddata

\tablecomments{\\ (1) Galaxy ID. \\ (2) R.A. in degree. \\ (3) Declination in degree. \\ 
(4) Spectroscopic redshift. \\ (5) Quality flag (a: multiple features, b: single feature). \\ 
(6) Photometric redshift. \\ 
(7) Subaru $R$-band magnitude. \\ 
(8) Cluster/group membership.}

\tablenotetext{a}{SDSS r-band magnitude}

\end{deluxetable}

\clearpage

\begin{deluxetable}{ccccccccccccc}
\rotate
\tablecolumns{13} \tabletypesize{\footnotesize} \tablewidth{0pc} \tablecaption{Properties of galaxy clusters/groups \label{tab2}}
\tablehead{ \colhead{Name} & \colhead{R.A. (J2000) }   &
\colhead{dec. (J2000)}   &
\colhead{$z_{cl}$}   &
\colhead{$Projected Radius$}   &
\colhead{$N_{spec}$}   &
\colhead{$w_i$}   &
\colhead{log ($M_{*,total}$)}  &
\colhead{log $M_{200}$}  &
\colhead{$N_{g}$}  &
\colhead{$F_{quies.}$} &
\colhead{$R_{200}$} &
\colhead{$Concent.$} \\
\colhead{(1)} & \colhead{(2)} & \colhead{(3)} & \colhead{(4)} & 
\colhead{(5)} & \colhead{(6)} & \colhead{(7)} & \colhead{(8)} &
\colhead{(9)} & \colhead{(10)} & \colhead{(11)} & \colhead{(12)} & 
\colhead{(13)}}

\startdata
 UDSOD1-a  &  34.84044 &  -4.88481 &  $0.8751 \pm 0.0080$  &  1.0  &  6  
 & --- & --- & --- & --- & --- & --- & --- \\
 ---   &  ---  &  ---  &  ---              &  2.0  &  7  & 
 0.35 & 11.7 & $13.7 \pm 0.3$  & 43 & 0.09 & 540 & 0.19 \\
 UDSOD1-a1 & ---  &  ---  &  --- &  1.0 & 4 & 
 0.20 & 11.0 & $13.0 \pm 0.3$ & 9 & 0.16 & --- & --- \\
 UDSOD1-a2 & --- & --- & --- & 2.0 & 3 & 
 0.15 & 11.3 & $13.3 \pm 0.3$ & 12 & 0.26 & --- & --- \\
 UDSOD1-b  &  34.84452 &  -4.88373 &  $0.9198 \pm 0.0018$  &  1.0  &  8  & 
 0.40 & 11.4 & $13.4 \pm 0.3$ & 11 & 0.27 & 420 & 0.19 \\
 UDSOD1-c  &  34.86235 &  -4.89081 &  $0.9626 \pm 0.0009$  &  1.0  &  5  & 
 0.25 & 10.5 & $12.5 \pm 0.3$ & 6 & 0.00 & 220 & 0.27 \\
 UDSOD2-a  &  34.52312 &  -4.73168 &  $0.8400 \pm 0.0017$  &  1.0  &  5  & 
 0.33 & 11.4 & $13.4 \pm 0.3$ & 11 & 0.44 & 460 & 0.35 \\
 UDSOD2-b  &  34.52580 &  -4.75497 &  $0.8648 \pm 0.0029$  &  1.0  &  4  & 
 --- & --- & --- & --- & --- & --- & --- \\
 ---   & ---  & ---  &  --- &  1.5  &  6  & 0.40 & 11.7 & $13.7 \pm 0.3$ & 22 & 0.53 & 540 & 0.22 \\
 UDSOD3    &  34.63638 &  -5.00953 &  $0.8731 \pm 0.0045$  &  1.0  & 11(16)$^{a}$  & 
 0.61 & 12.0 & $14.0 \pm 0.3$ & 56 & 0.51 & 690 & 0.34 \\
 UDSOD4    &  34.69148 &  -4.70580 &  $1.2940 \pm 0.0126$  &  1.0  &  7  & 
 --- & --- & --- & --- & --- & --- & --- \\
---   &  ---  & ---  & ---              &  2.0  &  9  & 
0.75 & 12.0 & $14.0 \pm 0.3$ & 59 & 0.40 & 590 & 0.37 \\
 UDSOD4$^{b}$ & --- & --- & --- & 2.0 & 7 &
 0.58 & 11.7 & $13.7 \pm 0.3$ & 24 & 0.27 & --- & --- \\
\enddata

\tablecomments{\\ (1) Galaxy cluster name. \\ (2) R.A. in degree. \\ (3) Declination in degree. \\ 
(4) Cluster Redshift. \\ (5) Projected radius in Mpc. \\ (6) Number of galaxies with spectroscopic redshift within the projected radius indicated in column (5). \\ 
(7) Weight computed as the number of spectroscopically confirmed members divided by the 
sum of the numbers of spectroscopic members and outliers. \\ 
(8) Total stellar mass (in M$_{\odot}$).
\\ (9) Halo mass calibrated from the correlation between total stellar mass and halo mass 
of model clusters derived in Appendix B (in M$_{\odot}$). \\
(10) Number of member galaxies corrected by applying weights 
for photo-$z$ selected members (see text for detailed explanation).
\\ (11) Quiescent galaxy fraction within $R_{200}$ (applying weights for 
photo-$z$ members).\\ 
(12) Halo radius calculated from $M_{200}$ (in kpc).\\
(13) Halo Concentration.}
\tablenotetext{a}{Number in the parenthesis is the total number including the 
spectroscopic redshifts from literature.}
\tablenotetext{b}{Properties when two outliers in UDSOD4 are treated as non-members.}

\end{deluxetable}

\clearpage

\begin{deluxetable}{cccccc}
\tablecolumns{6} \tablewidth{0pc} \tablecaption{Properties of galaxy cluster candidates \label{clcndprop}}
\tablehead{ \colhead{Name} & \colhead{$z_{phot}$}   & \colhead{log $M_{200}$}   &
 \colhead{$z_{spec}$}   & \colhead{log $M_{200}$}   & \colhead{Other Names} \\
 \colhead{} & \colhead{} & \colhead{} & \colhead{(G18)} &
 \colhead{(F10)} & \colhead{} \\
 \colhead{(1)} & \colhead{(2)} & \colhead{(3)} & \colhead{(4)} & 
 \colhead{(5)} & \colhead{(6)}}
\startdata
 UDSOD J0218-0515 & 0.626 & 13.9 & 0.6453 & --- & C2$^{c}$ \\
 UDSOD J0217-0524 & 0.629 & 13.7 & 0.6451 & 13.9 & SXDF08XGG$^{b}$,U9$^{c}$ \\
 UDSOD J0216-0509 & 0.633 & 13.8 & --- & --- & --- \\
 UDSOD J0217-0513 & 0.639 & 14.1 & 0.6470 & 14.2 & 1A$^{a}$,SXDF69XGG$^{b}$,C1$^{c}$ \\
 UDSOD J0218-0524 & 0.644 & 14.0 & 0.6459 & 14.0 & SXDF07XGG$^{b}$,U2$^{c}$ \\
 UDSOD J0218-0458 & 0.659 & 13.6 & 0.6476 & --- & 3$^{a}$,U3$^{c}$ \\
 UDSOD J0217-0527 & 0.677 & 14.0 & 0.6937 & 13.8 & SXDF04XGG$^{b}$,U1N$^{c}$ \\
 UDSOD J0218-0507 & 0.684 & 13.7 & --- & --- & --- \\
 UDSOD J0217-0441 & 0.685 & 13.5 & --- & --- & --- \\
 UDSOD J0219-0449 & 0.744 & 13.4 & --- & --- & --- \\
 UDSOD J0217-0530 & 0.774 & 13.6 & --- & --- & --- \\
 UDSOD J0217-0505 & 0.791 & 13.8 & --- & --- & --- \\
 UDSOD J0219-0505 & 0.865 & 13.8 & --- & 13.9 & SXDF52XGG$^{b}$ \\
 UDSOD J0216-0544 & 0.889 & 13.8 & --- & --- & --- \\ 
\enddata

\tablecomments{\\ (1) Name. \\ (2) Photometric redshift. \\ 
(3) Halo mass calibrated from total stellar mass. \\
(4) Spectroscopic redshift from G18. \\
(5) Halo mass from F10. \\
(6) Other names from various references.}

\tablenotetext{a}{Name from VB06.}
\tablenotetext{b}{Name from F10.}
\tablenotetext{c}{Name from G18.}

\end{deluxetable}

\clearpage

\end{document}